\documentclass[twocolumn]{aastex62}
\usepackage{graphicx}
\usepackage{rotating}

\begin{document}
\title{Atomic Absorption Line Diagnostics for the Physical Properties of Red Supergiants}

\author{Brooke Dicenzo}
\affil{University of Washington, Seattle, WA 98195}
\email{dicenzob@uw.edu}

\author{Emily M. Levesque}
\affil{Department of Astronomy, University of Washington, Seattle, WA 98195, USA}
\email{emsque@uw.edu}

\begin{abstract}
Red supergiants (RSGs) are evolved massive stars that represent extremes, in both their physical sizes and their cool temperatures, of the massive star population. Effective temperature ($T_{\rm eff}$) is the most critical physical property needed to place a RSG on the Hertzsprung-Russell Diagram, due to the stars' cool temperatures and resulting large bolometric corrections. Several recent papers have examined the potential utility of atomic line equivalent widths in cool supergiant spectra for determining $T_{\rm eff}$ and other physical properties \citep{dorda2016b, dorda2016a} and found strong correlations between Ti I and Fe I spectral features and $T_{\rm eff}$ in earlier-type cool supergiants (G and early K) but poor correlations in M-type stars, a spectral subtype that makes up a significant fraction of RSGs. We have extended this work by measuring the equivalent widths of Ti, Fe, and Ca lines in late K- and M-type RSGs in the Milky Way, Large Magellanic Cloud, and Small Magellanic Cloud, and compared these results to the predictions of the MARCS stellar atmosphere models. Our analyses show a poor correlation between $T_{\rm eff}$ and the Fe I and Ti I lines in our observations (at odds with strong correlations predicted by stellar atmosphere models), but do find statistically significant correlations between $T_{\rm eff}$ and the Ca II triplet (CaT) features of Milky Way RSGs, suggesting that this could be a potential diagnostic tool for determining $T_{\rm eff}$ in M type supergiants. We also examine correlations between these spectral features and other physical properties of RSGs (including metallicity, surface gravity, and bolometric magnitude), and consider the underlying physics driving the evolution of atomic line spectra in RSGs.
\end{abstract}

\section{Introduction}
Red supergiants (RSGs) represent a critical phase in massive stellar evolution. They are He-fusing evolved descendants of 10-25$M_{\odot}$ main sequence stars, the end result of a nearly horizontal evolution across the Hertzsprung-Russell (H-R) diagram as their blue H-fusing predecessors leave the main sequence and cross the ``yellow void". They are the largest (in physical size) and coldest ($\sim$3500-4500 K) members of the massive star population, representing a significant extreme in their evolution. These cool temperatures place them at the Hayashi limit for hydrostatic equilibrium \citep{hayashi1961}.
    
Effective temperature ($T_{\rm eff}$) is, along with bolometric luminosity ($M_{\rm bol}$), one of the two key physical properties needed to place a star on the Hertzsprung-Russell (H-R) diagram, and it is the most critical physical property that must be determined for RSGs. At these cool temperatures, the bolometric corrections for standard $UBVRI$ photometry are large (1-4 mag) and strongly dependent on $T_{\rm eff}$ (e.g. \citealt{massey2003, levesque2005}); as a result, accurately calculating the luminosity of a RSG requires a robust determination of the star's $T_{\rm eff}$. 
 
The scarcity of nearby RSGs has limited the use of interferometric data in ascertaining an accurate $T_{\rm eff}$ scale (see, for example, \citealt{dyck1996}). Alternatively, scales in the past have been determined by broad-band colors of RSGs with known diameters (\citealt{lee1970, johnson1964, johnson1966}) or by bolometric corrections derived from IR measurements under the assumption of a blackbody continuum (Flower 1975, 1977). However these methods are also limited because of the effects of line blanketing, which make color indices such as $B-V$ highly sensitive to surface gravity (log $g$). More recently, \cite{levesque2005, levesque2006} used the MARCS stellar atmosphere models to fit the strengths of the $T_{\rm eff}$-sensitive TiO bands for K-type and M-type stars in the Milky Way and Magellanic Clouds, creating a $T_{\rm eff}$ scale significantly warmer than previous works (\citealt{humphreys1984, massey2003}) and one that shows good agreement with the predictions of stellar evolutionary tracks (including the metallicity dependence of the Hayashi limit).  \cite{davies2013} determined warmer $T_{\rm eff}$ values for RSGs using broad SED fitting across the optical and near-IR. However, these results do not reproduce the correlation between spectral type and $T_{\rm eff}$ in RSGs or the metallicity dependence of RSG $T_{\rm eff}$s (see, for example, \citealt{levesque2006, tabernero2018}), and the work notes that 3D models (as opposed to the 1D MARCS models) are required to properly account for wavelength-dependent in the extended atmospheres of RSGs that would otherwise lead to determining a warmer $T_{\rm eff}$ at longer wavelengths.

Several recent papers have also examined the potential utility of atomic lines in these cool stars' spectra for determining $T_{\rm eff}$ and other physical properties. \cite{dorda2016b} compared the widths of several atomic lines (including Fe I and Ti I features and the Ca II triplet lines at 8498\AA, 8542\AA, and 8662\AA, hereafter CaT) observed in the spectra of a large sample of cool supergiants (CSGs, ranging from G0 to M7 in spectral type and thus encompassing the late-type yellow supergiant population as well as RSGs) in the Large and Small Magellanic Clouds; they found that the strength of the Ti I lines was strongly correlated with $T_{\rm eff}$ (though no similar correlation was seen for Fe I or CaT). This result was further supported by \cite{dorda2016a}, which successfully used a principal component analysis based on spectral features in the CaT region (the same region covered by the Gaia Radial Velocity Spectrograph) to automatically differentiate CSGs from other bright late-type stars. This is potentially a very exciting result, offering the possibility of determining $T_{\rm eff}$ for RSGs from data with relatively limited wavelength coverage (as opposed to existing methods which require optical+IR photometry or spectrophotometry with wide optical wavelength coverage). \cite{tabernero2018} studied the $T_{\rm eff}$ scale of CSGs in different metallicity environments; while their method adopted atomic line fitting as a means of determining $T_{\rm eff}$ they found a warmer and shallower scale than \cite{levesque2006}, with only a weak correlation in the LMC and no correlation in the SMC.

However, the utility of using atomic line features for determining $T_{\rm eff}$ in RSGs is still unclear. While the correlation between Ti I and spectral type presented in \cite{dorda2016b} is quite robust at earlier types (G and early K), the correlation is much weaker for the M-type stars in their sample, which represent a significant fraction of the RSG population. The potential dependence of these features on other physical properties is also a complicating factor. 
 For example, recent observations of RSG $J$-band spectra in nearby galaxies have revealed that, while atomic absorption features such as Ti I, Fe I, and Si I are not strongly sensitive to $T_{\rm eff}$, they serve as excellent probes of metallicity (e.g. \citealt{davies2010,davies2015}; \citealt{gazak2015}; \citealt{patrick2015,patrick2016,patrick2017}).

The CaT is widely cited as a potential tracer of luminosity class in cool stars due to its sensitivity to log $g$ effects (e.g. \citealt{cenarro2001a, cenarro2001b} and references therein), and is also sensitive to metallicity (e.g. \citealt{armandroff1991, sakari2016}). Non-LTE effects in the atmospheres of these stars can also impact the equivalent widths (EWs) of some lines; \cite{jennings2016} found that the H$\alpha$ absorption feature in cool stars is also effective as a luminosity class diagnostic, a consequence of the density-dependent overpopulation of the metastable 2s level and an effect that becomes stronger in the non-LTE conditions present in supergiant atmospheres. The \cite{jennings2016} study of the CaT feature indicated that while the feature in early M-type stars had a clear relationship with luminosity class, as supported by the literature, this relationship also broke down in late-type (beyond M3-3.5) supergiants.

Previous work on the CaT has studied its effectiveness as a diagnostic for several physical parameters such as luminosity, log $g$, metallicity, and $T_{\rm eff}$. The CaT is a near-IR feature and therefore is subject to contamination from multiple strong stellar features such as higher-order Paschen lines and the TiO absorption band at 8433\AA. \cite{ginestat1994} studied the relationship between the EW of absorption features between 8380--8780 {\AA} and spectral type, finding a positive correlation between CaT and luminosity for A to M type stars that was initially weak but began to increase for later types beginning at G0. Ginestat also proposed that the weak Ca I, Ti I, and Fe I lines of the {\it giants} in their study may be due to low metallicity.

\cite{e.mb1990}, hereafter EM\&B, used synthetic stellar atmosphere models from \cite{gustasson1975} to generate synthetic CaT lines in order to examine their variation with $T_{\rm eff}$, log $g$, and metallicity ([M/H]). EM\&B found the CaT to be primarily dependent on [M/H] and log $g$. Their study indicates that the CaT is sensitive to metallicity for stars with [M/H] $>$ -2.0 and sensitive to log $g$ for giants with [M/H] $>$ -1.0.  EM\&B also found that the relationship between CaT and temperature is only present in low log $g$ populations, making it applicable to giants and supergiants rather than dwarf stars, and becomes more pronounced as metallicity increases. This is in agreement with the results of \cite{smith1990} but it should be noted that their work was restricted to stars between 4000 and 5500 K. When compared against $T_{\rm eff}$, EM\&B found only a weak relationship between the CaT triplet and $T_{\rm eff}$, which they attributed to the increasing intensity of the 8433\AA\ TiO feature at cool temperatures. The increasing strength of this TiO band can lead to a decrease in the local continuum and a subsequent apparent weakening of the CaT (as noted by \citealt{ginestat1994}, who used a local continuum definition for measuring the CaT in stars later than M2 in order to account for this effect); however, this particular TiO band is only prominent in the spectra of RSGs with relatively late spectral types ($\sim$M4-M5, e.g. \citealt{levesque2005, levesque2017}), an effect in agreement with the evolution of the CaT seen in \cite{jennings2016}.

\cite{mallik1996} analyzed the CaT features of 146 stars spanning from F7 to M4 to determine the dependence of CaT on luminosity, $T_{\rm eff}$, and metallicity. They found a non-linear relationship for luminosity that became more pronounced with increased metallicity, and that was more apparent in supergiants than in dwarfs, but did not find a relationship between CaT and $T_{\rm eff}$ across the full sample. Mallik also found that at low log $g$ (0.0 to 2.0), the EW of the CaT in supergiants and giants decreased as log $g$ increased. This correlation - which is counter to the typical expectation that lines will get stronger at higher log $g$ due to increased collisional effects - has been explained as a continuum effect. An increase in the continuous absorption coefficient at higher log $g$ (due to an increased electron density in stars where H$^-$ is the dominant source of continuum opacity) leads to a lower apparent continuum level and subsequent weaker measurements of EW for the CaT.

\cite{cenarro2001a} presented a new stellar library of the near-IR spectral region based on 706 stars with 2750 K $< T_{\rm eff} < $ 38400 K, 0.0 $<$ log $g$ $<$ 5.12, and metallicities of $-3.45 <$ [Fe/H] $< +0.60$. Based on these data they offer a newly defined index for measuring the strength of the CaT features, the ``CaT*" index, developed with careful treatments of previously noted effects such as continuum definition and Paschen line contamination.

Collectively, the utility of the CaT feature has been extensively studied, but conclusions about its use as a diagnostic of log $g$, luminosity, and $T_{\rm eff}$ are conflicting and further complicated by the differing sample sizes and parameter spaces of previous works, with most samples of stars spanning from dwarfs to supergiants and covering a broad range of spectral types. In this work we specifically consider the utility of atomic absorption line features as potential $T_{\rm eff}$ diagnostics in the uniquely cool and low-density environments of M-type RSGs.

We present a study examining the strengths of Ca, Ti, and Fe absorption features in the spectra of M-type RSGs. Using echelle spectra of 25 Milky Way RSGs, 16 Large Magellanic Cloud RSGs, and 17 Small Magellanic Cloud RSGs along with a series of RSG model atmosphere spectra (Section 2), we present the EWs of a large sample of atomic lines, including features of Fe I, Ti I, Ca I, and the CaT, and compare these EWs to the $T_{\rm eff}$ determinations of Levesque et al.\ (2005; Section 3). We find a strong positive and statistically significant correlation between CaT and $T_{\rm eff}$ for Milky Way RSGs, but no similar correlation in the Magellanic Cloud samples, and no relationship between the Ti I and Fe I features and $T_{\rm eff}$ as a function of metallicity (Section 4). We discuss the implications of these results for understanding the physical properties of RSGs as well as potential future work in this area (Section 5).

\begin{figure*}[ht!]
\center
  \includegraphics[width=\textwidth] {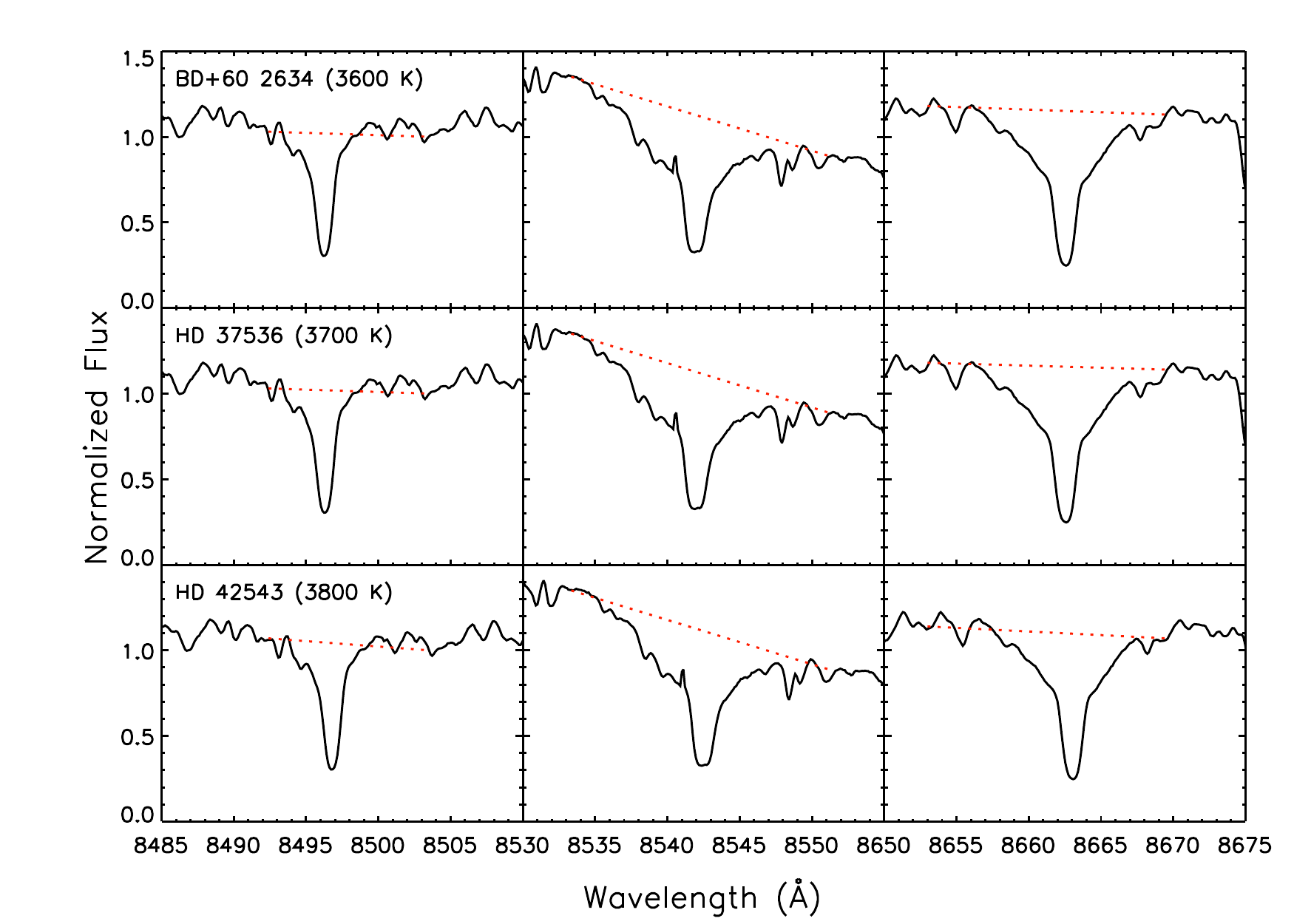}
 \caption{ Example normalized spectra of the Ca II 8498\AA\ (left), 8542\AA\ (center), and 8662\AA\ (right) absorption triplet features for three Milky Way RSGs in our sample, spanning a 200 K range in $T_{\rm eff}$. The spectra are shown in black, while the ``pseudocontinuua" - defined by the continuum points listed in Table 2 and described in \S3.1 - used for fitting the lines and determining their equivalent widths are illustrated as dashed red lines.}
\end{figure*}

\section{Samples and Observations}
The RSG echelle spectra used in these analyses were originally observed as part of a spectroscopic search for potential Thorne-\.Zytkow objects in the Milky Way and Magellanic Clouds \citep{levesque2014}. The sample of 25 Milky Way stars was selected from the coldest RSGs identified in \cite{levesque2005}; all have spectral types of K5-M0I or later. The spectra were observed using the Astrophysics Research Consortium Echelle Spectrograph (ARCES; \citealt{wang2003}) on the Apache Point Observatory 3.5-m telescope on 2011 February 11 and 12 (UT). The observations were taken using the default 1.6 arcsec $\times$ 3.2 arcsec slit, along with quartz lamps and ThAr lamps after each individual exposure to achieve precise flat-field and wavelength caliberations for each star. The spectra were reduced using standard IRAF\footnote{IRAF is distributed by NOAO, which is operated by AURA, Inc., under cooperative agreement with the NSF.} echelle routines, and each star's spectrum was corrected for radial velocity (RV) effects using the wavelengths of the CaT triplet. Examples of our spectra and the CaT triplet are shown in Figure 1.

Our Magellanic Cloud sample was drawn from late-type RSGs identified in \cite{levesque2006} and supplemented by additional stars with broadband colors consistent with RSGs (for a complete discussion see \citealt{levesque2014}). These stars were observed with the Magellan Inamori Kyocera Echelle (MIKE; \citealt{bernstein2003}) on the Magellan 6.5-m at Las Campanas Observatory during 2011 September 13–15. The spectra were taken using  the 0.7 arcsec × 5 arcsec slit with 2×2 binning, ‘slow’ readout, and the standard grating settings, and internal flats and ThAr lamps were observed for flat-fielding and wavelength calibration purposes. These data were reduced using a combination of standard IRAF echelle routines and the {\texttt{ mtools}} package. The Thorne-Zytkow object candidate HV2112 is not included in our sample. 

The physical properties of the observed stars in our sample that we adopt for our analyses - including $T_{\rm eff}$, log $g$, and $M_{\rm bol}$ - are drawn from \cite{levesque2005, levesque2006}, and are based on fitting observed spectrophotometry of the RSGs with MARCS stellar atmosphere models and determining the best-fit model based on the strengths of the TiO absorption bands and the overall fit of the SED.  This $T_{\rm eff}$ scale was chosen as it represents physical properties for RSGs determined based on the optical regime and optical absorption features, an appropriate choice for comparison with the optical atomic line features used in this work as it samples the same physical region of the RSG atmosphere (recalling these stars' extended geometries and wavelength-dependent optical depths, as discussed above). These $T_{\rm eff}$ scales also show good agreement with stellar evolutionary models (including those that treat both single and binary evolution; \citealt{levesque2018} and the effects of metallicity. For a complete list of the stars in our sample and their adopted physical properties, see Table 1.

\cite{tabernero2018} recently published a new $T_{\rm eff}$ scale for CSGs in the LMC and SMC; however, this scale is based on atomic line fitting and the assumption we wish to test here, namely that these lines $T_{\rm eff}$-sensitive. For late K- and M-type RSGs the \cite{tabernero2018} scale is slightly warmer than the \cite{levesque2006} scales as well as shallower (a weaker dependence on spectral type), but without knowing the dependence of the individual atomic features on $T_{\rm eff}$ in this very cool regime it is unclear whether this disagreement is due to a difference in method or a consequence of the lines' behavior (for further discussion see \S4.)

In addition to our sample of observed RSG spectra we also consider synthetic spectra produced by the MARCS stellar atmosphere models (e.g. \citealt{gustasson2008}). The spectra were generated for solar-metallicity 15M$_{\odot}$ RSGs and adopt a spherical atmosphere geometry, a microturbulence parameter of 5 km s$^{-1}$, and log $g=-0.5,0.0,0.5,1.0$. The $T_{\rm eff}$ of the models used in this work range from 3400-4000K in 100K increments.

It should be noted that we restrict our use of the MARCS models to $T_{\rm eff} \le$ 4000 K, and do the same for our observed sample (one LMC star from the original \citealt{levesque2014} sample, LMC 169754, was cut from this work due to its relatively high $T_{\rm eff}$ of 4100 K). This was done specifically to restrict our study of atomic line $T_{\rm eff}$ diagnostics to the collision-dominated regime of cool star atmospheres (at $T_{\rm eff} \gtrsim 4000 K$ the effects of photoionization increasingly dominate the abundance of neutral Fe and Ti) and a regime where non-LTE effects are minimal; for further discussion see Section 4.

\section{Analyses}
\subsection{Atomic Lines and Equivalent Widths}
 We measured the EWs of absorption line features of Fe I, Ti I, Ca I, and Ca II in each of our spectra, using the line profile fitting function contained in the {\texttt splot} task in IRAF's {\texttt kpnoslit} package to determine the best-fit Voigt profile. The Ca features include Ca I 6572\AA\ and the CaT, while the Fe I and Ti I absorption features measured are the same as those used in \cite{dorda2016b}. Upper and lower wavelength bounds were set by identifying the closest local maxima in the surrounding region of the spectrum to define a local continuum which could then be used to measure integrated line strengths; where possible the upper and lower bounds were selected to match the analyses of (\citealt{dorda2016b}; see Table 2). These local continuua were defined consistently across all of the stars in a given host galaxy, although slight variations in the local continuum definition were necessary between the three host galaxies in the case of the CaT lines.

The full set of absorption lines and their measured EWs are given in tables 3-5 for Ca, Fe, and Ti features respectively, and in Table 6 for the MARCS models.

\subsection{Correlation Coefficients}
We used the {\texttt matplotlib} Pylab software from SciPy to plot the measured EW data against the $T_{\rm eff}$, $M_{\rm bol}$, and log $g$ of each star as determined by \cite{levesque2005, levesque2006}, with EW as the dependent variable and the stellar parameters as the independent variable. For each variable pair we also calculated the Pearson's $r$ correlation coefficient and associated $p$-value for the sample, along with both linear (A$x$+B) and second-degree polynomial (A$x^2$+B$x$+C) functions of best fit. Given our small sample sizes (ranging from 16 to 25 stars in a given host galaxy, which we treat separately due to metallicity effects) we adopt a conservative significance threshold of $p < 0.01$, rather than the more typical $p < 0.05$, to decrease our likelihood of incorrectly rejecting the null hypothesis. The Pearson correlation coefficients and best-fit function coefficients are summarized in Table 7. A similar analysis was also done for the MARCS stellar atmosphere model spectra, with the results summarized in Table 8.

Below we consider each spectral feature and its potential diagnostic utility:

{\bf CaT}: In our Milky Way sample, the equivalent width of the CaT feature (Figure 2) shows a strong and statistically significant positive correlation with $T_{\rm eff}$, with Pearson's r=0.755 and $p$=0.00001. This is in good agreement with the MARCS stellar atmosphere models at Milky Way metallicity (Figure 8), which predict robust positive correlations between CaT and $T_{\rm eff}$ across the full range of supergiant surface gravities (e.g., Pearson's r=0.869 and $p$=0.0023 for the log $g=0.0$ models). However, no similar correlation is seen in the LMC and SMC samples, despite the LMC- and SMC-metallicity MARCS models also predicting strong positive correlations at all supergiant surface gravities (Figures 9 and 10).

In the Milky Way the CaT equivalent width is positively correlated with log $g$ and M$_{\rm bol}$. No correlation is seen in the LMC data; however, the SMC sample shows a positive correlation between the CaT equivalent width and M$_{\rm bol}$.

{\bf Ca I}: The MARCS models predict a correlation between the equivalent width of the Ca I 6572.0 line and $T_{\rm eff}$ for all but the lowest surface gravities and highest metallicities (e.g., the log $g=-0.5$ models at Milky Way metallicity and the log $g=-0.5$ and log $g=0.0$ models at LMC metallicity). However, this is not borne out by our observations (Figure 3); the LMC and SMC samples show no significant correlation between Ca I and $T_{\rm eff}$, while the Milky Way sample shows a moderately strong positive correlation ($r$=0.500) with a borderline $p$=0.01. It is also worth noting that the MARCS models predict a positive correlation between Ca I and $T_{\rm eff}$ at Milky Way metallicity, but a negative correlation at the lower LMC and SMC metallicities, with the Ca I line getting weaker at higher $T_{\rm eff}$.

{\bf Ti I/Fe I Ratio}: None of the observed data revealed significant correlations between the Ti I 8518.1/Fe I 8514.1 ratio (Figure 4) and RSG physical properties. By contract, the MARCS models predict significant negative correlations between this ratio and T$_{\rm eff}$ for all log $g$ values at Milky Way metallicity, while the LMC and SMC metallicity models show significant negative correlations at particular values of log $g$ (log $g=0.0$ in the LMC and log $g=-0.5$, 0.0, and 1.0 in the SMC). 

{\bf Ti I Sum}: None of the observed samples - at Milky Way, LMC, or SMC metallicity - show any evidence for statistically significant correlations between the sum of the Ti I line (Figure 5) equivalent widths and any RSG physical properties. 

The Milky Way metallicity MARCS models predict no statistically significant correlation between Ti I EW and $T_{\rm eff}$. However, in the LMC the Ti I sum showed a strong positive correlation with $T_{\rm eff}$ for the log $g=0.0$ models only (r=0.997, $p$=0.00186), and the SMC models showed a strong negative correlation with $T_{\rm eff}$ at log $g=-0.5$ (r=$-$0.933, $p$=0.00071) and a strong positive correlation at log $g=0.5$ (r=0.879, $p$=0.00404).

{\bf Fe I Sum}: Our observed Milky Way and SMC samples show no correlation between the sum of the Fe I line (Figure 6) equivalent widths and any RSG physical properties; however, the LMC sample shows evidence of a positive correlation between the Fe I sum and T$_{\rm eff}$. Both of these results are also at odds with the predictions of the MARCS models, which only predict a significant positive correlation between Fe I and T$_{\rm eff}$ for Milky Way supergiants with log $g=1.0$ (r=0.965, $p$=0.0001) and a significant {\it negative} correlation between Fe I and T$_{\rm eff}$ for SMC supergiants with log $g=0.5$ and 1.0 (r=$-$0.972, $p$=0.00005 and r=$-$0.975, $p$=0.00004, respectively).

{\bf Ca II 3-D Plots}: The CaT equivalent width showed the most promise in our observed data as an atomic line diagnostic of T$_{\rm eff}$; however, this feature is also well-known as a potential diagnostic of log $g$ and luminosity, calling the degeneracy of its T$_{\rm eff}$ correlation into question. To further examine this we created 3D plots for our observed data from all three host galaxies with $T_{\rm eff}$, Log $g$, and Ca II EW (see Figure 7) as the respective x, y, and z axes. \footnote{The base code to make the 3D plots can be found at \texttt{https://gist.github.com/amroamroamro/1db8d69b4b65e8bc66a6}} These data were then fitted with linear and quadratic planes of best fit. The full suite of correlation coefficients for the data in the 3D plots is given in Table 9, and equations for the linear and quadratic best fits are given in Table 10. The linear plane best fit equation is $Z = Ax + By + C$, while the quadratic plane fit is $Z = Ax + By + Cxy + Dx^{2} + Ey^{2} + F$.

\section{Discussion and Future Work} 
In light of the small sample sizes used in this work (25 stars in the Milky Way sample, 16 in the LMC, and 17 in the SMC) we are cautious about over-interpreting the statistical results drawn from our data. However, it is still interesting to examine areas where our observations and model results agree or diverge from each other and from past work, and to consider potential physical explanations for why this may be and the implications for future work.

Our observed spectra show a strong positive correlation ($r=0.755$, $p=0.00001$) between CaT and $T_{\rm eff}$ for MW RSGs, in agreement with the predictions of the MARCS stellar atmosphere models. However, it is difficult to discern whether this is primarily a consequence of $T_{\rm eff}$ or log $g$ effects on the CaT absorption features and surrounding continuum. In the MW there is a significant correlation ($r=0.549$, $p=0.0082$) between $T_{\rm eff}$ and log $g$, with cooler stars having lower surface gravities (an unsurprising consequence of the effect that a decreasing $T_{\rm eff}$ and constant or increasing $M_{\rm bol}$ will have on the stellar radii), and both log $g$ and $M_{\rm bol}$ are positively correlated with the CaT sum (see Table 9). Figure 7 compares the CaT equivalent width, $T_{\rm eff}$, and log $g$ of our Milky Way stars in 3-D space, along with the best quadratic plane fit to the sample, but - as also noted by previous work - it is unclear which physical property is primarily responsible for driving the evolution of CaT in RSG spectra.

We also do {\it not} see any correlation between CaT and $T_{\rm eff}$ in either the LMC or SMC observations. While this suggests that metallicity may also play a role in the evolution of the CaT with stellar properties (in agreement with previous work that found a metallicity dependence in the CaT equivalent width for supergiants (e.g. \citealt{armandroff1991, ginestat1994, mallik1996}) this is at odds with the predictions of the MARCS models, which predict a strong correlation between CaT and $T_{\rm eff}$ at all of the model metallicities. The models do, however, predict an expected overall decrease in the strength of the CaT with metallicity. Figures 7-9 compare the MARCS models and observed data, highlighting the decrease in EW with metallicity as well as comparing the EWs predicted by the models to those observed in the data. Note that Dorda et al.\ (2016b) also directly compare LMC- and SMC-metallicity MARCS model equivalent widths to $T_{\rm eff}$ (though their data span a broader $T_{\rm eff}$ range of 3300-4500 K to better encompass the warmer F-, G-, and K-type supergiants in their observed sample) and find similar results.

Our observed data also blend a range of surface gravities that sample the lower end of RSG surface gravities (the mean log $g$ of the MW, LMC, and SMC samples is 0.15, $-$0.275, AND $-$0.247 respectively) while the models with different log $g$ are considered separately. If we combine the results from the MARCS models across {\it all} surface gravities, the correlations between CaT and $T_{\rm eff}$ get weaker in the Milky Way (r=0.467, $p$=0.007) and LMC (r=0.712, $p$=0.0004), while the SMC data fails to satisfy our $p<0.01$ significance threshold (r=0.356, $p$=0.046). Considering these results, it is possible that decreased metallicity combined with a mix of surface gravities in our observed samples could contribute to the lack of statistically robust correlation between CaT and $T_{\rm eff}$ in the LMC and SMC data. 

Quadratic best fits to the CaT EW, $T_{\rm eff}$ and log $g$ data in 3D space show a stronger relationship between the three parameters for the lowest $T_{\rm eff}$ and log $g$ values. The CaT EW is consistently high at high log $g$ values ($\geq0.6$) for all values of $T_{\rm eff}$, but decreases nearly linearly with log $g$ at low $T_{\rm eff}$. Above $T_{\rm eff}$ $\geq$ 3750 K, the relationship between EW and log $g$ becomes more complex. By comparison, both the LMC and SMC quadratic best fits in 3D show a ``concave" shape, with the evolution of CaT as a function of log $g$ and $T_{\rm eff}$ that is hard to quantify and not well-fit by a linear relation (for example, both high $T_{\rm eff}$ + low log $g$ and low $T_{\rm eff}$ + high log $g$ correspond to CaT EW minimums.

The relationships are simplified (but also more poorly fit) in linear plane best fits to the data, given in Table 10. In this case the MW and LMC results broadly align with predictions from \cite{ginestat1994} asserting that the intensity of the CaT is correlated with log $g$, but the SMC does not, suggesting a more complex relationship for this sample. \cite{e.mb1990} concluded that at high metallicity Ca II was inversely related to log $g$; as the SMC has the lowest metallicity of our sample it suggests that this difference could in part be attributable to changes in this relationship as a function of metallicity.

The \cite{dorda2016b} sample consisted of early G to M3 stars, with a small sample of later-type M stars, as it was thought that the TiO band would significantly erode the continuum in the latest M type stars stars and make reliable EW measurements difficult beginning at a spectral type of M0. However, as found both by \cite{ginestat1994} and this work, the stronger TiO band at later spectral types (corresponding to cool $T_{\rm eff}$, e.g. \citealt{levesque2005, tabernero2018}) reliably corresponds to a decrease in the local continuum, and a subsequent apparent weakening of the CaT, in agreement with the evolution of the CaT seen at warmer $T_{\rm eff}$. This effect therefore improves rather than weakens the utility of the CaT as a $T_{\rm eff}$ diagnostic. For a direct comparison between the models results and the MW, LMC, and SMC results see Figures 8-10.

Moving beyond the CaT feature, it is interesting to note that there is no correlation in any of our observed data between Ca I or the Ti/Fe ratio and any RSG physical properties, at odds with what the models predict. Beyond that, most of the observed {\it and} model samples predict no correlation between the sum of the Ti I or Fe I absorption features and the physical properties of RSGs, with a few noted exceptions. For example, the observed LMC spectra show a positive correlation between the Fe I sum and $T_{\rm eff}$. However, this is at odds with the predictions of the MARCS stellar evolution models, which only predict correlations between Fe I and $T_{\rm eff}$ for the highest log $g$ models (notably higher than the average log $g$ of our observed RSGs), and is almost certainly a consequence of small number statistics in our 16-star LMC sample. Still, since \cite{dorda2016b} predict a strong correlation between Ti I and Fe I and spectral type for cool supergiants - which we would expend to extend to a correlation between these features and $T_{\rm eff}$ and be observable even in a small sample - it is worth considering some of the physical phenomena that may impact the formation and evolution of these lines in RSG spectra.

Our MARCS model results do not highlight Ti I and Fe I as robust diagnostic lines for M-type RSGs; however, these models also assume LTE. How might non-LTE conditions affect these predictions? \cite{bergemann2012} studied individual RSGs to determine the impact of non-LTE on Ti I and Fe I spectral features. They found that the significance of non-LTE corrections was dependent on $T_{\rm eff}$, metallicity, and log $g$, noting that for both Ti I and Fe I, non-LTE corrections in order to align results with observations were lower, or near zero, at lower temperatures - 3400 K $\geq$ $T_{\rm eff}$ $\leq$ 3800 K. At higher temperatures, the formation of Fe I lines remains largely unaffected by non-LTE, while the Ti I line does show some variation, with Ti I EWs underestimated by LTE as compared to non-LTE models. However, for M-type RSGs non-LTE effects on these lines do not play a significant role and as such cannot be considered a variable for the disagreements between the observed data and the models.

As these are all neutral lines, it is also worth considering the excitation potential of these features. Both \cite{dorda2016b} and our models predict a relationship between Fe I and $T_{\rm eff}$ and between Ti I and $T_{\rm eff}$, with \cite{dorda2016b} arguing that this is due to their low excitation potentials (6.82 eV for Ti I, 7.87 eV for Fe I), thus rendering the neutral abundances of these elements particularly sensitive to $T_{\rm eff}$. However, while this reasoning is robust for warmer stars it breaks down for M-type RSGs. At these low temperatures ($\lesssim$4000 K) photoionization is no longer the primary means of producing Ti II and Fe II, and instead collision becomes the dominant means of excitation (for more discussion see \citealt{bergemann2012}). This decouples the Ti I/Ti II and Fe I/Fe II fractions from $T_{\rm eff}$ in the cool and low-density atmospheres of RSGs. Taken as a whole, the Ti I and Fe I absorption features are not effective diagnostics of $T_{\rm eff}$ for the coolest RSGs (a result in agreement with Dorda et al.\ (2016b)'s figures 7 and 8, which shower a weak correlation between these features and spectral type for the M0-M3 stars in their sample).

The Ca I absorption feature remains a puzzle. The MARCS models predict a positive correlation between Ca I EW and $T_{\rm eff}$ at Milky Way metallicity, but a {\it negative} correlation for SMC metallicities and the higher surface gravity LMC models. The latter is what would naively be expected based simply on the evolution of the CaT absorption feature: as the Ca II abundance (and the CaT EW increases), we would expect a corresponding decrease in the Ca I abundance and EW. It is unclear why this is predicted at lower metallicities (and higher surface gravities) but not at solar metallicity. It is possible that at higher metallicities the Ca I abundance is high enough to saturate, resulting in a non-linear evolution of the Ca I absorption feature at higher metallicities. As in the case of Ti and Fe, the relative contributions from photoionization and collisional excitation could also play a role. Finally, none of these expected correlations appear in our observed data, suggesting that additional effects (including the impact of non-LTE) could further complicate the formation and evolution of the Ca I with $T_{\rm eff}$.

While these results are based on only a small sample of M-type RSGs, it is nevertheless important to consider whether these or other atomic lines can be used to directly infer the stars' physical properties. To take just one example, spectra from {\it Gaia} span only a narrow wavelength range ($\sim$8450-8750\AA), but this critical regime includes the CaT absorption feature as well as all of the Ti I absorption lines and 7 of the Fe I absorption lines included in this work. Identifying - or excluding - useful $T_{\rm eff}$ diagnostics in this wavelength regime represents a potentially powerful tool for leveraging the wealth of potential RSG data available in current and future {\it Gaia} data releases (which extends throughout the Milky Way and to the Large and Small Magellanic Clouds), and may make it possible to greatly improve the accuracy of the physical properties determined for these stars.

\acknowledgements
We would like to thank Trevor Dorn-Wallenstein, Philip Massey, and George Wallerstein for useful discussions regarding this research, as well as the staff of Apache Point Observatory and Las Campanas Observatory for their support in acquiring the observed spectra used in this work. These efforts were supported in part by a fellowship from the Alfred P. Sloan Foundation. \software{IPython \citep{PER-GRA:2007}, SciPy \citep{jones_scipy_2001}, NumPy \citep{van2011numpy}, IRAF \citep{tody1986, tody1993}, Matplotlib \citep{Hunter2007} }.

\startlongtable
 \begin{deluxetable*}{lccccc}
 \tablewidth{0pt}
 \tablecolumns{6}
 \tablecaption{Sample Stars} 
 \tablehead{
     \colhead{Star} & \colhead{Spectral Type} & \colhead{T$\rm_{eff}$} & \colhead{log $g$} & \colhead{M$\rm_{bol}$} & \colhead{OB Assoc.} 
    } 
     \startdata
     MW & & & & &  \\
     \hline
      BD+59 38 & M2 I & 3650 & 0.1 & -7.17 & Cas OB4 \\
      BD+56 595 & M1 I & 3800 & 0.4 & -6.31 & Per OB1-D  \\
      BD+57 647 & M2 I & 3650 & 0.0 & -7.51 & Per OB1-D? \\
      BD+59 274 & M1 I & 3750 & 0.4 & -6.14 & Cas OB8/NGC581 \\
      BD+59 372 & K5-M0 I & 3825 & 0.6 & -5.77 & Per OB1-A \\
      BD+60 335 & M4 I & 3525 & 0.1 & -7.05 & Cas OB8/NGC663 \\
      BD+60 2613 & M3 I & 3600 & -0.7/-0.4 & -9.64/8.57 & Cas OB5 \\
      BD+60 2634 & M3 I & 3600 & -0.1 & -7.73 & Cas OB5 \\
      Case 23 & M3 I & 3600 & 0.3 & -6.28 & Cas OB7 \\
      Case 80 & M3 I & 3600 & 0.1 & -7.00 & Cas OB2 \\
      Case 81 & M2 I & 3700 & 0.1 & -7.19 & Cas OB2  \\
      HD 14469 & M3-4 I & 3575 & -0.1 & -7.64 & Per OB1-D \\
      HD 14488 & M4 I & 3550 & -0.3 & -8.15 & Per OB1-D/NGC884 \\
      HD 23475 & M2.5 II & 3625 & --- & --- & --- \\
      HD 35601 & M1.5 I & 3700 & 0.2 & -6.81 &  Aur OB1 \\
      HD 36389 & M2 I & 3650 & --- & --- & --- \\
      HD 37536 & M2 I & 3700 & 0.1 & -7.33 & Aur OB1 \\
      HD 42475 & M1 I & 3700 & -0.1 & -7.76 & Gem OB1 \\
      HD 42543 & M0 I & 3800 & 0.0 & -7.55 & Gem OB1 \\
      HD 44537 & M0 I & 3750 & --- & --- & --- \\
      HD 219978 & M1 I & 3750 & 0.4 & -6.44 & Cep OB3 \\
      HD 236697 & M1.5 I & 3700 & 0.4 & -6.25 & NGC 457 \\
      HD 236871 & M2 I & 3625 & 0.2 & -6.80 & Cas OB8 \\
      HD 236915 & M2 I & 3650 & 0.3 & -6.40 & Per OB1-A \\
      W Per & M4.5 I & 3550 & 0.1 & -7.09 & Per OB1-D? \\
       \hline
      LMC & & & & & \\
      \hline
       LMC 064048 & M2.5 I & 3500 & -0.2 & -7.81 & \nodata\\
       LMC 109106 & M2.5 I & 3550 & -0.2 & -7.89 & \nodata \\
       LMC 116895 & M0 I & 3750 & -0.2 & -8.10 & \nodata \\
       LMC 141430 & M1 I & 3700 & -0.3 & -8.55 &\nodata  \\
       LMC 142202 & M1.5 I & 3650 & -0.3 & -8.36 & \nodata  \\
       LMC 146126 & K5 I & 3875 & -0.2 & -8.62 & \nodata  \\
       LMC 061753 & M2 I & 3600 & 0.0 & -7.80 & \nodata \\
       LMC 170452 & M2.5 I & 3550 & -0.5 & -8.67 & \nodata  \\
       WOH S274 & M1.5 I & 3650 & 0.0 & -8.24 & \nodata  \\
       HV 12802 & M1 I & 3700 & -0.5 & -8.28 & \nodata \\
       LMC 170079 & M2 I & 3625 & -0.5 & -8.80 & \nodata  \\
       LMC 054365 & M2.5 I & 3525 & -0.2 & -7.88 & \nodata \\
       LMC 068125 & M4 I & 3475 & -0.3 & -8.21 & \nodata \\
       LMC 135720 & M4.5 I & 3425 & -0.4 & -8.38 & \nodata  \\
       LMC 174714 & M1.5 I & 3625 & -0.3 & -8.39 & \nodata \\
       LMC 175746 & M3 I & 3500 & -0.3 & -8.35 & \nodata \\
       \hline
     SMC  & & & & &   \\
     \hline
      SMC 005092 & M2 I & 3475 & -0.4 & -8.48 &\nodata  \\
      SMC 008930 & M0 I & 3625 & -0.3 & -8.38 & \nodata   \\
       SMC 018136 & M0 I & 3575 & -0.4 & -8.76 & \nodata \\
       SMC 020133 & M0 I & 3625 & -0.3 & -8.39 & \nodata  \\
       SMC 025879 & M0 I & 3700 & -0.3 & -8.44 & \nodata  \\
       SMC 050840 & M1 I & 3625 & -0.2 & -8.12 & \nodata  \\
       SMC 060447 & K2 I & 3900 & 0.1 & -7.37 &\nodata \\
     SMC 069886 & M2 I & 3750 & -0.3 & -8.76 & \nodata   \\
       SMC 078282 & M3 I & 3600 & -0.5 & -9.23 & \nodata   \\
       SMC 055275 & M2 I & 3650 & 0.0 & -7.88 & \nodata \\
       SMC 056389 & M2 I & 3675 & -0.5 & -8.56 & \nodata  \\
       J00534794-7202095 & M3 I & 3575 & -0.5 & -8.77 & \nodata \\
       SMC 011709 & K5-M0 I & 3725 & -0.1 & -7.93 & \nodata   \\
       SMC 046497 & K5-M0 I & 3700 & -0.2 & -8.30 & \nodata  \\
       SMC 049478 & K5-M0 I & 3700 & -0.3 & -8.49 & \nodata  \\
      SMC 052334 & K5-M0 I & 3675 & 0.0 & -7.82 & \nodata   \\
       SMC 056732 & K5-M0 I & 3725 & 0.0 & -7.66 & \nodata  \\
     \enddata
\end{deluxetable*}

\begin{deluxetable*}{cccccccc}
 \tablewidth{0pt}
 \tablecolumns{8}
 \tablecaption{Measured Absorption Features and Upper and Lower Bounds for Local Continuum}
 \tablehead{
     \colhead{Wavelength (\AA)} & \colhead{Chem. Species} & \colhead{Min. (MW)} & \colhead{Max. (MW)} & \colhead{Min. (LMC)} & \colhead{Max. (LMC)} & \colhead{Min. (SMC)} & \colhead{Max. (SMC)} }
     \startdata
     6572.0 & Ca I & 6571.0 & 6572.35 & 6572.35 & 6573.48 & 6571.91 & 6573.43 \\
     8498.0 & Ca II & 8492.40 & 8503.55 & 8494.5 & 8501.00 & 8495.00 & 8501.50 \\ 
     8514.1 & Fe I & 8513.40 & 8514.80 & 8513.28 & 8514.97 & 8513.19 & 8514.64  \\
     8518.1 & Ti I & 8516.60 & 8520.25 & 8517.40 & 8518.96 & 8517.68 & 8619.03 \\
     8542.0 & Ca II & 8533.35 & 8551.50 & 8533.5 & 8547.20 & 8533.75 & 8547.40 \\
     8582.0 & Fe I & 8581.73 & 8583.40 & 8581.274 & 8583.745 & 8581.24 & 8583.60 \\
     8611.0 & Fe I & 8609.95 & 8611.45 & 8611.085 & 8612.740 & 8611.03 & 8612.30 \\
     8662.0 & Ca II & 8653.00 & 8669.50 & 8655.30 & 8666.50 & 8655.50 & 8666.50 \\
     8679.4 & Fe I & 8679.2 & 8680.3 & 8678.0 & 8679.4 & 8678.6 & 8679.4 \\
     8683.0 & Ti I & 8681.3 & 8684.7 & 8681.6 & 8683.7 & \nodata & \nodata \\
     8688.5 & Fe I & 8687.0 & 8689.3 & 8687.6 & 8689.5 & 8685.6 & 8690.7 \\
     8692.0 & Ti I & 8690.4 & 8693.0 & \nodata & \nodata & \nodata & \nodata \\
     8710.2 & Fe I & 8709.8 & 8710.4 & 8709.9 & 8711.2 & 8709.9 & 8711.7 \\
     8712.8 & Fe I & 8711.0 & 8713.5 & 8711.6 & 8713.8 & 8712.0 & 8715.4 \\
     8730.5 & Ti I & 8728.2 & 8730.5 & \nodata & \nodata & \nodata & \nodata \\
     8757.0 & Fe I & 8754.8 & 8759.6 & 8756.3 & 8757.9 & 8756.5 & 8758.8 \\
     8793.2 & Fe I & 8791.4 & 8795.5 & 8792.6 & 8794.2 & 8792.4 & 8794.2 \\
     8805.0 & Fe I & 8802.6 & 8805.7 & 8804.0 & 8805.8 & 8804.0 & 8805.9 \\
     8824.0 & Fe I & 8823.0 & 8825.3 & 8823.2 & 8825.6 & 8823.4 & 8825.7 \\
     8838.0 & Fe I & 8837.7 & 8838.6 & 8836.7 & 8840.3 & 8837.3 & 8840.6 \\
     \enddata
\end{deluxetable*}

\startlongtable
 \begin{deluxetable*}{lcccc}
 \tablewidth{0pt}
 \tablecolumns{5}
 \tablecaption{Ca I and CaT Equivalent Widths} 
 \tablehead{
     \colhead{Star} & \colhead{6572} & \colhead{8498} & \colhead{8542} & \colhead{8662} 
    }
     \startdata 
     Milky Way & & & & \\
     \hline
     BD+59 38 & 0.311 & 2.14 & 3.61 & 3.30 \\
     BD+56 595 & 0.38 & 2.50 & 5.00 & 4.11 \\
     BD+57 647 & 0.388 & 2.19 & 4.61 & 3.79 \\
     BD+59 274 & 0.363 & 2.36 & 4.79 & 4.25 \\
     BD+59 372 & 0.350 & 2.37 & 5.01 & 3.96 \\
     BD+60 335 & 0.378 & 2.28 & 4.11 & 3.40 \\
     BD+60 2613 & 0.304 & 1.92 & 2.35 & 2.59 \\
     BD+60 2634 & 0.314 & 2.08 & 2.59 & 3.16 \\
     Case 23 & 0.356 & 2.31 & 3.63 & 3.76 \\
     Case 80 & 0.359 & 2.26 & 3.84 & 3.69 \\
     Case 81 & 0.373 & 2.38 & 4.48 & 4.00 \\
     HD 14469 & 0.347 & 2.28 & 3.67 & 3.32 \\
     HD 14488 & 0.295 & 2.07 & 1.76 & 2.34 \\
     HD 23475 & 0.370 & 2.25 & 4.25 & 3.42 \\
     HD 35601 & 0.381 & 2.36 & 4.48 & 3.81 \\
     HD36309 & 0.395 & 2.47 & 3.83 & 3.69 \\
     HD 37536 & 0.369 & 2.30 & 4.17 & 3.55 \\
     HD 42475 & 0.343 & 2.40 & 4.64 & 3.87 \\
     HD 42543 & 0.379 & 2.23 & 5.21 & 3.74 \\
     HD 44537 & 0.402 & 2.85 & 5.08 & 4.49 \\
     HD 219978 & 0.376  & 2.37 & 5.29 & 4.23 \\
     HD 236697 & 0.380 & 2.48 & 5.23 & 4.18 \\
     HD 236871 & 0.368 & 2.10 & 4.57 & 3.74 \\
     HD 236915 & 0.370 & 2.33 & 4.56 & 3.81 \\
     W Per & 0.291 & 2.13 & 2.95 & 2.85 \\
     \hline
     LMC & & & & \\
     \hline
     LMC 064048 & 0.377 & 2.367 & 4.611 & 3.619 \\
     LMC 109106 & 0.4133 & 2.59 & 4.72 & 4.50 \\
     LMC 116895 & 0.425 & 2.59 & 4.92 & 4.40 \\
     LMC 141430 & 0.383 & 2.21 & 3.64 & 3.21 \\
     LMC 142202 & 0.381 & 2.44 & 4.23 & 4.01 \\
     LMC 146126 & 0.389 & 2.40 & 3.95 & 3.63 \\
     LMC 061753 & 0.338 & 1.83 & 4.02 & 3.07 \\
     LMC 170452 & 0.285 & 2.03 & 2.93 & 3.27 \\
     WOH S274 & 0.430 & 2.38 & 3.73 & 3.71 \\
     HV 12802 & 0.397 & 2.24 & 3.11 & 3.30 \\
     LMC 170079 & 0.366 & 2.32 & 3.61 & 3.50 \\
     LMC 054365 & 0.410 & 2.42 & 4.65 & 4.03 \\
     LMC 068125 & 0.368 & 2.21 & 3.23 & 3.22 \\
     LMC 135720 & 0.300 & 1.99 & 2.41 & 2.50 \\
     LMC 174714 & 0.430 & 2.07 & 3.50 & 3.11 \\
     LMC 175746 & 0.347 & 2.27 & 3.60 & 3.55 \\
     \hline
     SMC & & & & \\
    \hline
     SMC 005092 & 0.385 & 1.53 & 2.75 & 2.79 \\
     SMC 008930 & 0.353 & 1.92 & 3.76 & 3.62 \\
     SMC 018136 & 0.360 & 2.27 & 4.16 & 3.38 \\
     SMC 020133 & 0.334 & 1.89 & 3.81 & 3.29 \\
     SMC 025879 & 0.312 & 2.06 & 4.04 & 3.71 \\
     SMC 050840 & 0.365 & 2.11 & 4.32 & 3.56 \\
     SMC 060447 & 0.36 & 2.01 & 4.37 & 3.54 \\
     SMC 069886 & 0.343 & 1.44 & 2.30 & 2.78 \\
     SMC 078282 & 0.305 & 1.51 & 2.85 & 3.17 \\
     SMC 055275 & 0.355 & 1.93 & 4.00 & 3.088 \\
     SMC 056389 & 0.334 & 1.92 & 3.42 & 3.15 \\
     J00534794-7202095 & 0.322 & 1.87 & 3.25 & 3.23  \\
     SMC 011709 & 0.355 & 2.08 & 4.21 & 3.72 \\
     SMC 046497 & 0.352 & 1.97 & 3.86 & 3.31 \\
     SMC 049478 & 0.352 & 1.96 & 3.87 & 3.04 \\
     SMC 052334 & 0.346 & 1.98 & 4.28 & 3.51 \\
     SMC 056732 & 0.343 & 1.97 & 4.31 & 3.41 \\
     \enddata
 \end{deluxetable*}

\startlongtable
   \begin{deluxetable*}{lcccccccccccc}
   \tablewidth{0pt}
   \tablecolumns{13}
   \tablecaption{Fe I Equivalent Widths}
   \tablehead{
      \colhead{Star} & \colhead{8514.1} & \colhead{8582.0} & \colhead{8611.0} & \colhead{8679.4} & \colhead{8688.5} & \colhead{8710.2} & \colhead{8712.8} & \colhead{8757.0} & \colhead{8793.2} & \colhead{8805.0} & \colhead{8824.0} & \colhead{8838.0} 
      }
      \startdata
       Milky Way & & & & & & & & & & & & \\
    \hline
     BD+59 38 & 0.319 & 0.148 & 0.100 & 0.034 & 0.453 & 0.025 & 0.470 & 0.443 & 0.325 & 0.282 & 0.477 & 0.052 \\
     BD+56 595 & 0.299 & 0.192 & 0.146 & 0.051 & 0.444 & 0.030 & 0.493 & 0.454 & 0.343 & 0.267 & 0.424 & 0.069 \\
     BD+57 647 & 0.320 & 0.171 & 0.119 & 0.039 & 0.488 & 0.022 & 0.471 & 0.516 & 0.399 & 0.243 & 0.476 & 0.052 \\
     BD+59 274 & 0.306 & 0.171 & 0.132 & 0.057 & 0.408 & 0.029 & 0.406 & 0.388 & 0.299 & 0.241 & 0.373 & 0.060 \\
     BD+59 372 & 0.306 & 0.185 & 0.140 & 0.061 & 0.413 & 0.031 & 0.409 & 0.454 & 0.308 & 0.212 & 0.380 & 0.064 \\
     BD+60 335 & 0.264 & 0.158 & 0.108 & 0.038 & 0.401 & 0.025 & 0.449 & 0.454 & 0.303 & 0.259 & 0.455 & 0.065 \\
     BD+60 2613 & 0.251 & 0.262 & 0.235 & 0.087 & 0.508 & 0.017 & 0.379 & 0.498 & 0.357 & 0.279 & 0.439 & 0.066 \\
     BD+60 2634 & 0.236 & 0.169 & 0.092 & 0.038 & 0.492 & 0.029 & 0.411 & 0.391 & 0.301 & 0.224 & 0.391 & 0.031 \\
     Case 23 & 0.305 & 0.160 & 0.119 & 0.046 & 0.498 & 0.033 & 0.450 & 0.425 & 0.331 & 0.266 & 0.426 & 0.045 \\
     Case 80 & 0.297 & 0.211 & 0.118 & 0.050 & 0.496 & 0.029 & 0.480 & 0.400 & 0.292 & 0.241 & 0.385 & 0.047 \\
     Case 81 & 0.330 & 0.186 & 0.152 & 0.053 & 0.563 & 0.033 & 0.489 & 0.457 & 0.333 & 0.271 & 0.423 & 0.047 \\
     HD 14469 & 0.251 & 0.155 & 0.103 & 0.047 & 0.603 & 0.033 & 0.415 & 0.484 & 0.296 & 0.281 & 0.468 & 0.048 \\
     HD 14488 & 0.238 & 0.144 & 0.069 & 0.034 & 0.453 & 0.023 & 0.091 & 0.452 & 0.246 & 0.243 & 0.460 & 0.049 \\
     HD 23475 & 0.317 & 0.191 & 0.155 & 0.068 & 0.361 & 0.027 & 0.339 & 0.353 & 0.242 & 0.247 & 0.434 & 0.067 \\
     HD 35601 & 0.319 & 0.179 & 0.179 & 0.042 & 0.459 & 0.035 & 0.502 & 0.450 & 0.294 & 0.284 & 0.418 & 0.030 \\
     HD 36389 & 0.301 & 0.194 & 0.128 & 0.050 & 0.411 & 0.035 & 0.546 & 0.466 & 0.266 & 0.279 & 0.433 & 0.047 \\
     HD 37536 & 0.288 & 0.194 & 0.124 & 0.040 & 0.390 & 0.027 & 0.456 & 0.457 & 0.278 & 0.287 & 0.342 & 0.033 \\
     HD 42475 & 0.253 & 0.192 & 0.096 & 0.031 & 0.602 & 0.018 & 0.535 & 0.546 & 0.353 & 0.302 & 0.463 & 0.040 \\
     HD 42543 & 0.275 & 0.182 & 0.114 & 0.026 & 0.579 & 0.029 & 0.463 & 0.542 & 0.379 & 0.338 & 0.472 & 0.052 \\
     HD 44537 & 0.294 & 0.208 & 0.150 & 0.038 & 0.541 & 0.034 & 0.470 & 0.590 & 0.414 & 0.372 & 0.460 & 0.044 \\
     HD 219978 & 0.312 & 0.188 & 0.155 & 0.061 & 0.496 & 0.038 & 0.484 & 0.442 & 0.287 & 0.270 & 0.417 & 0.077 \\
     HD 236697 & 0.305 & 0.224 & 0.138 & 0.062 & 0.482 & 0.031 & 0.441 & 0.402 & 0.277 & 0.233 & 0.412 & 0.083 \\
     HD 236871 & 0.303 & 0.167 & 0.124 & 0.046 & 0.518 & 0.033 & 0.476 & 0.415 & 0.351 & 0.260 & 0.435 & 0.065 \\
     HD 236915 & 0.302 & 0.207 & 0.117 & 0.059 & 0.496 & 0.024 & 0.423 & 0.379 & 0.235 & 0.225 & 0.393 & 0.077 \\
     W Per & 0.243 & 0.167 & 0.079 & 0.021 & 0.706 & 0.012 & 0.398 & 0.562 & 0.270 & 0.306 & 0.459 & 0.061 \\
     \hline 
     LMC & & & & & & & & & & & & \\
    \hline
     LMC 064048 & 0.323 & 0.204 & 0.266 & 0.047 & 0.655 & 0.182 & 0.248 & 0.288 & 0.292 & 0.554 & 0.605 & 0.964 \\
     LMC 109106 & 0.356 & 0.243 & 0.328 & 0.065 & 0.712 & 0.150 & 0.290 & 0.333 & 0.310 & 0.343 & 0.690 & 0.738 \\
     LMC 116895 & 0.320 & 0.213 & 0.353 & 0.064 & 0.747 & 0.187 & 0.333 & 0.515 & 0.336 & 0.684 & 0.784 & 1.031 \\
     LMC 141430 & 0.275 & 0.359 & 0.309 & 0.045 & 0.706 & 0.167 & 0.323 & 0.513 & 0.335& 0.597 & 0.720 & 0.799 \\
     LMC 142202 & 0.324 & 0.407 & 0.334 & 0.036 & 0.751 & 0.133 & 0.377 & 0.499 & 0.371 & 0.621 & 0.730 & 0.779 \\
     LMC 146126 & 1.103 & 0.446 & 0.501 & 0.026 & 1.100 & 0.170 & 0.4139 & 0.609 & 0.475 & 0.803 & 1.001 & 0.763 \\
     LMC 061753 & 0.897 & 0.368 & 0.361 & 0.085 & 0.974 & 0.283 & 0.400 & 0.566 & 0.449 & 0.746 & 1.438 & 0.725 \\
     LMC 170452 & 0.226 & 0.340 & 0.349 & 0.067 & 1.017 & 0.129 & 0.425 & 0.351 & 0.247 & 0.562 & 0.714 & 0.650 \\
     WOH S274 & 0.314 & 0.455 & 0.386 & 0.067 & 1.153 & 0.141 & 0.491 & 0.480 & 0.271 & 0.667 & 0.832 & 0.689 \\
     HV 12802 & 0.618 & 0.401 & 0.400 & 0.055 & 1.119 & 0.146 & 0.459 & 0.468 & 0.259 & 0.668 & 0.880 & 0.700 \\
     LMC 170079 & 0.290 & 0.366 & 0.253 & 0.054 & 0.675 & 0.161 & 0.239 & 0.296 & 0.298 & 0.591 & 0.682 & 0.680 \\
     LMC 054365 & 0.333 & 0.452 & 0.323 & 0.058 & 0.728 & 0.153 & 0.236 & 0.478 & 0.311 & 0.636 & 0.709 & 0.767 \\
     LMC 068125 & 0.290 & 0.289 & 0.227 & 0.048 & 0.597 & 0.136 & 0.198 & 0.251 & 0.242 & 0.467 & 0.539 & 0.644 \\
     LMC 135720 & 0.252 & 0.218 & 0.190 & 0.035 & 0.600 & 0.092 & 0.154 & 0.222 & 0.162 & 0.476 & 0.633 & 0.701 \\
     LMC 174714 & 0.865 & 0.382 & 0.405 & 0.036 & 1.205 & 0.153 & 0.423 & 0.491 & 0.438 & 0.720 & 0.938 & 0.815 \\
     LMC 175746 & 0.312 & 0.446 & 0.279 & 0.052 & 0.379 & 0.156 & 0.278 & 0.292 & 0.289 & 0.565 & 0.683 & 0.727 \\
     \hline
     SMC & & & & & & & & & & & & \\
    \hline
     SMC 005092 & 0.255 & 0.332 & 0.374 & 0.038 & 1.056 & 0.149 & 0.417 & 0.376 & 0.246 & 0.319 & 0.718 & 0.717 \\
     SMC 008930 & 0.329 & 0.238 & 0.357 & 0.047 & 0.725 & 0.157 & 0.454 & 0.319 & 0.225 & 0.300 & 0.687 & 0.626 \\
     SMC 018136 & 0.332 & 0.439 & 0.387 & 0.039 & 0.694 & 0.139 & 0.257 & 0.325 & 0.212 & 0.297 & 0.760 & 0.586 \\
     SMC 020133 & 0.227 & 0.397 & 0.369 & 0.038 & 0.616 & 0.116 & 0.385 & 0.326 & 0.241 & 0.277 & 0.738 & 0.510 \\
     SMC 025879 & 0.286 & 0.336 & 0.357 & 0.035 & 0.663 & 0.168 & 0.294 & 0.327 & 0.229 & 0.277 & 0.694 & 0.631 \\
     SMC 050840 & 0.328 & 0.251 & 0.328 & 0.032 & 0.666 & 0.119 & 0.247 & 0.304 & 0.199 & 0.288 & 0.743 & 0.522 \\
     SMC 060447 & 0.317 & 0.258 & 0.324 & 0.056 & 0.628 & 0.124 & 0.263 & 0.303 & 0.214 & 0.249 & 0.699 & 0.652 \\
     SMC 069886 & 0.190 & 0.094 & 0.263 & 0.061 & 0.459 & 0.183 & 0.104 & 0.298 & 0.176 & 0.283 & 0.649 & 0.539 \\
     SMC 078282 & 0.343 & 0.215 & 0.292 & 0.059 & 0.583 & 0.103 & 0.176 & 0.275 & 0.176 & 0.255 & 0.617 & 0.477 \\
     SMC 055275 & 0.229 & 0.254 & 0.333 & 0.044 & 0.656 & 0.145 & 0.191 & 0.361 & 0.235 & 0.291 & 0.634 & 0.691 \\
     SMC 056389 & 0.318 & 0.396 & 0.376 & 0.055 & 0.731 & 0.124 & 0.450 & 0.349 & 0.257 & 0.313 & 0.780 & 0.571 \\
     J00534794-7202095 & 0.301 & 0.193 & 0.341 & 0.040 & 0.665 & 0.137 & 0.451 & 0.325 & 0.217 & 0.256 & 0.697 & 0.547 \\
     SMC 011709 & 0.309 & 0.263 & 0.345 & 0.046 & 0.704 & 0.133 & 0.476 & 0.309 & 0.264 & 0.372 & 0.722 & 0.563 \\
     SMC 046497 & 0.295 & 0.237 & 0.313 & 0.025 & 0.643 & 0.122 & 0.337 & 0.304 & 0.222 & 0.284 & 0.820 & 0.508 \\
     SMC 049478 & 0.246 & 0.371 & 0.386 & 0.040 & 1.032 & 0.117 & 0.385 & 0.499 & 0.205 & 0.306 & 0.813 & 0.590 \\
     SMC 052334 & 0.290 & 0.354 & 0.344 & 0.044 & 0.660 & 0.121 & 0.364 & 0.327 & 0.217 & 0.285 & 0.747 & 0.504 \\
     SMC 056732 & 0.327 & 0.222 & 0.310 & 0.038 & 0.617 & 0.115 & 0.253 & 0.284 & 0.197 & 0.271 & 0.796 & 0.614 
      \enddata 
   \end{deluxetable*}{}

\startlongtable
   \begin{deluxetable*}{lccccc}
   \tablewidth{0pt}
   \tablecolumns{6}
   \tablecaption{Ti I Equivalent Widths}
   \tablehead{
      \colhead{Star} & \colhead{8518.1} & \colhead{8683.0} & \colhead{8692.0} & \colhead{8730.5} & \colhead{8734.5} 
      }
      \startdata
      Milky Way & & & & & \\
      \hline
      BD+59 38 & 0.422 & 0.497 & 0.261 & 0.170 &  0.345 \\
      BD+56 595 & 0.412 & 0.430 & 0.277 & 0.202 & 0.305 \\    
      BD+57 647 & 0.511 & 0.582 & 0.292 & 0.205 & 0.305 \\
      BD+59 274 & 0.473 & 0.434 & 0.253 & 0.189 & 0.285 \\
      BD+59 372 & 0.463 & 0.422 & 0.226 & 0.179 & 0.274 \\
      BD+60 335 & 0.459 & 0.536 & 0.238 & 0.181 & 0.348 \\
      BD+60 2613 & 0.367 & 0.531 & 0.198 & 0.123 & 0.278 \\
      BD+60 2634 & 0.374 & 0.701 & 0.287 & 0.160 & 0.340 \\
      Case 23 & 0.480 & 0.453 & 0.211 & 0.184 & 0.348 \\
      Case 80 & 0.459 & 0.426 & 0.215 & 0.181 & 0.319 \\
      Case 81 & 0.523 & 0.443 & 0.268 & 0.205 & 0.314 \\
      HD 14469 & 0.395 & 0.532 & 0.239 & 0.199 & 0.274 \\
      HD 14488 & 0.279 & 0.443 & 0.202 & 0.136 & 0.333 \\
      HD 23475 & 0.447 & 0.420 & 0.247 & 0.173 & 0.291 \\
      HD 35601 & 0.508 & 0.444 & 0.201 & 0.199 & 0.317 \\
      HD 36389 & 0.528 & 0.493 & 0.233 & 0.209 & 0.346 \\
      HD 37536 & 0.484 & 0.437 & 0.185 & 0.177 & 0.326 \\
      HD 42475 & 0.474 & 0.524 & 0.220 & 0.148 & 0.273 \\
      HD 42543 & 0.536 & 0.567 & 0.225 & 0.208 & 0.328 \\
      HD 44537 & 0.543 & 0.549 & 0.293 & 0.212 & 0.288 \\
      HD 219978 & 0.491 & 0.465 & 0.256 & 0.192 & 0.298 \\
      HD 236697 & 0.480 & 0.481 & 0.201 & 0.178 & 0.312 \\
      HD 236871 & 0.458 & 0.523 & 0.257 & 0.204 & 0.335 \\
      HD 236915 & 0.453 & 0.487 & 0.279 & 0.190 & 0.311 \\
      W Per & 0.412 & 0.653 & 0.296 & 0.187 & 0.362 \\ 
      \hline
       LMC & & & & & \\
      \hline
     LMC 064048 & 0.261 & 0.319 & 0.346 & 0.105 & 0.274 \\
     LMC 109106 & 0.284 & 0.339 & 0.374 & 0.143 & 0.283 \\
     LMC 116895 & 0.267 & 0.342 & 0.437 & 0.164 & 0.273 \\
     LMC 141430 & 0.232 & 0.280 & 0.348 & 0.095 & 0.247 \\
     LMC 142202 & 0.200 & 0.335 & 0.4308 & 0.118 & 0.254 \\
     LMC 146126 & 0.055 & 0.446 & 0.168 & 0.134 & 0.055 \\
     LMC 061753 & 0.373 & 0.342 & 0.247 & 0.100 & 0.132 \\
     LMC 170452 & 0.153 & 0.400 & 0.292 & 0.171 & 0.296 \\
     WOH S274 & 0.231 & 0.346 & 0.359 & 0.191 & 0.279 \\
     HV 12802 & 0.291 & 0.306 & 0.334 & 0.176 & 0.227 \\
     LMC 170079 & 0.229 & 0.368 & 0.367 & 0.120 & 0.295 \\
     LMC 054365 & 0.277 & 0.331 & 0.344 & 0.129 & 0.308 \\
     LMC 068125 & 0.252 & 0.322 & 0.344 & 0.117 & 0.303 \\
     LMC 135720 & 0.190 & 0.386 & 0.463 & 0.092 & 0.275 \\
     LMC 174714 & 0.339 & 0.394 & 0.439 & 0.106 & 0.265 \\
     LMC 175746 & 0.236 & 0.369 & 0.378 & 0.148 & 0.304 \\
     \hline
      SMC & & & & & \\
      \hline
     SMC 005092 & 0.297 & 0.348 & 0.296 & 0.017 & 0.222 \\
     SMC 008930 & 0.206 & 0.279 & 0.222 & 0.071 & 0.196 \\
     SMC 018136 & 0.222 & 0.286 & 0.402 & 0.112 & 0.264 \\
     SMC 020133 & 0.193 & 0.225 & 0.293 & 0.074 & 0.194 \\
     SMC 025879 & 0.173 & 0.276 & 0.199 & 0.040 & 0.164 \\
     SMC 050840 & 0.222 & 0.259 & 0.345 & 0.078 & 0.272 \\
     SMC 060447 & 0.216 & 0.216 & 0.288 & 0.015 & 0.233 \\
     SMC 069886 & 0.118 & 0.201 & 0.381 & 0.020 & 0.217 \\
     SMC 078282 & 0.245 & 0.288 & 0.338 & 0.011 & 0.263 \\
     SMC 055275 & 0.223 & 0.234 & 0.205 & 0.016 & 0.191 \\
     SMC 056389 & 0.179 & 0.259 & 0.238 & 0.039 & 0.169 \\
     J00534794-7202095 & 0.192 & 0.296 & 0.263 & 0.014 & 0.210 \\
     SMC 011709 & 0.211 & 0.265 & 0.267 & 0.076 & 0.187 \\
     SMC 046497 & 0.193 & 0.250 & 0.284 & 0.013 & 0.212 \\
     SMC 049478 & 0.195 & 0.229 & 0.327 & 0.091 & 0.213 \\
     SMC 052334 & 0.203 & 0.260 & 0.289 & 0.065 & 0.218 \\
     SMC 056732 & 0.223 & 0.249 & 0.329 & 0.014 & 0.228 
      \enddata 
   \end{deluxetable*}{}
   
   \startlongtable
   \begin{deluxetable*}{llcccccccc}
   \tablewidth{0pt}
   \tablecolumns{9}
   \tablecaption{MARCS Model Equivalent Widths}
   \tablehead{
      \colhead{log $g$} &\colhead{Feature} & \colhead{3300 K} & \colhead{3400 K} & \colhead{3500 K} & \colhead{3600 K} & \colhead{3700 K} & \colhead{3800 K} & \colhead{3900 K} & \colhead{4000 K}
      }
      \startdata
      Milky Way & & & & & & & & \\
      \hline
      $-$0.5 & Ca II Sum & 9.99 & 10.93 & 11.7 & 12.0 & 12.4 & 12.7 & 13.0 & 13.0 \\
      & Ca I 6575.5 & 0.703 & 0.871 & 1.14 & 1.26 & 1.30 & 1.33 & 1.34 & 1.32 \\ 
      & Ti I 8520.0 & 0.536 & 0.615 & 0.635 & 0.631 & 0.614 & 0.575 & 0.536 & 0.490 \\ 
      & Ti I Sum & 2.02 & 2.23 & 2.39 & 2.50 & 2.54 & 2.60 & 2.57 & 2.47 \\ 
      & Fe I 8516.5 & 0.716 & 0.818 & 0.880 & 0.920 & 0.946 & 0.962 & 0.970 & 0.967  \\ 
      & Fe I Sum & 7.27 & 8.14 & 8.49 & 8.97 & 9.23 & 9.70 & 9.61 & 9.36 \\   
     \hline
     0.0 & Ca II Sum & 9.95 & 11.07 & 12.36 & 13.57 & 13.99 & 14.62 & 14.93 & 15.68 \\
      & Ca I 6575.5 & 0.678 & 0.807 & 1.07 & 1.19 & 1.25 & 1.28 & 1.31 & 1.32 \\
      & Ti I 8520.0 & 0.457 & 0.533 & 0.561 & 0.563 & 0.516 & 0.526 & 0.492 & 0.451 \\
      & Ti I Sum & 1.73 & 1.95 & 2.13 & 2.24 & 2.27 & 2.33 & 2.35 & 2.28 \\
      & Fe I 8516.5 & 0.647 & 0.760 & 0.828 & 0.871 & 0.900 & 0.919 & 0.928 & 0.928 \\
      &Fe I Sum & 6.49 & 7.40 & 7.77 & 8.15 & 8.44 & 8.73 & 8.80 & 8.87 \\
     \hline 
     0.5 & Ca II Sum & 6.85 & 7.43 & 7.97 & 8.45 & 8.79 & 9.11 & 9.29 & 9.81 \\
      & Ca I 6575.5 & 0.650 & 0.694 & 0.953 & 1.10 & 1.19 & 1.27 & 1.31 & 1.30 \\ 
      & Ti 8520.0 & 0.386 & 0.457 & 0.488 & 0.496 & 0.473 & 0.473 & 0.443 & 0.411 \\ 
      & Ti I Sum & 1.44 & 1.63 & 1.74 & 2.51 & 2.90 & 3.08 & 3.24 & 1.98 \\ 
      & Fe I 8516.5 & 0.580 & 0.693 & 0.762 & 0.810 & 0.844 & 0.865 & 0.879 & 0.886 \\ 
      & Fe I Sum & 5.64 & 6.23 & 6.76 & 7.28 & 7.70 & 7.92 & 7.99 & 7.94 \\ 
      \hline  
     1.0 & Ca II Sum & 6.62  & 7.19 & 7.71 & 8.22 & 8.73 & 9.09 & 9.44 & 9.528 \\
      & Ca I 6575.5 & 0.622 & 0.683 & 0.828 & 1.04 & 1.14 & 1.20 & 1.24 & 1.24  \\ 
      & Ti 8520.0 & 0.324 & 0.389 & 0.421 & 0.432 & 0.434 & 0.419 & 0.401 & 0.370 \\ 
      & Ti I Sum & 1.40 & 1.57 & 1.80 & 1.97 & 2.11 & 2.21 & 2.29 & 2.29 \\ 
      & Fe I 8516.5 & 0.523 & 0.620 & 0.690 & 0.741 & 0.777 & 0.812 & 0.822 & 0.832 \\ 
      & Fe I Sum & 3.96 & 4.43 & 4.70 & 4.96 & 5.20 & 5.37 & 5.46 & 5.50 \\ 
      \hline
      LMC & & & & & & & & \\
      \hline
      $-$ 0.5 & Ca II Sum &7.48 &8.28 &9.75 &10.36 &10.84 &11.19 &11.48 &11.50 \\ 
       & Ca I 6575.5 &0.134 &0.145 &0.079 &0.130 &0.150 &0.170 &0.178 &0.182 \\
       & Ti I 8520.0 &0.414 &0.349 &0.322 &0.269 &0.250 &0.231 &0.207 &0.183 \\
       & Ti I Sum &2.279 &2.08 &2.005 &2.12 &2.19 &2.26 &2.29 &2.29 \\
       & Fe I 8516.5 &0.854 &0.662 &0.569 &0.572 &0.582 &0.584 &0.587 &0.569 \\
       & Fe I Sum &3.898 &3.26 &3.346 &3.386 &3.556 &3.664 &3.68 &3.67 \\
      \hline
       0.0 & Ca II Sum &6.81 &7.90 &8.44 &8.91 &9.41 &9.89 &10.23 &10.54 \\ 
       & Ca I 6575.5 &0.657 &0.454 &0.442 &0.309 &0.269 &0.329 &0.316 &0.291 \\ 
       & Ti I 8520.0 &0.394 &0.317 &0.282 &0.265 &0.255 &0.203 &0.182 &0.152 \\
       & Ti I Sum &1.83 &1.71 &1.66 &1.71 &1.76 &1.75 &1.77 &1.74 \\
       & Fe I 8516.5 &3.67 &3.60 &3.53 &3.72 &3.94 &4.016 &4.37 &3.96 \\
       & Fe I Sum &3.67 &3.60 &3.53 &3.72 &3.94 &4.016 &4.37 &3.96 \\
      \hline
       0.5 & Ca II Sum &6.21 &6.63 &6.45 &7.90 &9.50 &10.14 &10.28 &10.71 \\ 
       & Ca I 6575.5 &0.689 &0.570 &0.389 &0.254 &0.294 &0.254 &0.240 &0.219 \\
       & Ti I 8520.0 &0.352 &0.289 &0.246 &0.223 &0.185 &0.168 &0.152 &0.131  \\
       & Ti I Sum &2.22 &2.00 &1.89 &1.85 &1.88 &1.95 &1.99 &2.01 \\
       & Fe I 8516.5 &0.845 &0.680 &0.556 &0.523 &0.538 &0.546 &0.547 &0.537 \\
       & Fe I Sum &3.53 &3.44 &3.32 &3.24 &3.51 &3.56 &3.66 &3.03 \\
      \hline
       1.0 & Ca II Sum &5.50 &5.95 &6.28 &6.64 &7.25 &7.61 &7.89 &8.02 \\ 
       & Ca I 6575.5 &0.096 &0.113 &0.109 &0.096 &0.080 &0.043 &0.052 &0.044 \\
       & Ti I 8520.0 &0.363 &0.288 &0.224 &0.194 &0.166 &0.133 &0.131 &0.115 \\
       & Ti I Sum &2.28 &1.96 &1.77 &1.77 &1.80 &1.80 &1.83 &1.83 \\
       & Fe I 8516.5 &0.827 &0.661 &0.556 &0.512 &0.488 &0.472 &0.484 &0.491 \\
       & Fe I Sum &3.35 &3.04 &2.88 &2.82 &2.81 &2.87 &2.86 &2.91 \\
      \hline
      SMC & & & & & & & & \\
      \hline
       $-$0.5 & Ca II Sum &10.84 &11.88 &12.75 &13.58 &14.11 &14.26 &14.29 &14.38 \\ 
       & Ca I 6575.5 &0.131 &0.112 &0.091 &0.077 &0.060 &0.046 &0.034 &0.025 \\ 
       & Ti I 8520.0 &0.299 &0.243 &0.209 &0.185 &0.164 &0.138 &0.109 &0.087 \\
       & Ti I Sum &1.99 &1.97 &1.96 &1.86 &1.88 &1.88 &1.81 &1.72 \\
       & Fe I 8516.5 &0.646 &0.534 &0.495 &0.508 &0.512 &0.516 &0.502 &0.488 \\
       & Fe I Sum &3.26 &3.21 &3.24 &3.28 &3.42 &3.37 &3.38 &3.24 \\
      \hline
       0.0 & Ca II Sum &9.13 &9.68 &10.15 &10.86 &11.27 &11.6 &11.67 &11.9 \\ 
       & Ca I 6575.5 &0.122 &0.100 &0.087 &0.029 &0.054 &0.044 &0.031 &0.026 \\ 
       & Ti I 8520.0 &0.272 &0.225 &0.175 &0.150 &0.124 &0.105 &0.0878 &0.0792 \\
       & Ti I Sum &1.35 &1.45 &1.39 &1.38 &1.37 &1.37 &1.33 &1.27 \\
       & Fe I 8516.5 &0.843 &0.660 &0.562 &0.524 &0.544 &0.549 &0.553 &0.537 \\
       & Fe I Sum &3.58 &3.49 &3.46 &3.58 &3.53 &3.69 &3.75 &3.75 \\
      \hline
       0.5 & Ca II Sum &7.35 &7.85 &8.34 &8.70 &9.12 &9.33 &9.53 &9.77 \\ 
       & Ca I 6575.5 &0.115 &0.115 &0.053 &0.067 &0.054 &0.039 &0.020 &0.024 \\ 
       & Ti I 8520.0 &0.261 &0.208 &0.157 &0.129 &0.114 &0.089 &0.0724 &0.0613 \\
       & Ti I Sum &1.85 &1.72 &1.66 &1.63 &1.66 &1.63 &1.61 &1.55 \\
       & Fe I 8516.5 &0.1561 &0.091 &0.0452 &0.0205 &0.0267 &0.0359 &0.0329 &0.0469\\
       & Fe I Sum &2.33 &2.34 &2.20 &2.17 &2.16 &2.11 &2.07 &2.01 \\
       \hline
       1.0 & Ca II Sum &6.46 &6.36 &6.89 &7.31 &7.59 &7.86 &7.87 &8.05 \\ 
       & Ca I 6575.5 &0.096 &0.113 &0.109 &0.096 &0.080 &0.043 &0.052 &0.044 \\ 
       & Ti I 8520.0 &0.233 &0.176 &0.149 &0.122 &0.103 &0.0797 &0.0608 &0.0537 \\
       & Ti I Sum &1.51 &1.47 &1.58 &1.51 &1.49 &1.43 &1.35 &1.27 \\
       & Fe I 8516.5 &0.587 &0.514 &0.453 &0.427 &0.419 &0.427 &0.432 &0.425 \\
       & Fe I Sum &2.84 &2.72 &2.73 &2.62 &2.59 &2.55 &2.52 &2.39 
    \enddata 
   \end{deluxetable*}{}
   
\startlongtable
  \begin{deluxetable*}{lccccccc}
 \tablewidth{0pt}
 \tabletypesize{\footnotesize}
 \tablecolumns{8}
 \tablecaption{Correlation and Best-Fit Coefficients - Observations}
 \tablehead{
     \colhead{EW} & \colhead{R1:} & \colhead{P1:} & \colhead{Linear} & \colhead{Linear} & \colhead{Quadratic} & \colhead{Quadratic} & \colhead{Quadratic} \\
     \colhead{} & \colhead{$T_{\rm eff}$ vs EW} & \colhead{$T_{\rm eff}$ vs EW} & \colhead{$A \times 10^{-3}$} & \colhead{B} & \colhead{$A \times 10^{-4}$} & \colhead{$B \times 10^{-2}$} & \colhead{C} 
     } 
    \startdata
    Milky Way & & & & & & & \\
    \hline
     Ca II Sum & 0.755 & 0.00001 & 14.2 & -41.99 & -3.6 & 28.11 & -532.2 \\
     Ca I 6572.0 & 0.500 & 0.01 & 0.019 & -0.3407 & -0.09979 & 0.75 & -13.81 \\
     Ti I Sum & -0.0115 & 0.9 & -0.02 & -1.755 & 0.0083 & -0.0063 & 2.874 \\
     Ti I/Fe I Ratio & 0.232 & 0.3 & 0.0005 & -0.1956 & -0.00271 & 24.76 & -385.29 \\ 
     Fe I Sum & 0.234 & 0.261 & 0.8 & 0.1800 & -0.1071 & 7.953 & -144.4 \\
     \hline
     LMC & & & & & & & \\ 
     \hline
     CaT Sum & 0.320 & 0.3 & 370 & -3.76 & -1.57 & 11.8 & -211.6 \\ 
     Ca I 6572.0 & 0.455 & 0.08 & 0.20 & -0.246 & -0.0691 & 0.520 & -9.39 \\
     Ti I Sum & -0.508 & 0.04 & -0.70 & 3.86 & -0.00048 & 0.034 & -60.09 \\
     Ti I/Fe I Ratio & -0.594 & 0.05 & -1.2 & 4.927 & -0.377 & 2.63 & -44.98 \\ 
     Fe I Sum & 0.712 & 0.002 & 2.00 & -19.98 & -0.691 & 5.73 & -111.42 \\
     \hline
     SMC & & & & & & & \\ 
     \hline
     CaT Sum & 0.344 & 0.2 & 4.0 & -5.63 & -0.1446 & 1.105 & -0.0202 \\
     Ca I 6572.0 & -0.0847 & 0.7 & -0.02 & 0.412 & 0.06027 & -0.0446 & 8.5192 \\
     Ti I Sum & -0.490 & 0.05 & -0.6 & 3.35 & 0.2461 & -0.1876 & 36.75 \\
     Ti I/Fe I Ratio & -0.465 & 0.06 & -0.8 & 3.52 & 0.4702 & -0.354 & 67.33 \\  
     Fe I Sum & -0.3054 & 0.2 & -1.4 & 9.56 & 0.5332 & -0.407 & 81.91 
    \enddata
 \end{deluxetable*}
 
\startlongtable
  \begin{deluxetable*}{lcccccccc}
 \tablewidth{0pt}
 \tabletypesize{\footnotesize}
 \tablecolumns{9}
 \tablecaption{Correlation and Best-Fit Coefficients - MARCS Models}
 \tablehead{
     \colhead{log $g$} & \colhead{EW} & \colhead{R1:} & \colhead{P1:} & \colhead{Linear} & \colhead{Linear} & \colhead{Quadratic} & \colhead{Quadratic} & \colhead{Quadratic} \\
     \colhead{} & \colhead{} & \colhead{$T_{\rm eff}$ vs EW} & \colhead{$T_{\rm eff}$ vs EW} & \colhead{$A \times 10^{-2}$} & \colhead{B} & \colhead{$A \times 10^{-4}$} & \colhead{B} & \colhead{C}
     } 
    \startdata
    Milky Way & & & & & & & & \\
    \hline
     $-$0.5 & CaT Sum & 0.937 & 0.0002 & 0.34 & -0.5974 & -0.04081 & -0.03412 & -57.97 \\
       & Ca I 6575.5 & 0.780 & 0.0131 & 0.05918 & -1.0271 & -0.01515 & 0.01985 & -22.32 \\
     - & Ti I Sum & 0.0301 & 0.939 & 0.002432 & 2.2700 & -0.02727 & 0.02053 & -36.07 \\
     - & Ti I/Fe I Ratio & -0.979 & 0.000004 & -0.04 & 2.2208 & -0.000323 & 0.00199 & -2.29 \\
     - & Fe I Sum & 0.518 & 0.153 & 0.1361 & 3.7286 & -0.07506 & 0.05782 & -101.88 \\
     \hline
    0.0 & CaT Sum & 0.869 & 0.0023 & 0.67 & -7.875 & -0.1269 & 0.1022 & -186.3 \\
     - & Ca I 6575.5 & 0.834 & 0.0052 & 0.06 & 1.275 & -0.01383 & 0.01105 & -20.7247 \\
     - & Ti I Sum & 0.347 & 0.3598 & 0.03 & 1.198 & -0.02288 & 0.01746 & -30.97 \\
     - & Ti I/Fe I Ratio & -0.980 & 0.000004 & -0.04 & 2.0135 & -0.002211 & 0.001278 & -1.0956 \\
     - & Fe I Sum & 0.604 & 0.0850 & 0.0016 & 224.74 & -0.06842 & 0.05302 & -93.95 \\
     \hline
     0.5 & CaT Sum & 0.926 & 0.0003 & 0.3 & -2.664 & -0.03986 & 0.033005 & -58.70 \\
     - & Ca I 6575.5 & 0.874 & 0.0048 & 0.07 & -1.541 & -0.0149 & 0.01908 & -22.48 \\
     - & Ti I Sum & 0.409 & 0.275 & 0.08988 & -1.0564 & 0.05430 & 0.04175 & -77.40 \\
     - & Ti I/Fe I Ratio & -0.974 & 0.001 & -0.0004 & 1.913 & - & - & - \\ 
     - & Fe I Sum & 0.785 & 0.0203 & 0.204 & -0.4100 & -0.06111 & 0.0481 & -86.33 \\
    \hline
   1.0 & CaT Sum & 0.818 & 0.0071 & 0.27 & -1.7245 & -0.06196 & 0.04931 & -88.34 \\
     - & Ca I 6575.5 & 0.888 & 0.0014 & 0.07236 & -1.6663 & -0.01196 & 0.009721 & -18.48  \\
     - & Ti I Sum & 0.779 & 0.0134 & 0.08 & -1.0700 & -0.02192 & 0.01731 & -31.89 \\
     - & Ti I/Fe I Ratio & -0.970 & 0.00007 & -0.027 & 1.52 & -0.00031 & 0.00199 & -2.59 \\
     - & Fe I Sum & 0.965 & 0.0001 & 0.216 & -2.95 & -0.0092 & 0.0234 & -41.65 \\
    \hline
    LMC & & & & & & & & \\
    \hline
    $-$0.5 & CaT Sum &0.945 &0.00039 &58.3 &-11.16 &-0.00096 &0.076 &-138 \\
           & Ca I 6575.5 &0.694 &0.0561 &1.000 &-0.2009 &0.0000231 &-0.00159 &2.867 \\
           & Ti I Sum &0.526 &0.1805 &2.4 &1.325 &0.00012 &-0.0086 &17.46 \\
           & Ti I/Fe I Ratio &0.357 &0.386 &0.00003 &90.65 &-0.000000989 &0.000748 &-0.4062 \\
           & Fe I Sum &0.222 &0.5962 &1.900 &2.851 &0.00025 &-0.0179 &35.68 \\
    0.0 & CaT Sum &0.983 &0.00001 &50.70 &-9.479 &-0.000419 &0.0357 &-65.17 \\
           & Ca I 6575.5 &-0.8156 &0.01358 &-4.3 &-0.8156 &0.0001309 &-0.00989 &19.33 \\
           & Ti I Sum &0.997 &0.00186 &0.000004 &1.74 &0.000000554 &-0.00405 &9.097 \\
           & Ti I/Fe I Ratio &-0.843 &0.0086 &-3.00 &1.51 &-0.000847 &0.00589 &-9.73 \\
           & Fe I Sum &0.604 &0.079 &0.01784 &0.0009 &0.5647 &0.00000254 &-0.000957 \\
    0.5 & CaT Sum &0.926 &0.00013 &74.30 &-18.65 &-0.00000842 &0.0136 &29.83 \\
           & Ca I 6575.5 &-0.8847 &0.00356 &-6.300 &2.666 &0.0000149 &-0.0115 &22.43 \\
           & Ti I Sum &-0.3407 &0.409 &-1.6 &2.562 &0.00002027 &-0.0149 &2946 \\
           & Ti I/Fe I Ratio &0.409 &0.3142 &-0.00005 &1.133 &0.0000000678 &-0.00106 &2.729 \\
           & Fe I Sum &0.672 &0.0680 &5.8 &1.40 &0.000295 &-0.0210 &40.60 \\
    1.0 & CaT Sum &0.9917 &0.000001 &38.05 &-6.991 &-0.00001579 &0.0153 &-27.95 \\
           & Ca I 6575.5 &-0.887 &0.0033 &-1.1 &0.463 &-0.00000115 &-1.066 & - \\
           & Ti I Sum &-0.625 &0.0974 &-4.1 &3.386 &0.0000227 &-0.0170 &33.52 \\
           & Ti I/Fe I Ratio &-0.453 &0.259 &-0.3 &1.119 &-0.00000012 &0.000059 &0.955 \\
           & Fe I Sum &-0.663 &0.0731 &-4.9 &4.719 &0.0000257 &-0.0192 &38.77 \\
    \hline
    SMC & & & & & & & & \\
    \hline
    $-$0.5 & CaT Sum &0.927 &0.00093 &0.498 &-4.926 &-0.0999 &0.0779 &-137.46 \\
           & Ca I 6575.5 &-0.994 &0.000001 &-0.015 &0.630 &0.000824 &-0.000754 &1.723 \\
           & Ti I Sum &-0.933 &0.00071 &-0.034 &3.113 &-0.000319 &0.00199 &-1.107 \\
           & Ti I/Fe I Ratio & -0.912 & 0.00158 & -0.043 & 2.266 & 0.000659 & -0.00524 & 11.00 \\
           & Fe I Sum &0.458 &0.255 &0.015 &2.765 &-0.000736 &0.00552 &-7.00 \\
    0.0 & CaT Sum &0.974 &0.00004 &0.406 &-4.0513 &0.000436 &0.0359 &-61.862 \\
           & Ca I 6575.5 &-0.892 &0.00287 &-0.0130 &0.000216 &-0.00170 &3.415 & - \\
           & Ti I Sum &-0.7132 &0.0469 &-0.015 &1.919 &-0.000526 &0.00367 &-5.054 \\
           & Ti I/Fe I Ratio & -0.970 & 0.000071 & -0.0304 & 1.357 & -0.000169 & 0.000930 & -0.886 \\
           & Fe I Sum &0.799 &0.01732 &0.0370 &2.257 &0.000926 &-0.00639 &14.543 \\
    0.5 & CaT Sum &0.983 &0.00001 &0.342 &-3.720 &-0.0288 &0.244 &-41.93 \\
           & Ca I 6575.5 &-0.9244 &0.00102 &-0.000104 &0.569 &0.000133 &-0.00115 &2.343 \\
           & Ti I Sum &0.879 &0.00404 &-3.300 &2.852 &0.000477 &-0.00381 &9.185 \\
           & Ti I/Fe I Ratio &0.391 &0.330 &0.0964 &-23.157 &0.906 &0.670 &-1225.411 \\  
           & Fe I Sum &-0.972 &0.00005 &-4.600 &3.862 &0.000107 &-0.00120 &5.277 \\
    1.0 & CaT Sum &0.967 &0.00009 &0.260 &-2.204 &-0.000179 &0.0157 &-25.924 \\
           & Ca I 6575.5 &-0.887 &0.0033 &-0.011 &0.463 &-0.000115 &0.00753 &-1.0659 \\
           & Ti I Sum &-0.806 &0.0157 &-0.033 &2.653 &-0.000157 &0.00739 &-11.375 \\
           & Ti I/Fe I Ratio &-0.993 &0.000001 &-0.0402 &1.725 &-0.0000362 &-0.000138 &1.244 \\
           & Fe I Sum &-0.975 &0.00004 &-0.050 &4.662 &0.00000121 &-0.000648 &4.284 \\ 
    \enddata
 \end{deluxetable*}

\startlongtable
  \begin{deluxetable*}{cccccc}
 \tablewidth{0pt}
 \tabletypesize{\footnotesize}
 \tablecolumns{6}
 \tablecaption{3D CaT Correlation Coefficients }
 \tablehead{
     \colhead{Galaxy} & \colhead{EW} & \colhead{R2:} & \colhead{P2:} & \colhead{R3:} & \colhead{P3:}  \\
     \colhead{} & \colhead{} & \colhead{EW v. $Log_{\rm g}$} & \colhead{EW v. $Log_{\rm g}$} & \colhead{EW v. $M_{\rm bol}$} & \colhead{EW v. $M_{\rm bol}$} 
     } 
    \startdata
     MW & CaT Sum & 0.665 & 0.00030 & 0.721 & 0.0002 \\
     MW & Ca I 6572.0 & 0.558 & 0.0070  & 0.526 & 0.0119 \\
     MW & Ti I Sum & 0.162 & 0.472 & 0.1593 & 0.4789 \\
     MW & Fe I Sum & -0.0700 & 0.757 & -0.079 & 0.727 \\
     LMC & CaT Sum & 0.443 & 0.0856 & 0.4471 & 0.0825 \\ 
     LMC & Ca I 6572.0 & 0.427 & 0.104 & 0.296 & 0.266 \\
     LMC & Ti I Sum & -0.149 & 0.5808 & 0.209 & 0.438 \\ 
     LMC & Fe I Sum & 0.398 & 0.127 & 0.018 & 0.948 \\
     SMC & CaT Sum & 0.5564 & 0.0204 & 0.631 & 0.0066 \\ 
     SMC & Ca I 6572.0 & 0.3601 & 0.156 & 0.436 & 0.0799 \\
     SMC & Ti I Sum & -0.2147 & 0.4079 & -0.239 & 0.356 \\
     SMC & Fe I Sum & 0.0866 & 0.7411 & 0.0866 & 0.741 \\
    \enddata
 \end{deluxetable*}
 
\startlongtable
  \begin{deluxetable*}{ccccccccccc}
 \tablewidth{0pt}
 \tabletypesize{\footnotesize}
 \tablecolumns{11}
 \tablecaption{CaT 3D Best Fits}
 \tablehead{
     \colhead{Galaxy} & \colhead{EW} & \colhead{Linear} & \colhead{Linear} & \colhead{Linear} & \colhead{Quad} & \colhead{Quad} & \colhead{Quad} & \colhead{Quad} & \colhead{Quad} & \colhead{Quad}  \\ \colhead{} & \colhead{} & \colhead{$A \times 10^{-2}$} & \colhead{B} & \colhead{C} & \colhead{A} & \colhead{B} & \colhead{C} & \colhead{D} & \colhead{$E \times 10^{-4}$} & \colhead{F}
     } 
    \startdata
     MW & CaT Sum & 0.8 & 2.927 & 20.46 & -23.21 & 0.0358 & -8.26 & 0.0027 & -0.0869 & -0.0741 \\ 
     LMC & CaT Sum & 6572.0 & 0.308 & 3.50 & -0.580 & -219.29 & 0.128 & 73.02 & -0.022 & -0.180 \\
     SMC & CaT Sum & -0.02246 & 3.1687 & 10.5622 & -1485.5 & 0.7967 & -303.03 & 0.08187 & -10.6 & -11.340 \\ 
    \enddata 
 \end{deluxetable*}
 
\begin{figure*}
  \includegraphics[width=.3\textwidth] {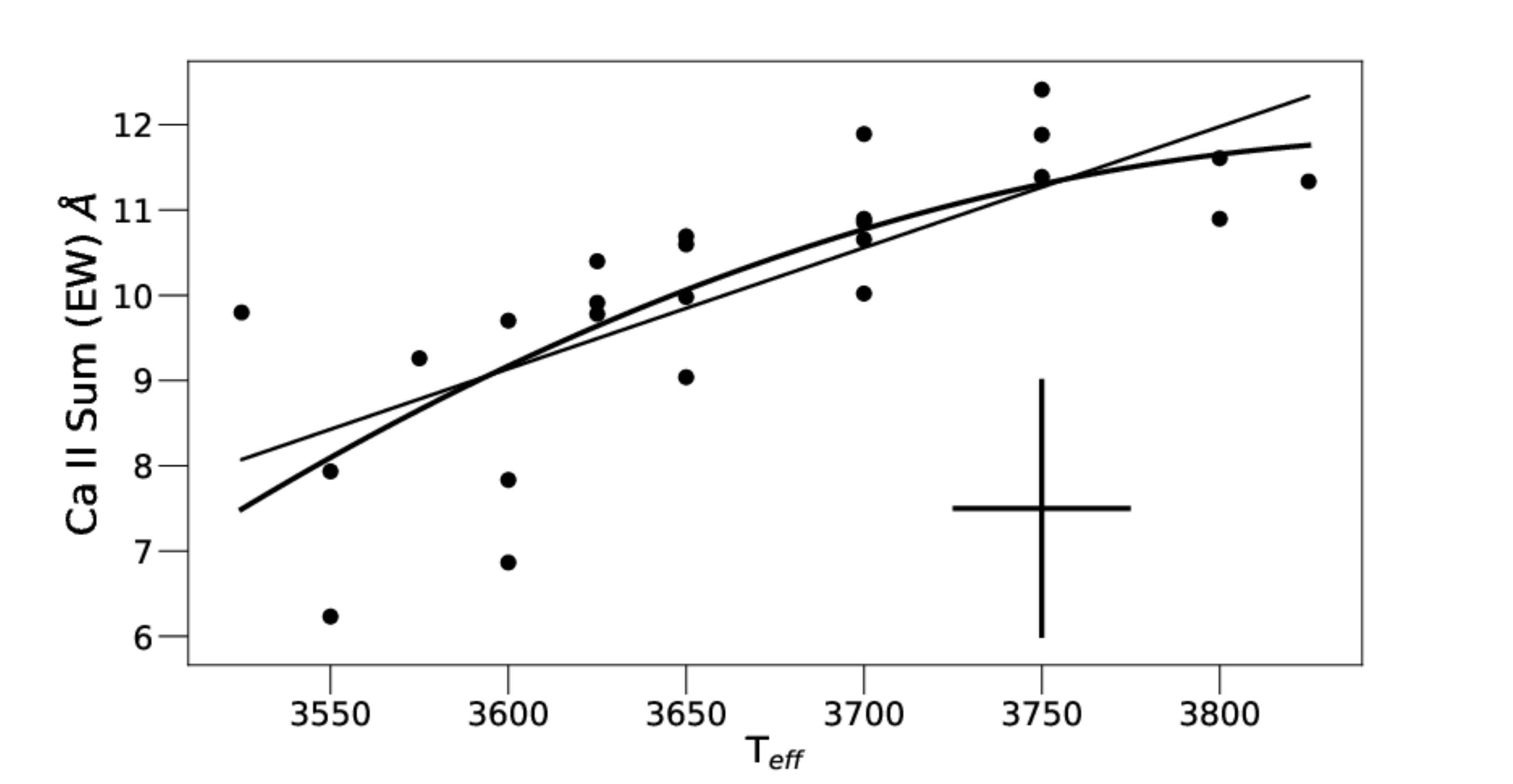}
  \includegraphics[width=.3\textwidth] {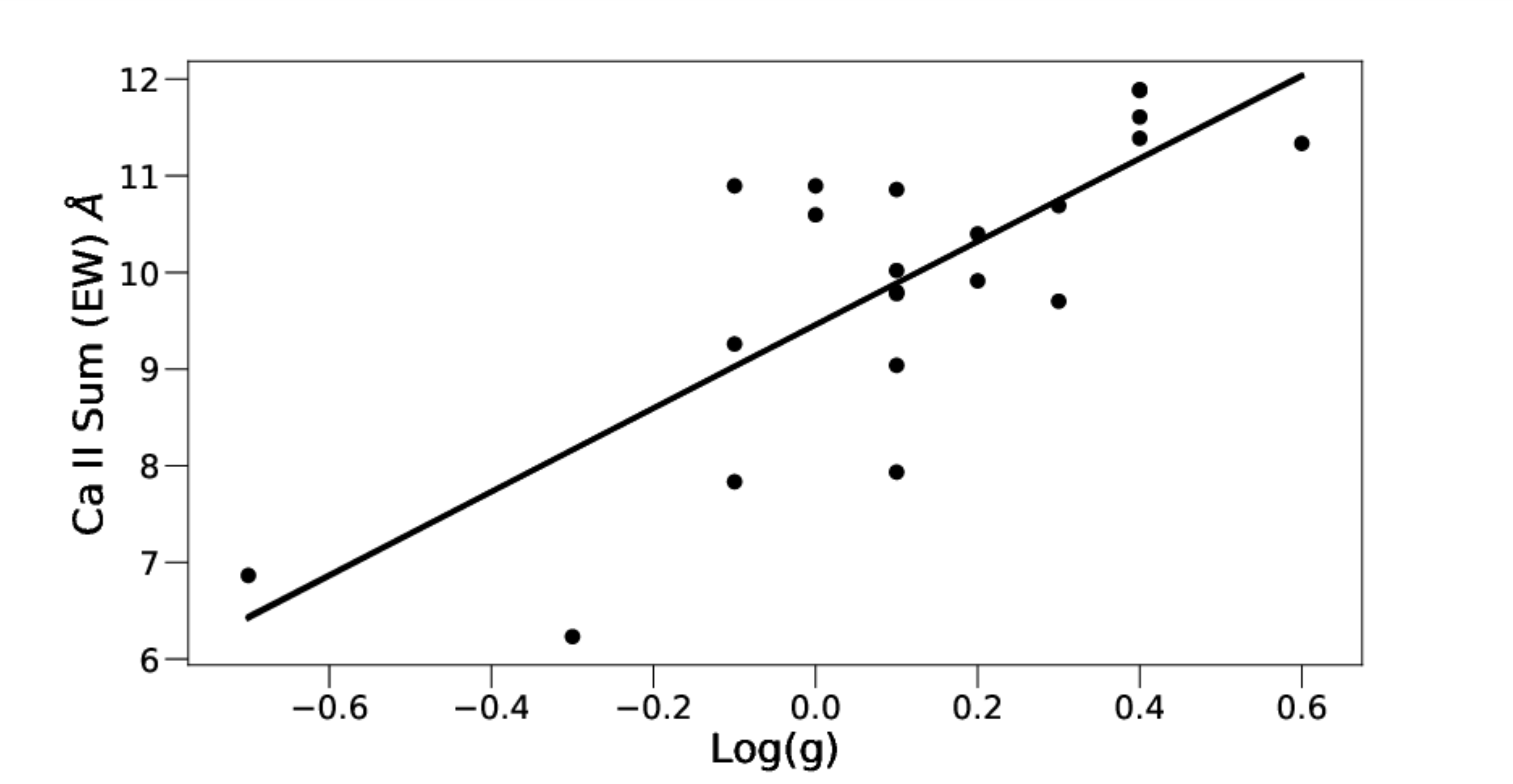}
  \includegraphics[width=.3\textwidth] {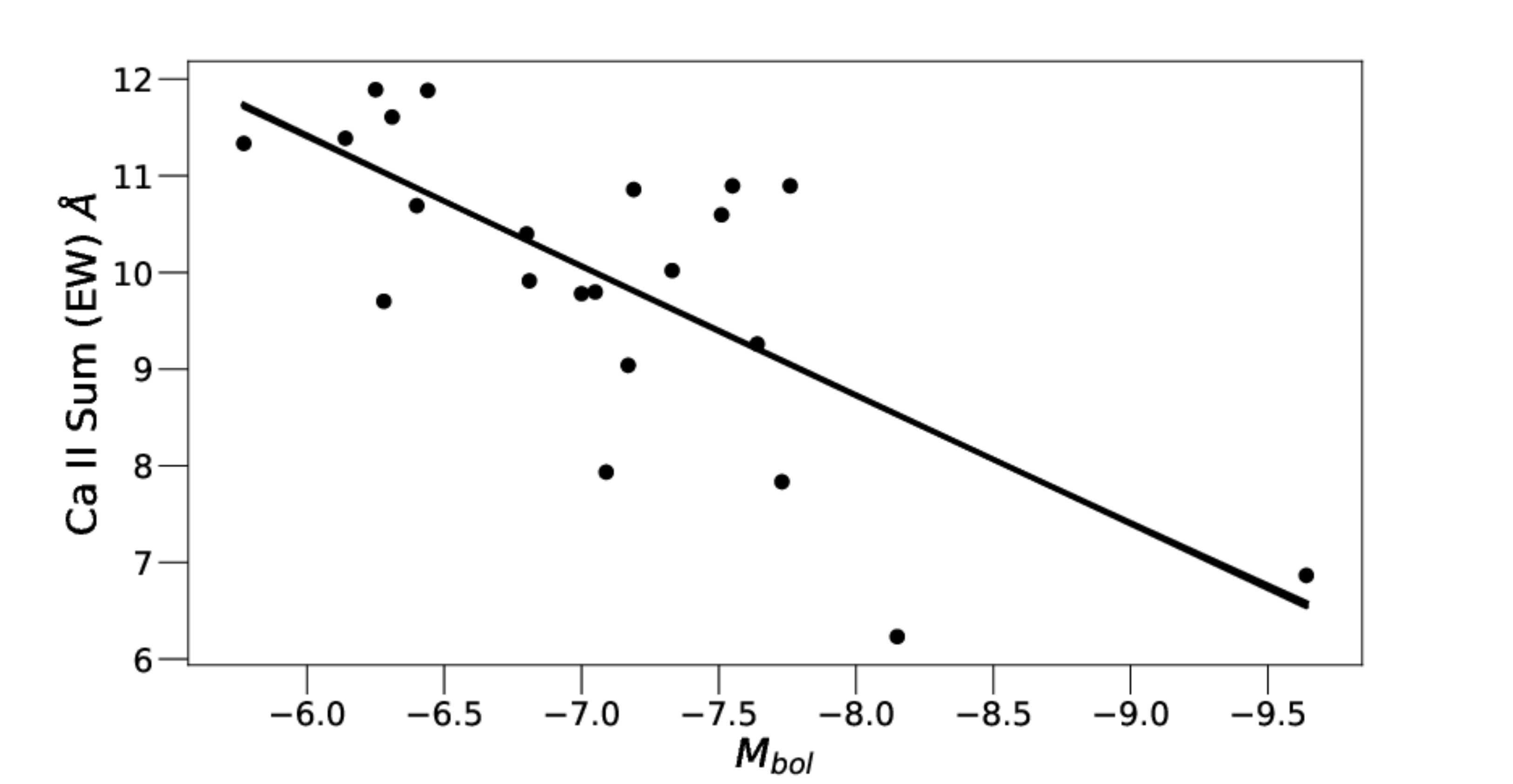}
  \includegraphics[width=.3\textwidth] {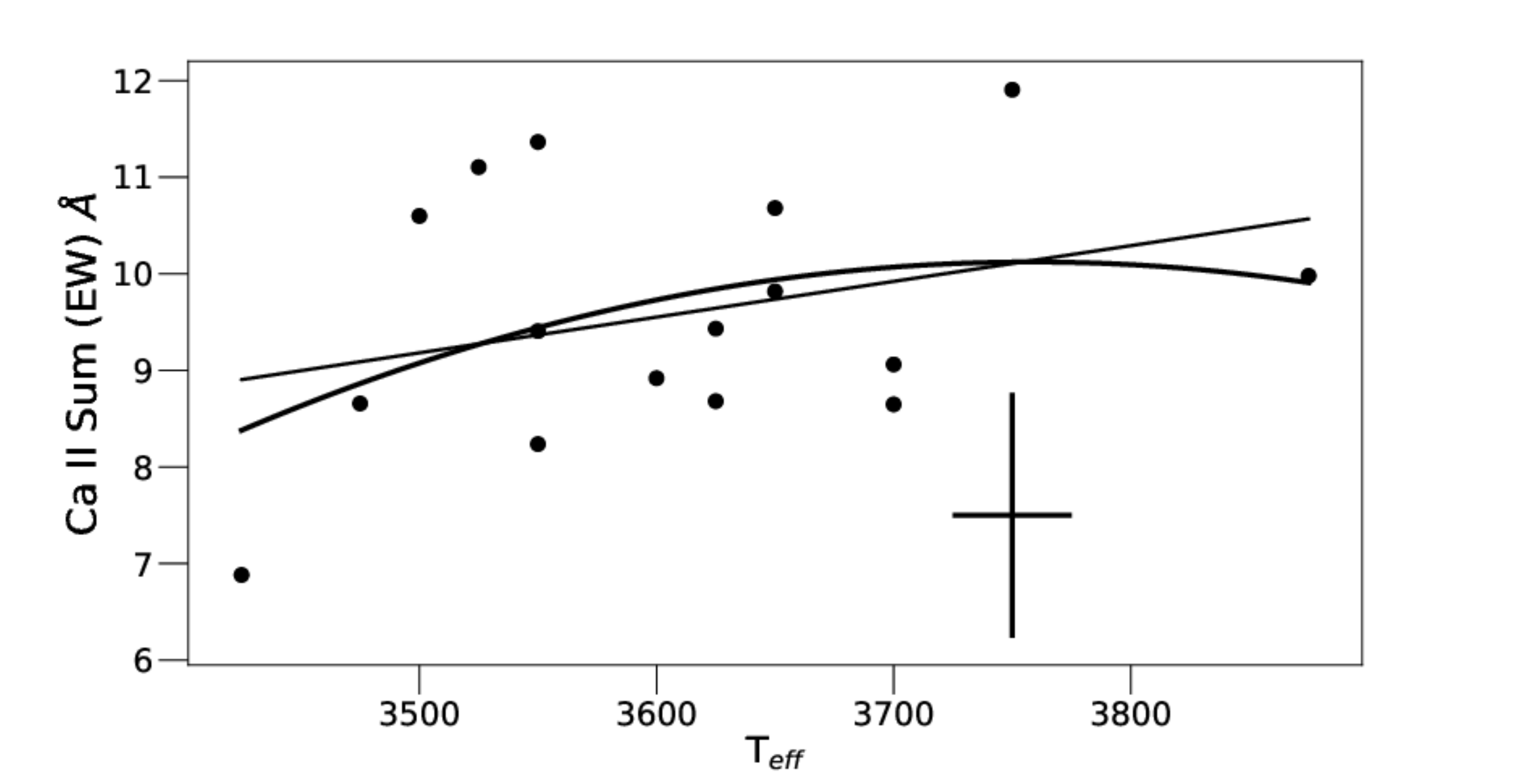}
  \includegraphics[width=.3\textwidth] {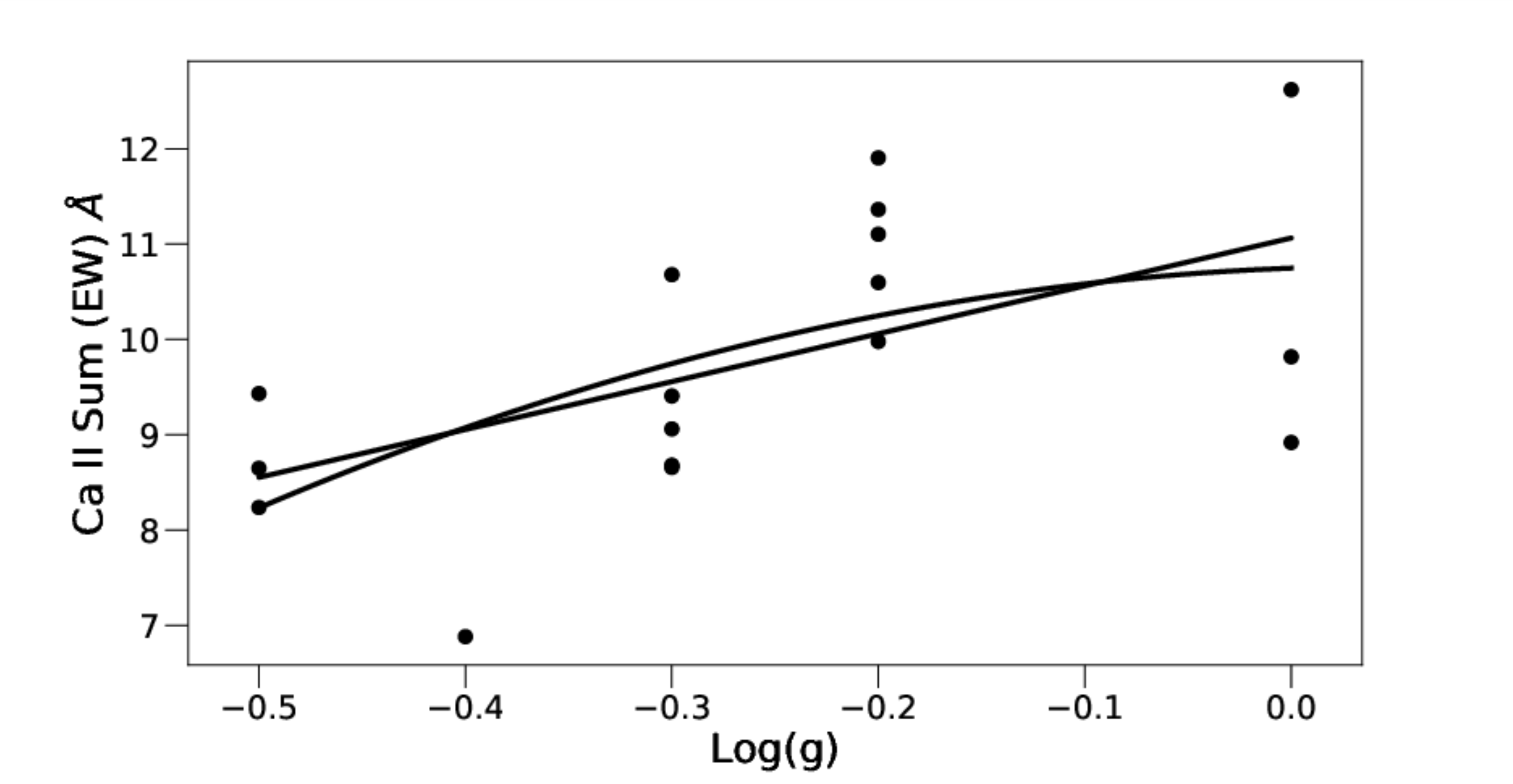}
  \includegraphics[width=.3\textwidth] {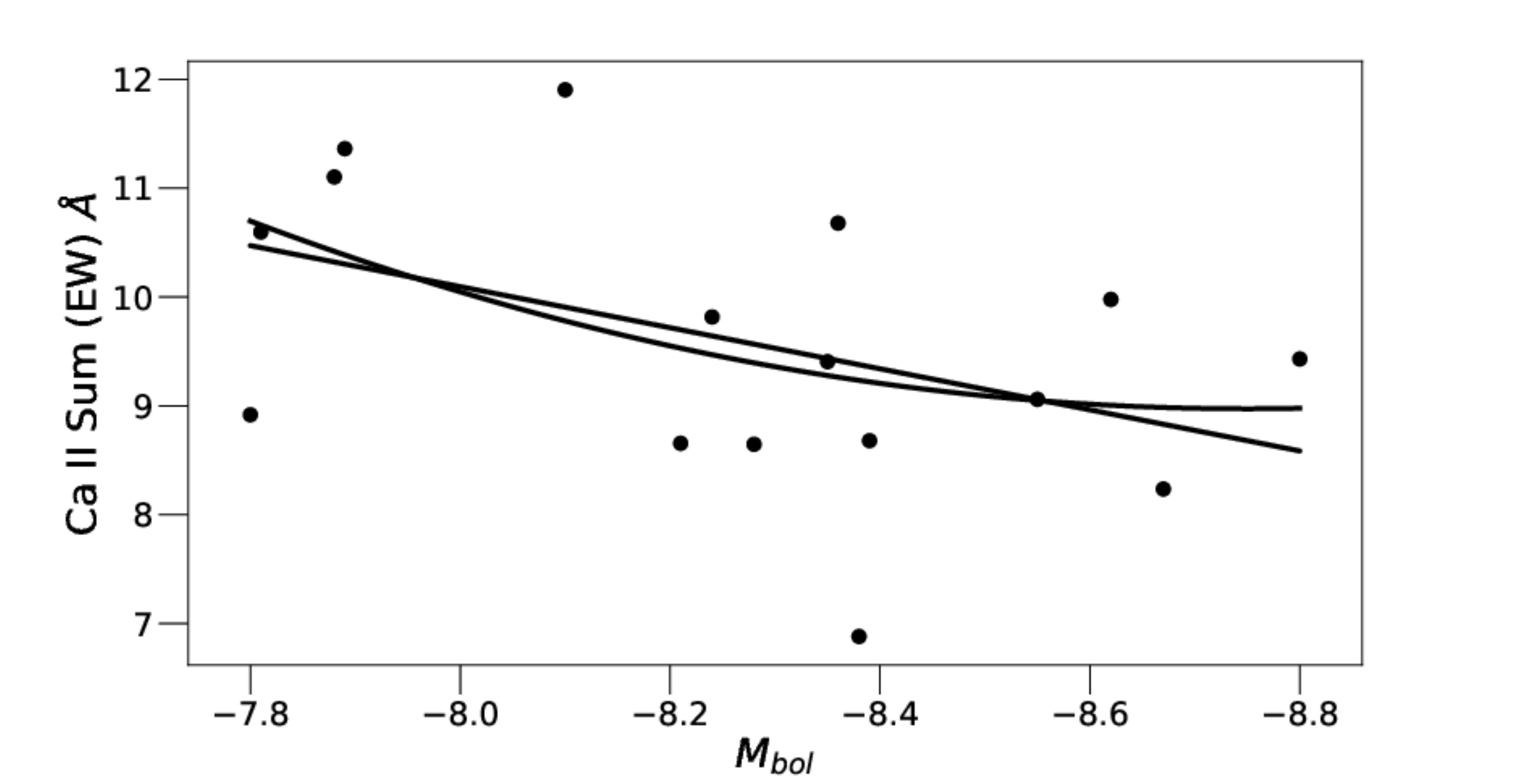}
  \includegraphics[width=.3\textwidth] {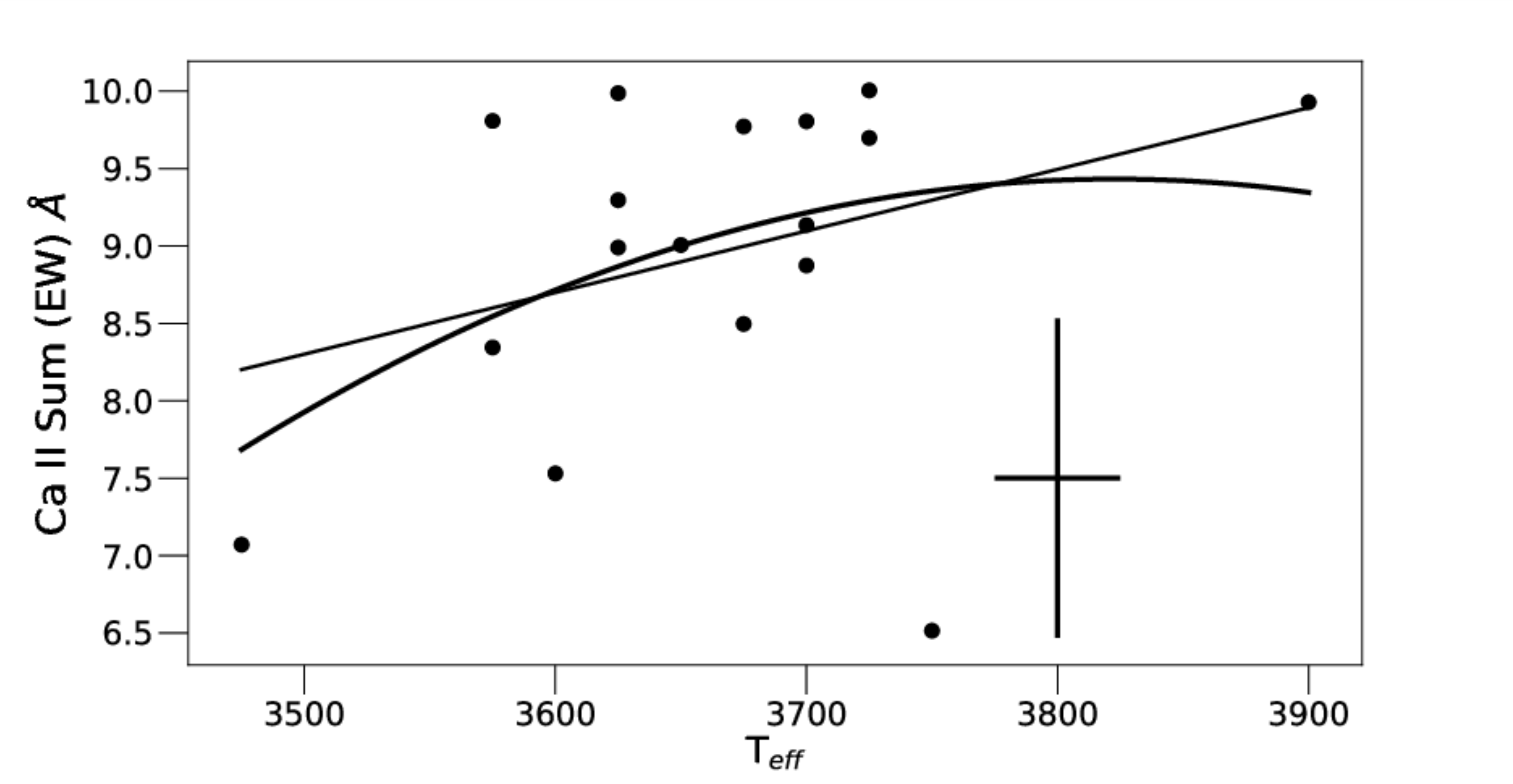}
  \includegraphics[width=.3\textwidth] {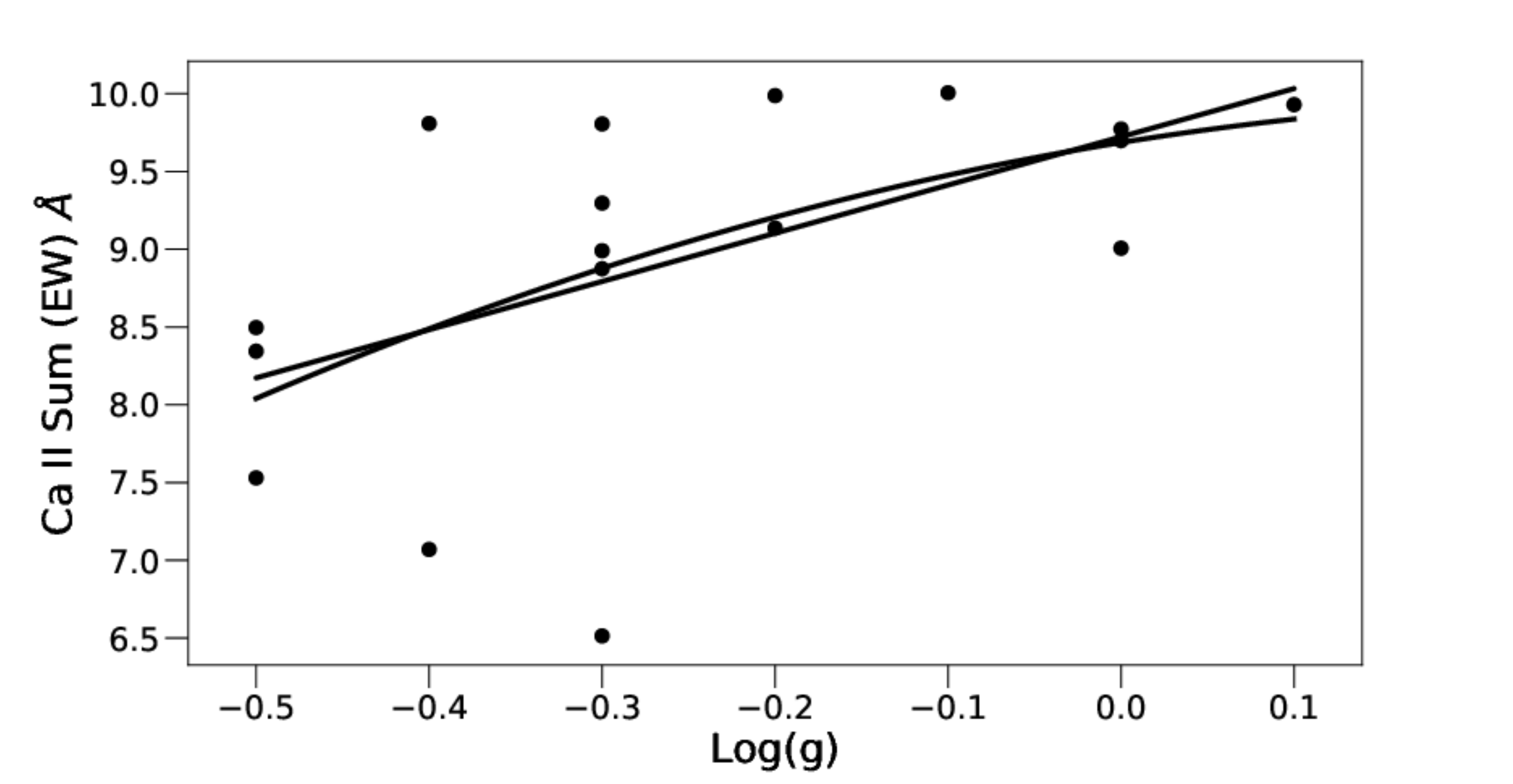}
  \includegraphics[width=.3\textwidth] {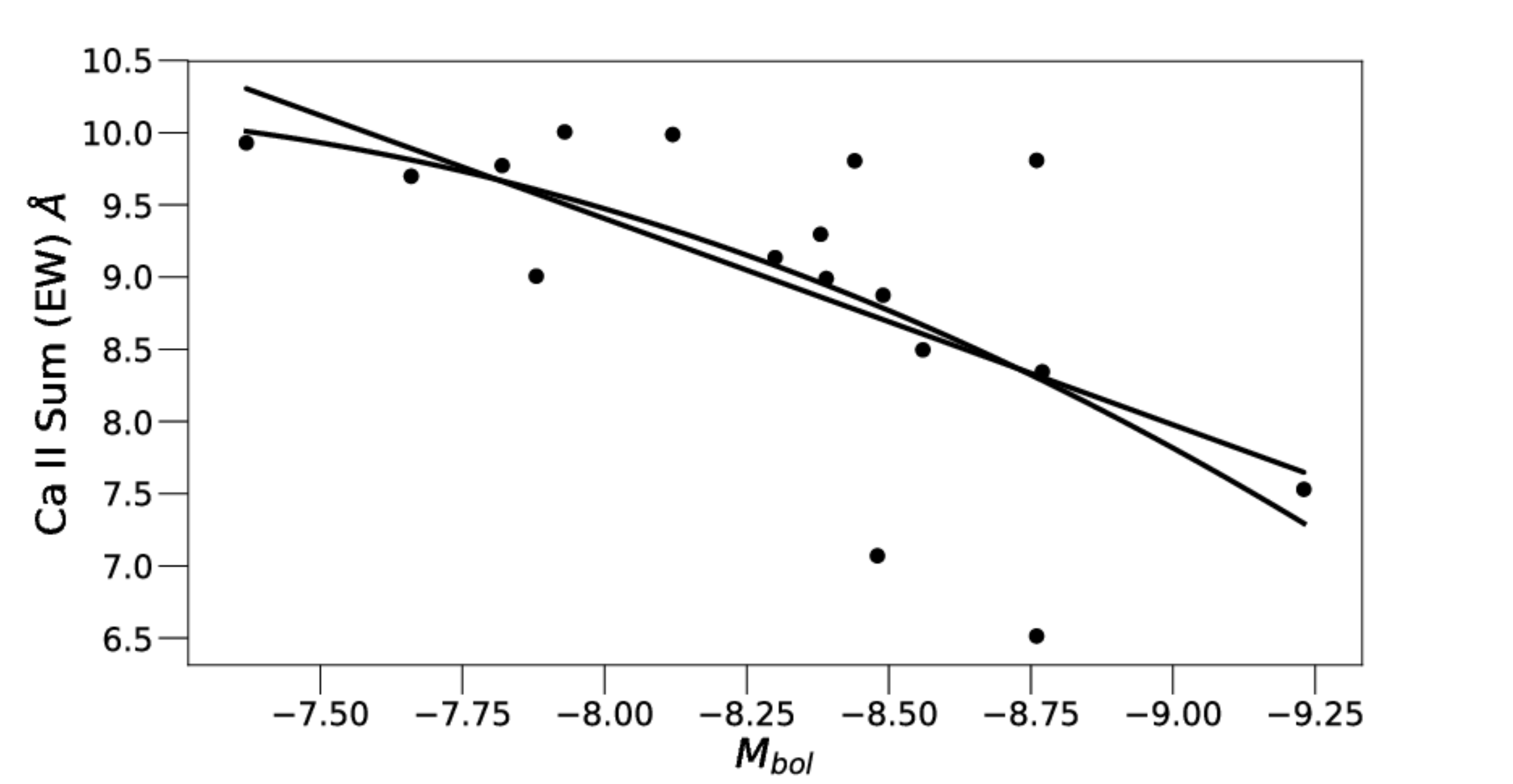}
 \caption{CaT EWs (points) measured in our MW (left), LMC (center), and SMC (bottom) samples, compared to the stars' $T_{\rm eff}$ (left), log $g$ (center), and $M_{\rm bol}$ (right). For each dataset we have plotted linear and quadratic best fits (solid lines). Systematic error bars for our EW measurements (this work) and errors on the stars' physical properties as given in \protect\cite{levesque2005} are indicated by crosses (left).}
\end{figure*}

\begin{figure*}
  \includegraphics[width=.3\textwidth] {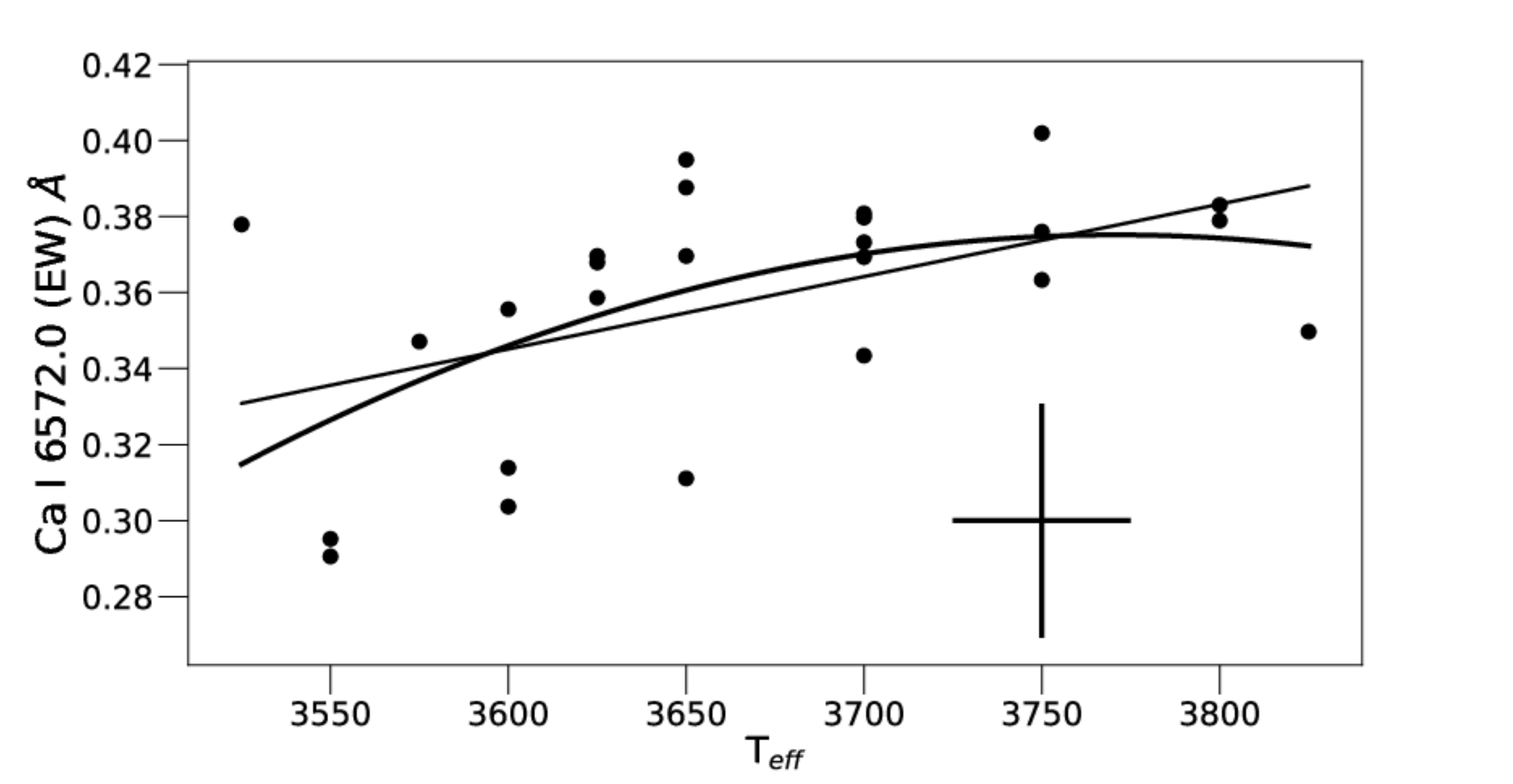}
  \includegraphics[width=.3\textwidth] {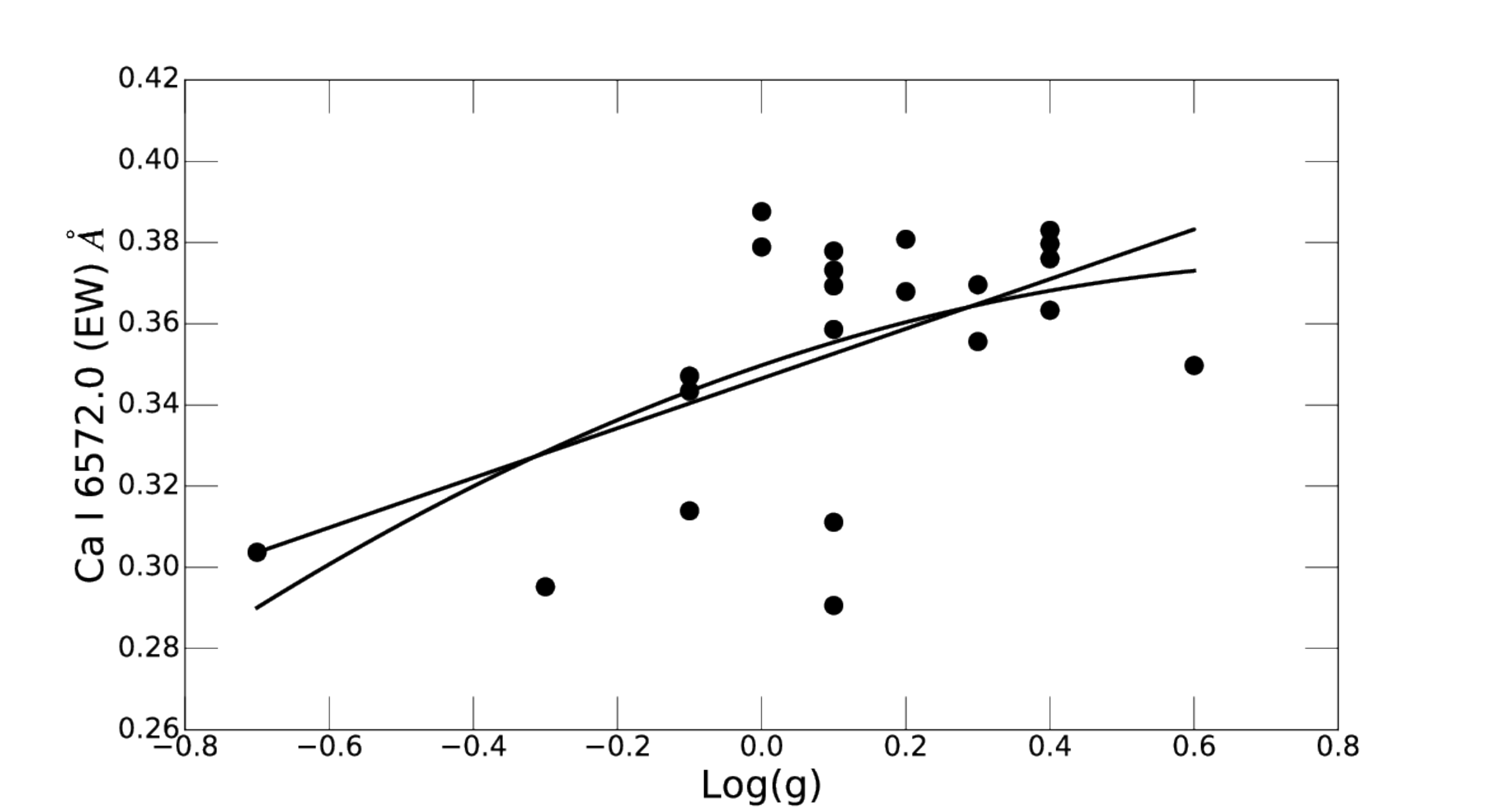}
  \includegraphics[width=.3\textwidth] {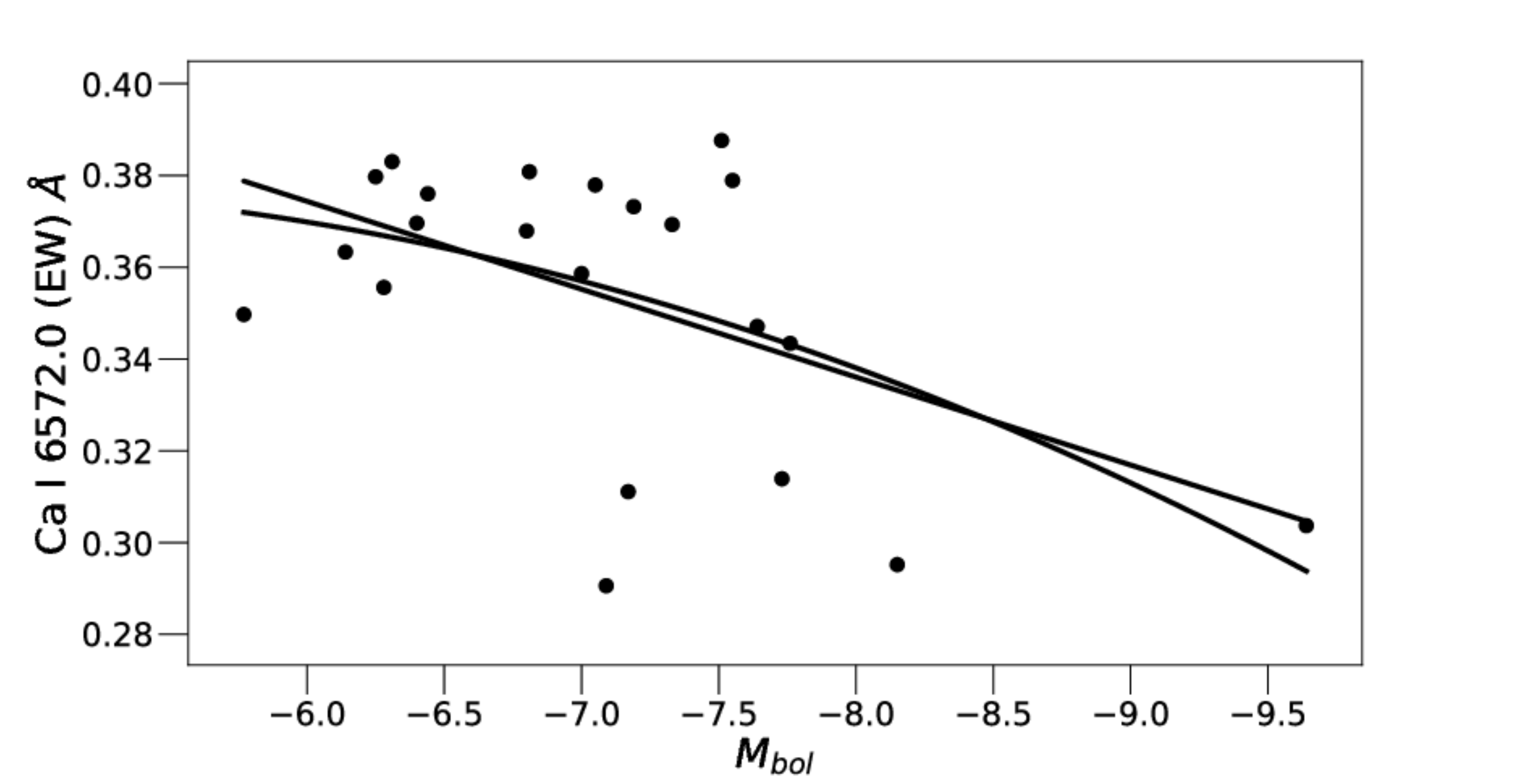}
  \includegraphics[width=.3\textwidth] {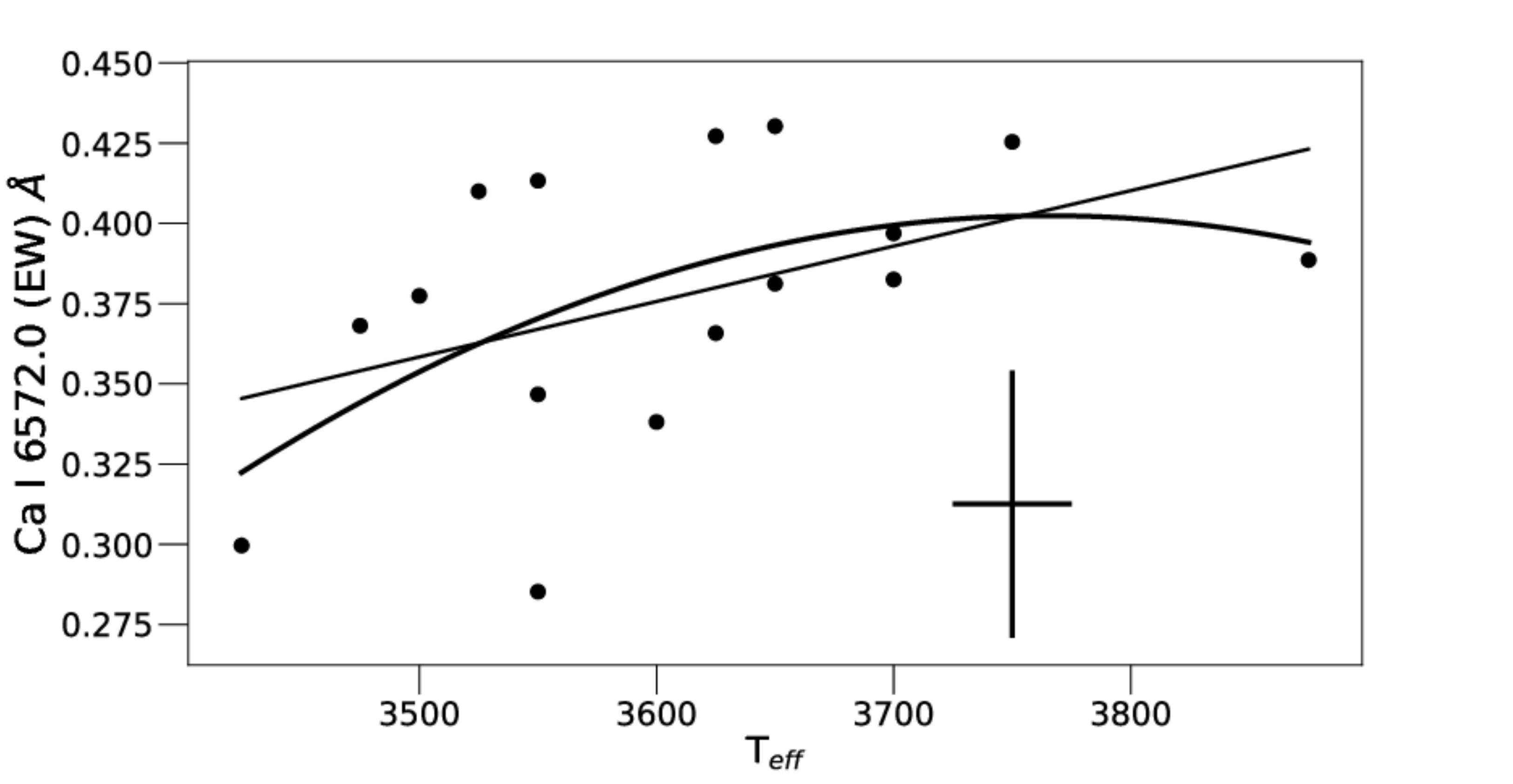}
  \includegraphics[width=.3\textwidth] {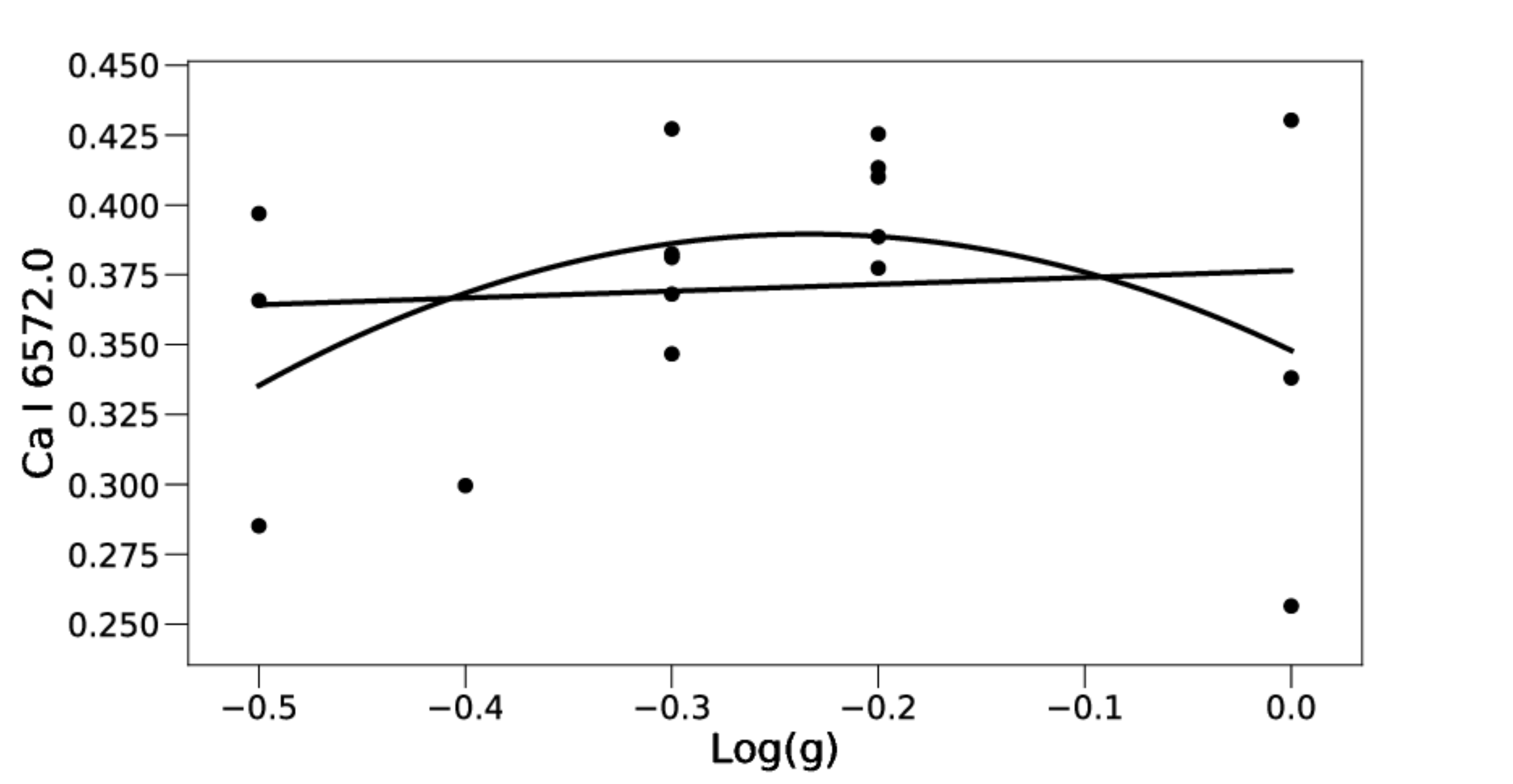}
  \includegraphics[width=.3\textwidth] {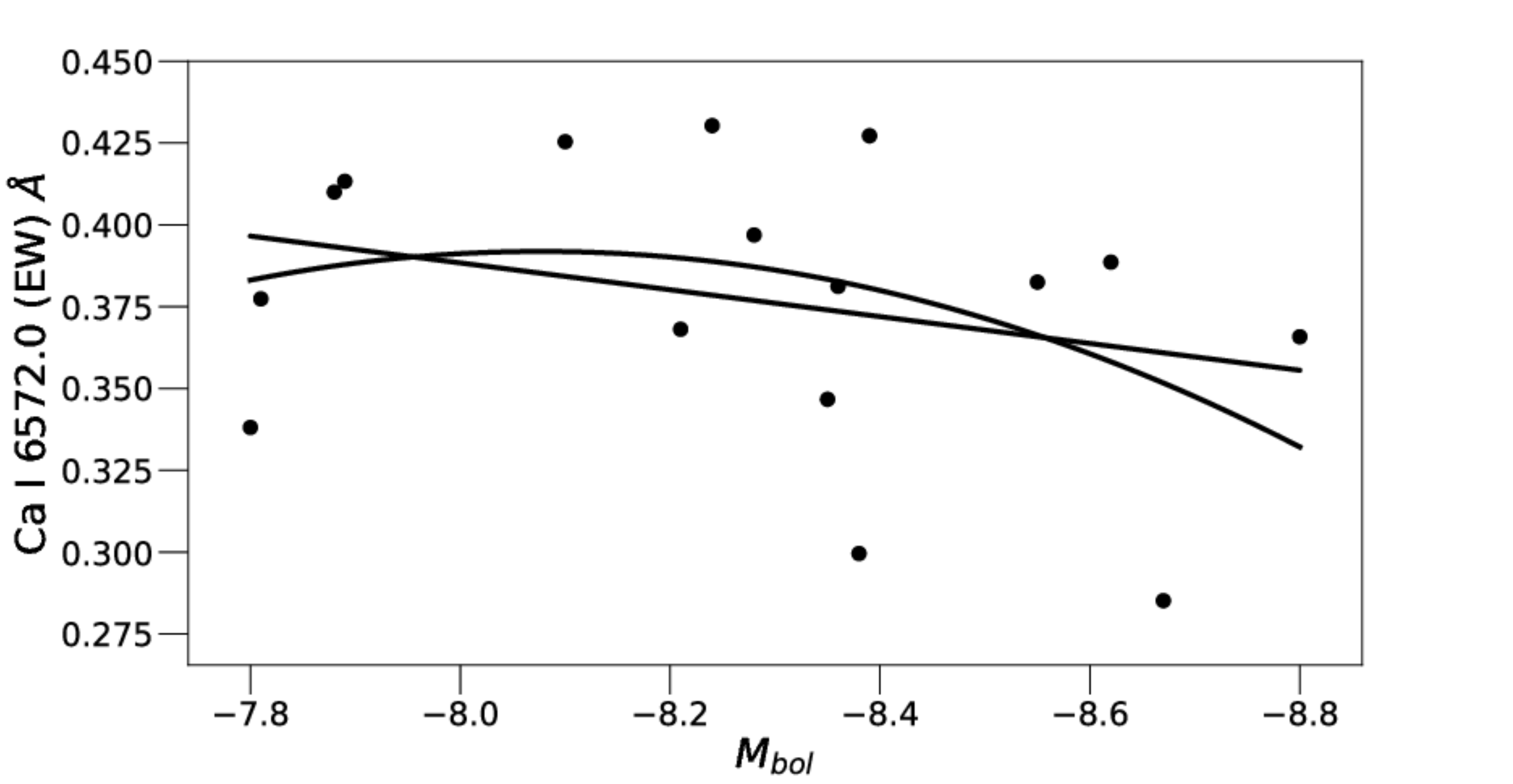}
  \includegraphics[width=.3\textwidth] {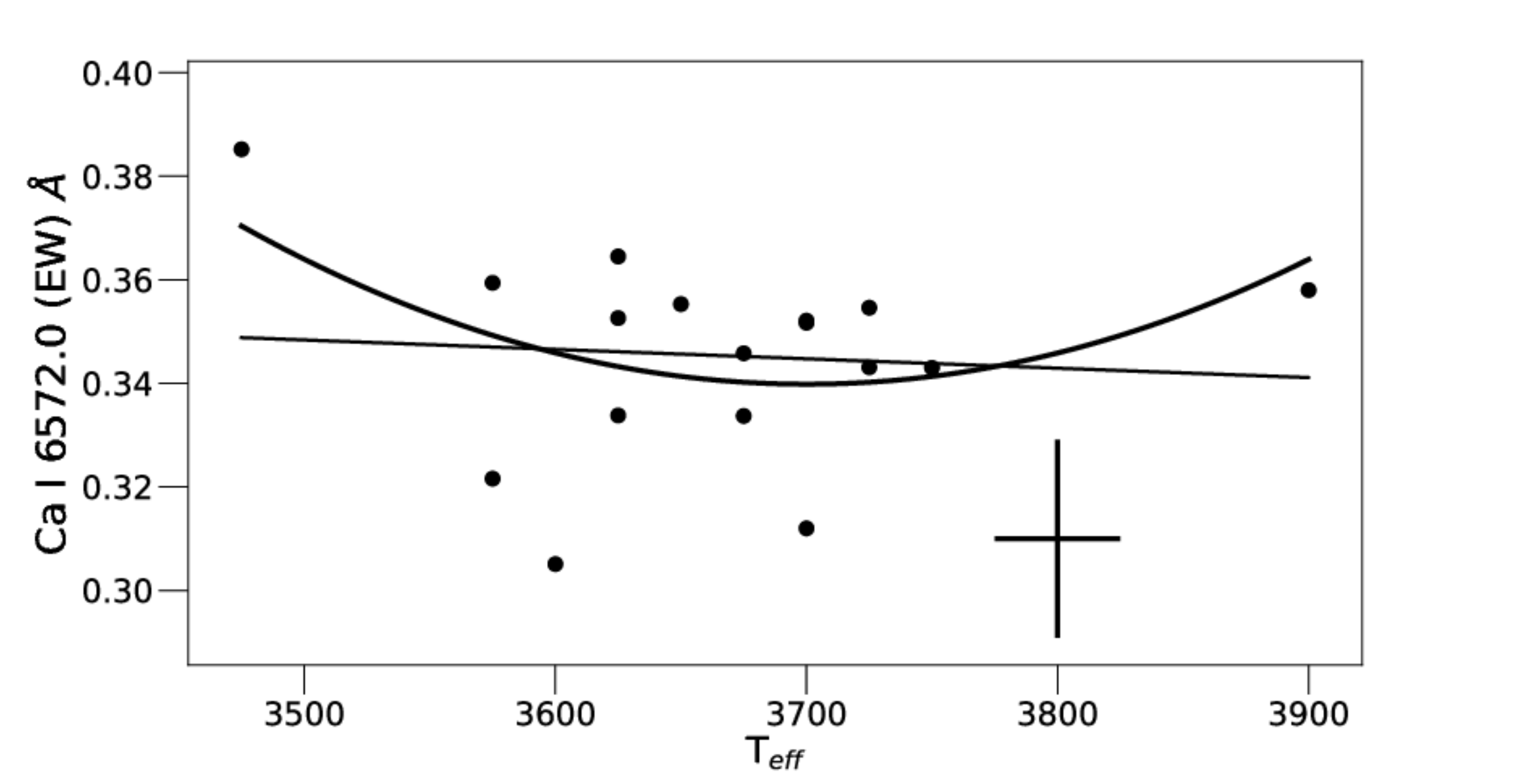}
  \includegraphics[width=.3\textwidth] {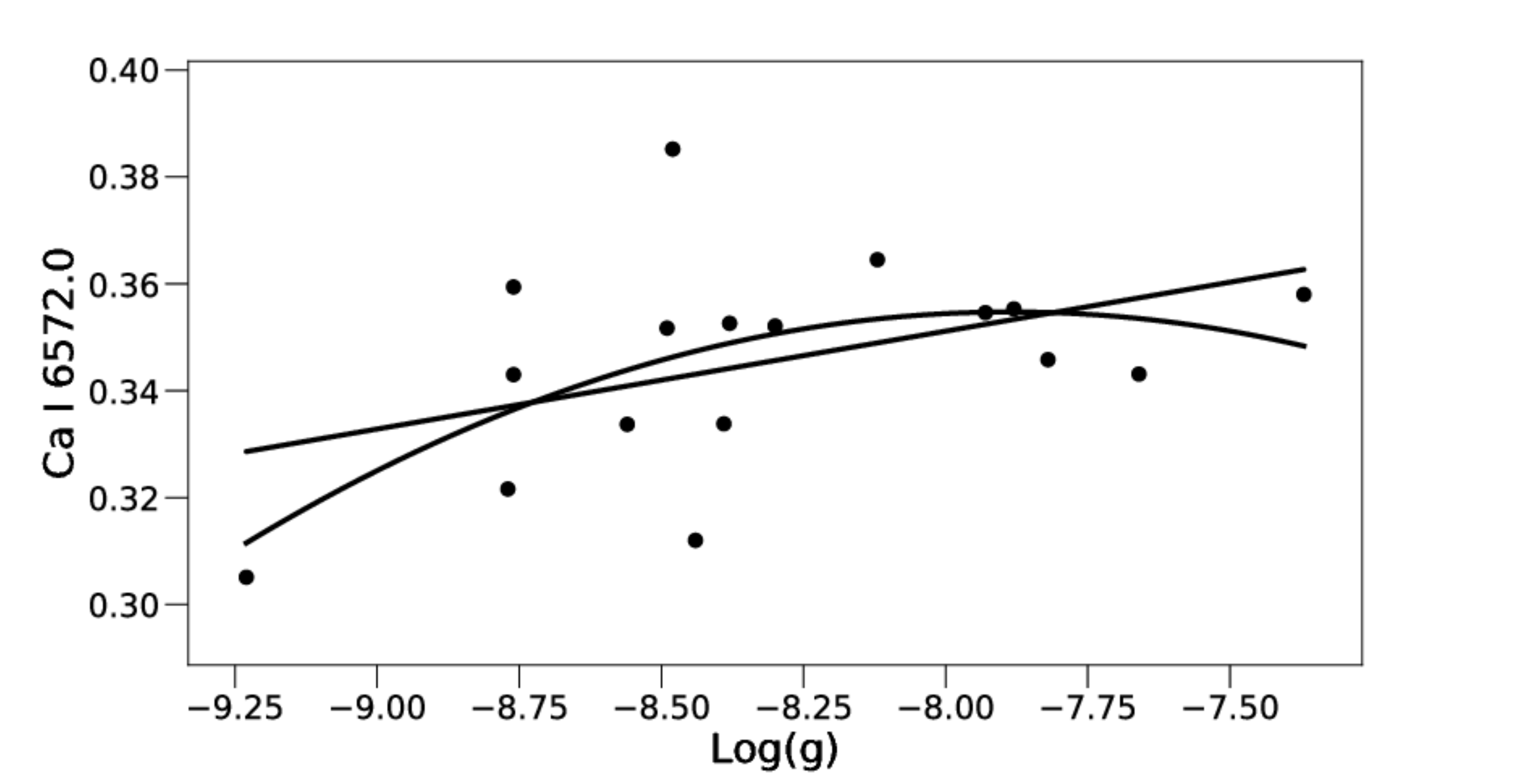}
  \includegraphics[width=.3\textwidth] {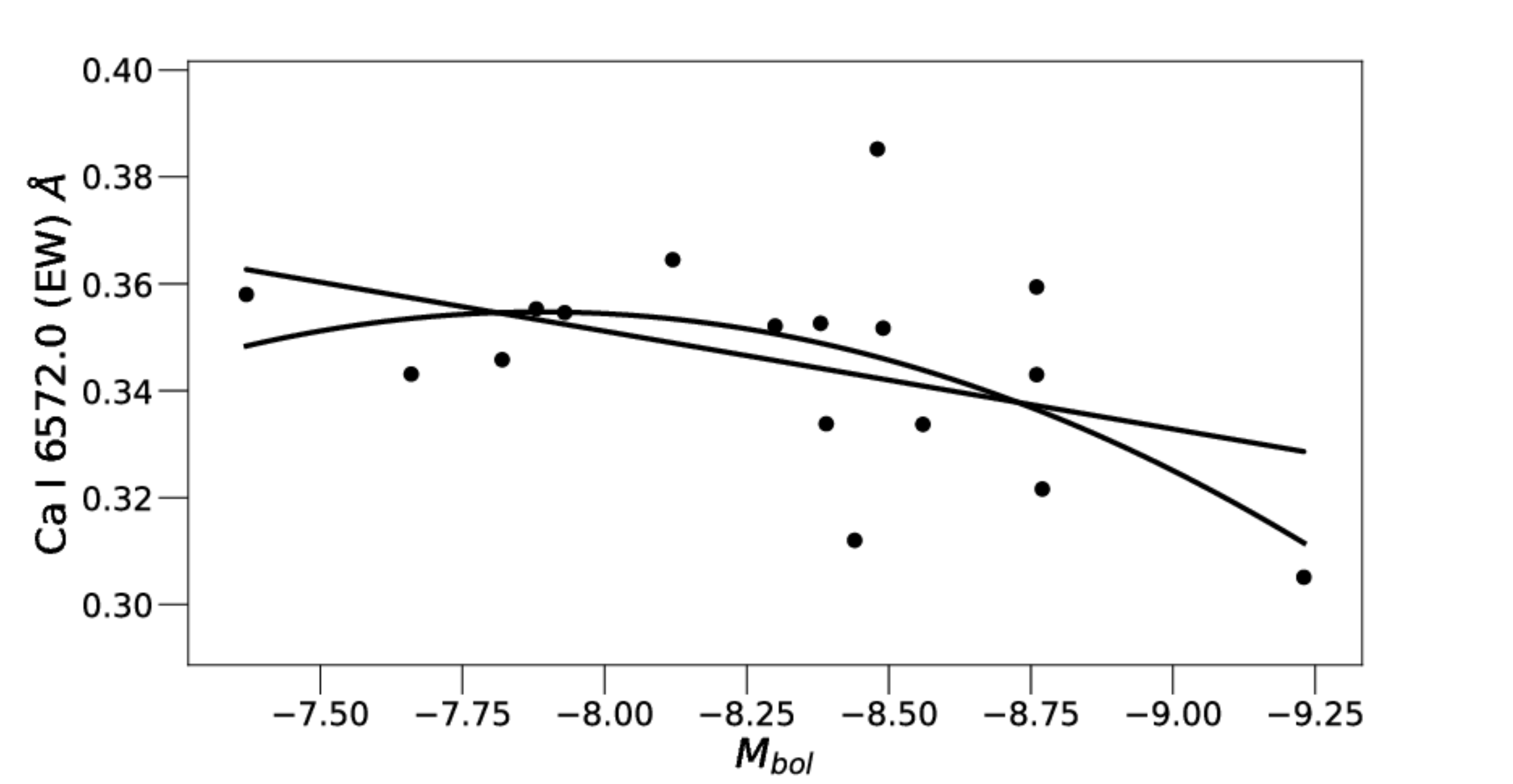}
 \caption{As in Figure 2, but for the Ca I 6572\AA\ absorption feature.}
\end{figure*}

\begin{figure*}
  \includegraphics[width=.3\textwidth] {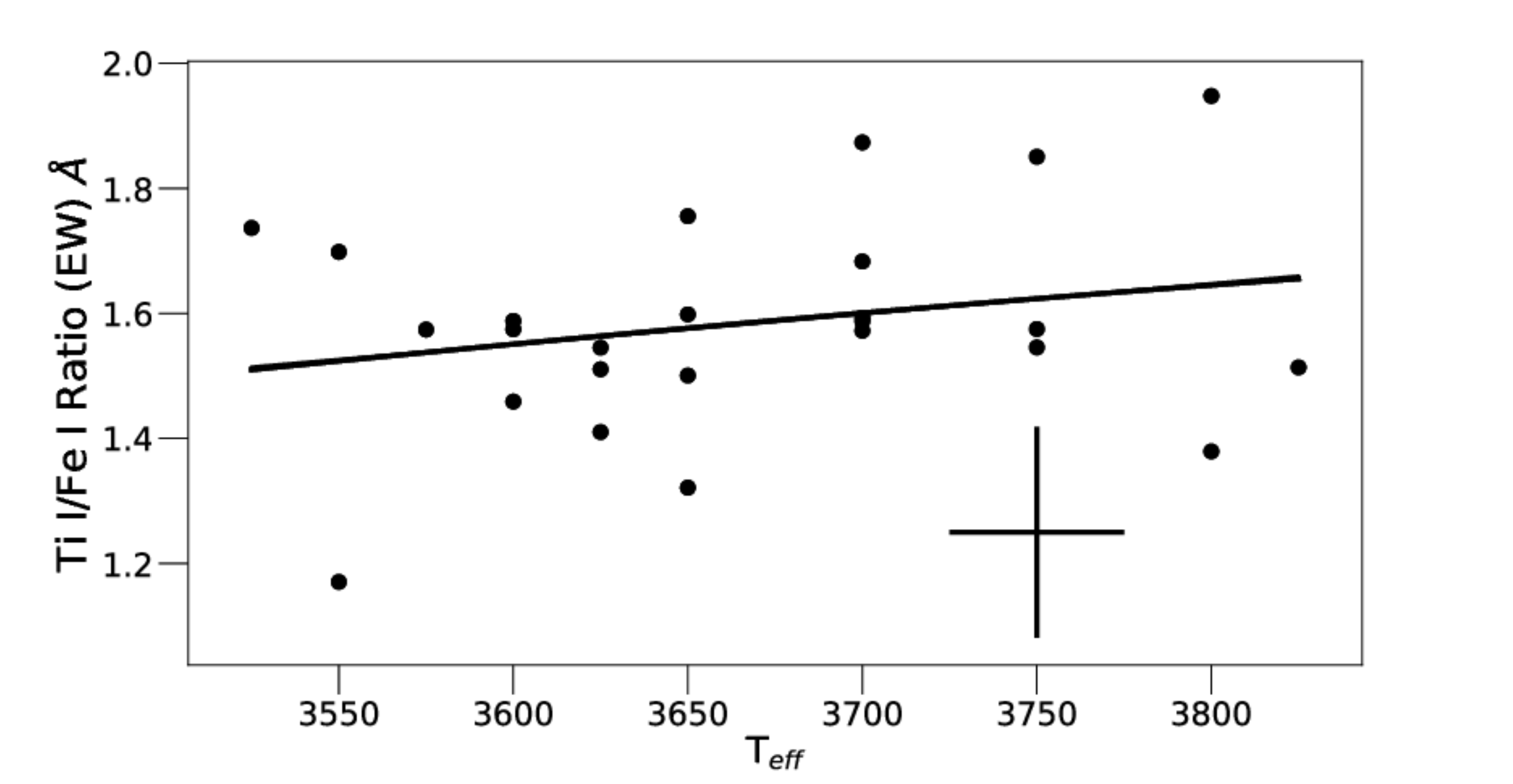}
  \includegraphics[width=.3\textwidth] {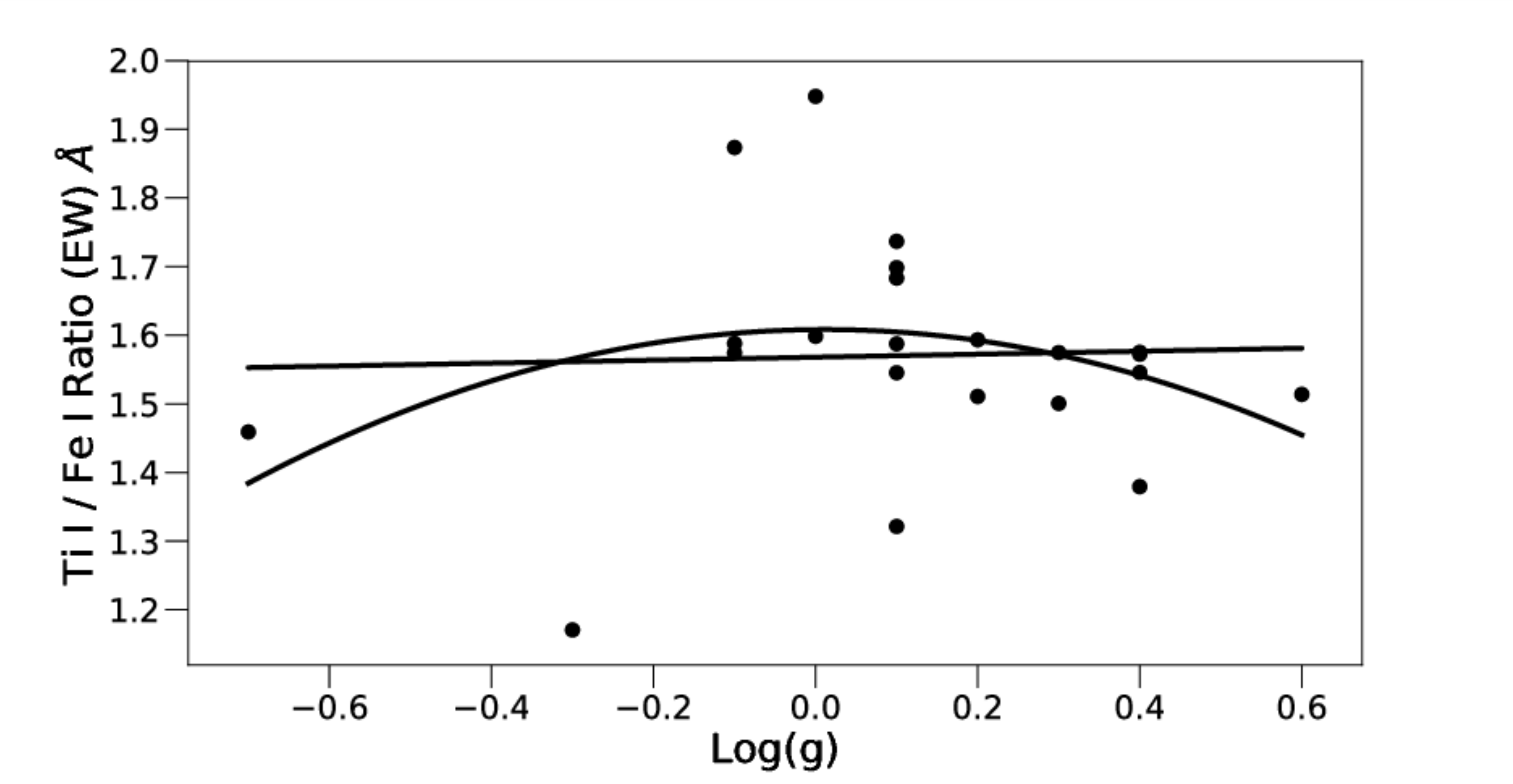}
  \includegraphics[width=.3\textwidth] {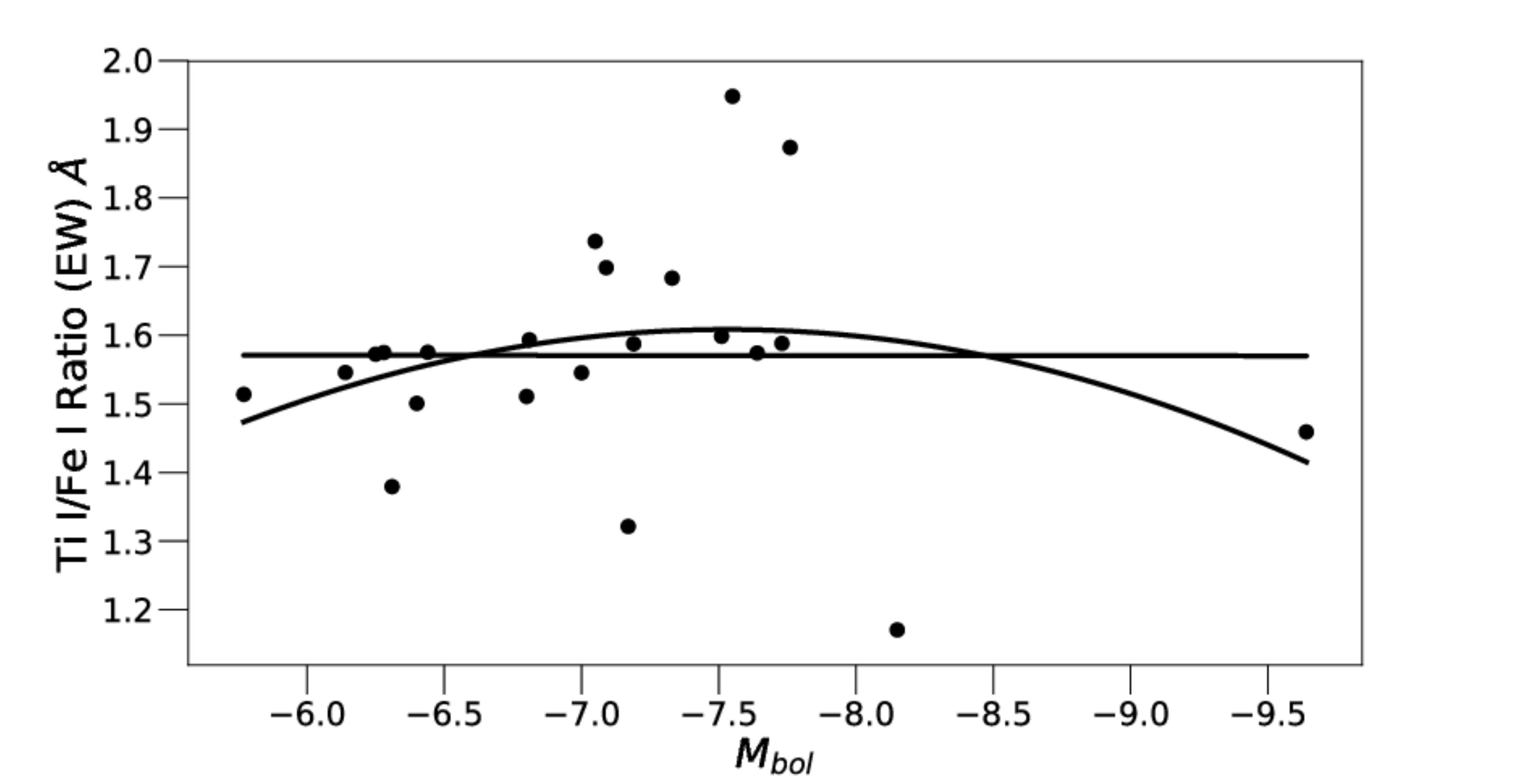}
  \includegraphics[width=.3\textwidth] {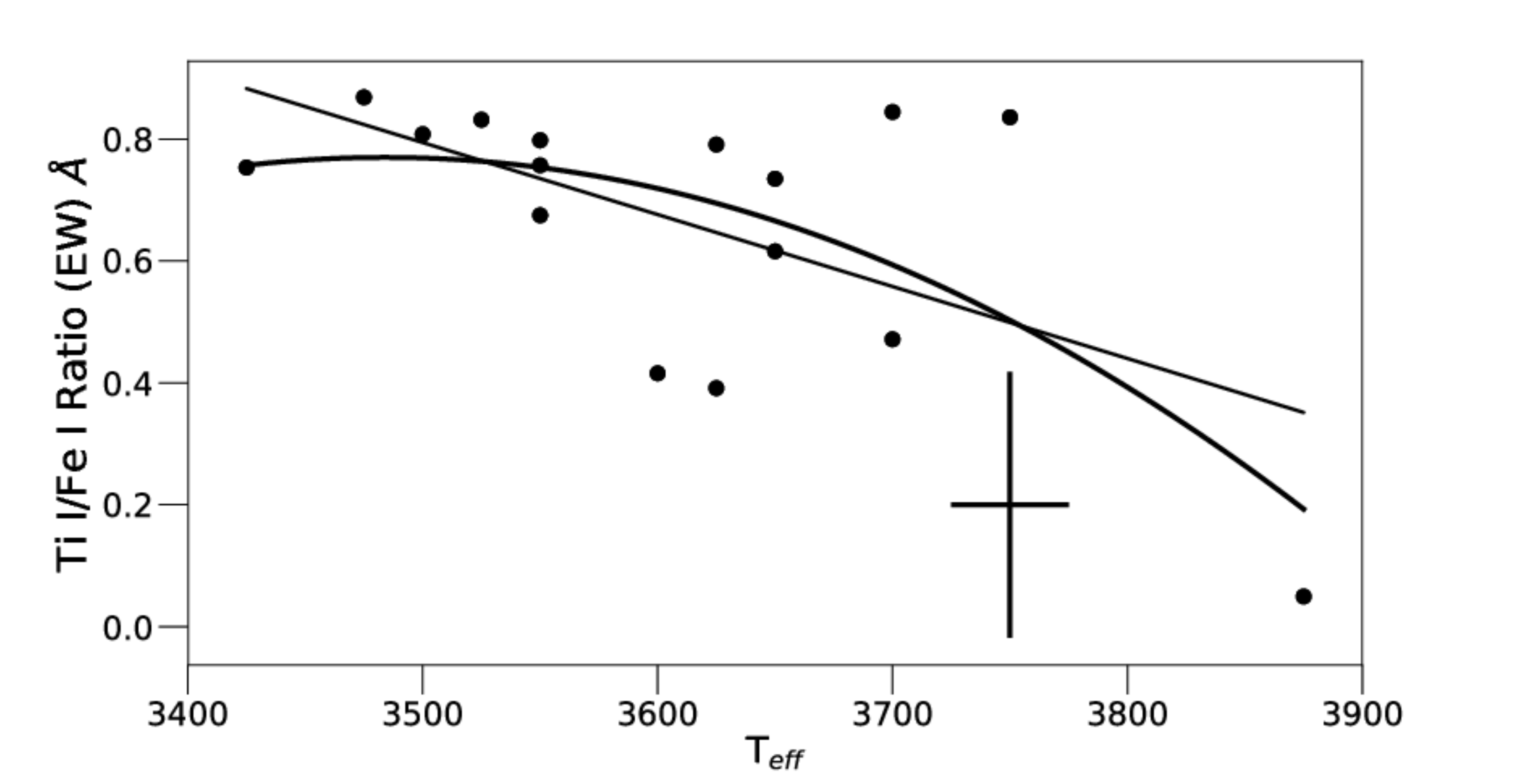}
  \includegraphics[width=.3\textwidth] {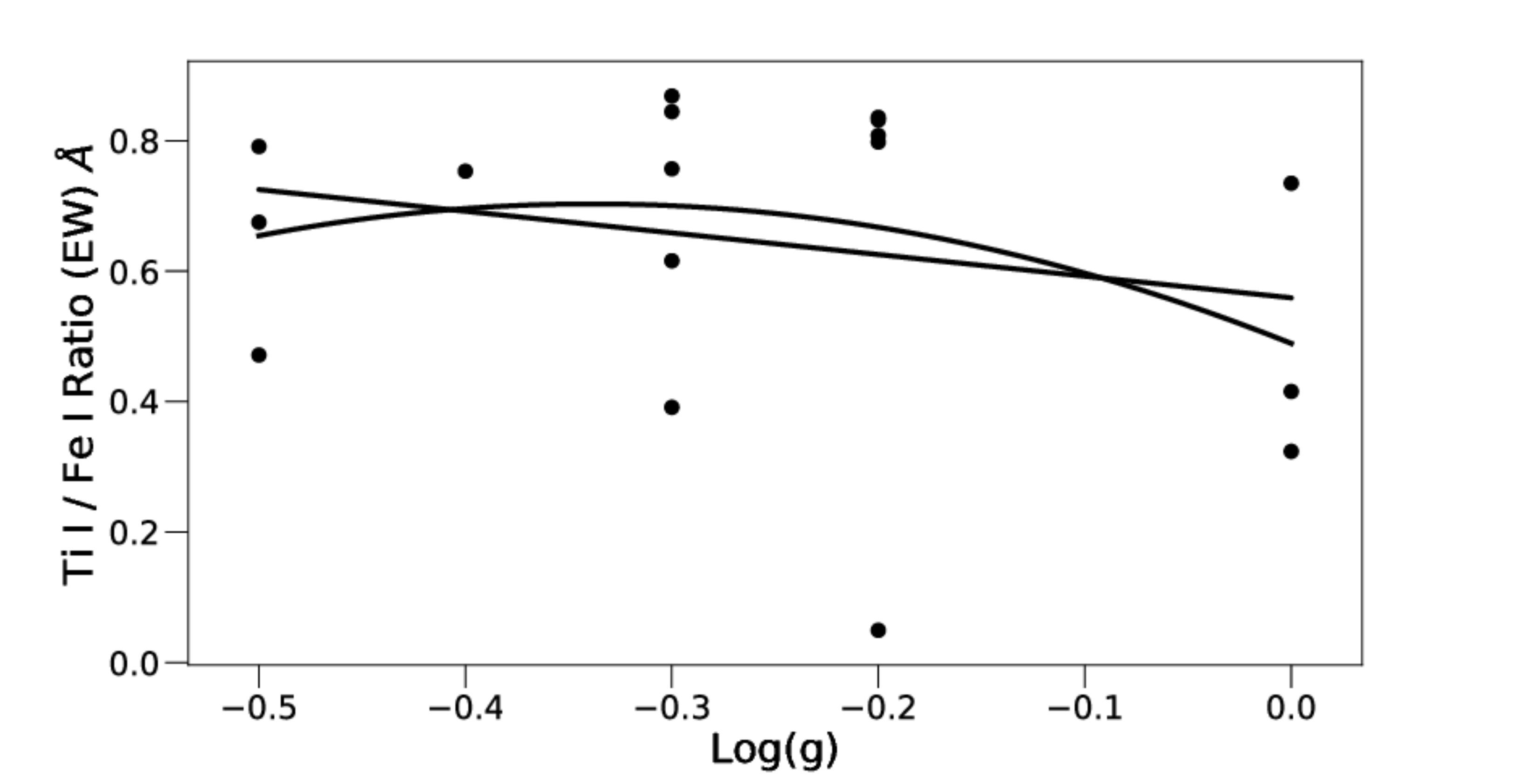}
  \includegraphics[width=.3\textwidth] {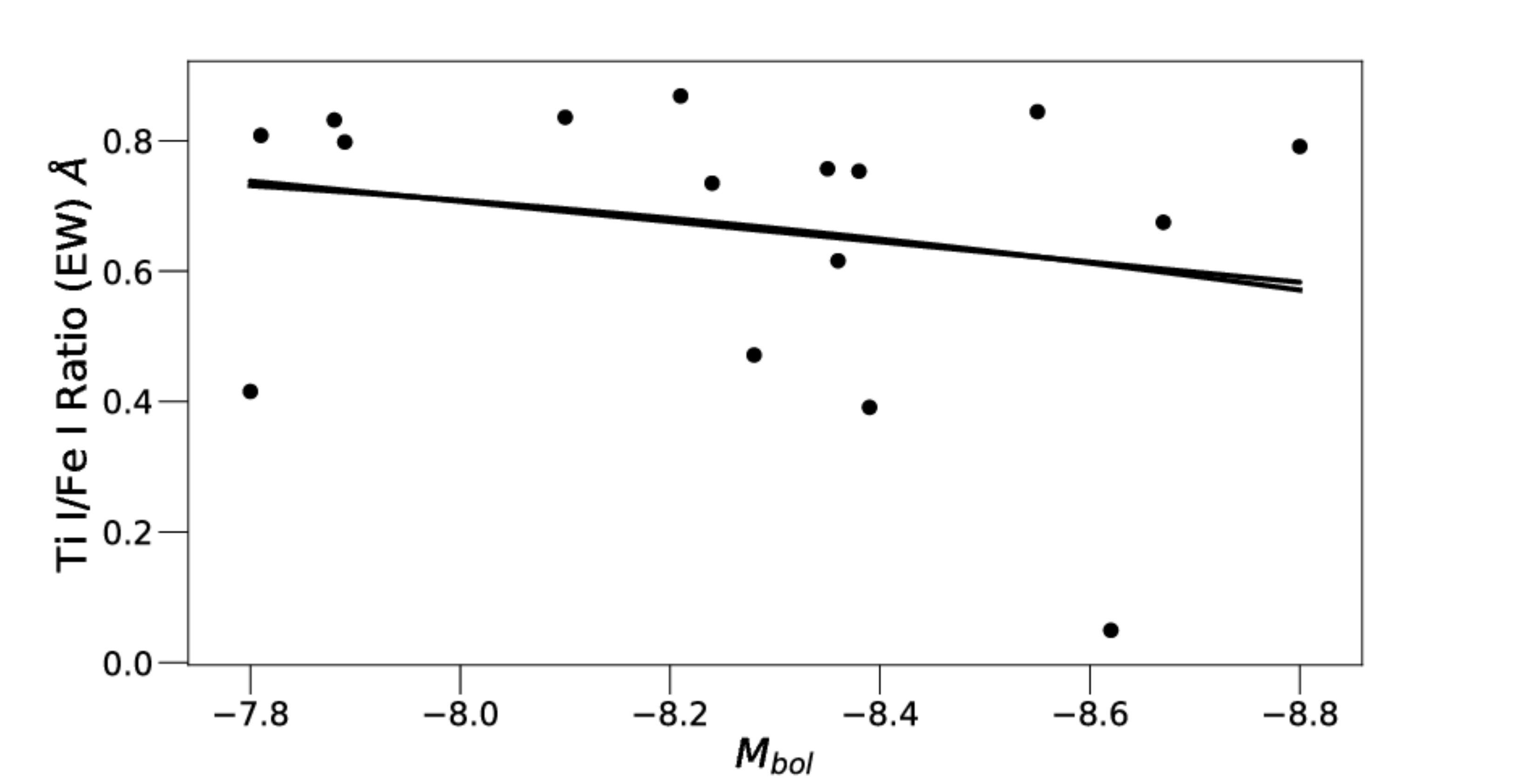}
  \includegraphics[width=.3\textwidth] {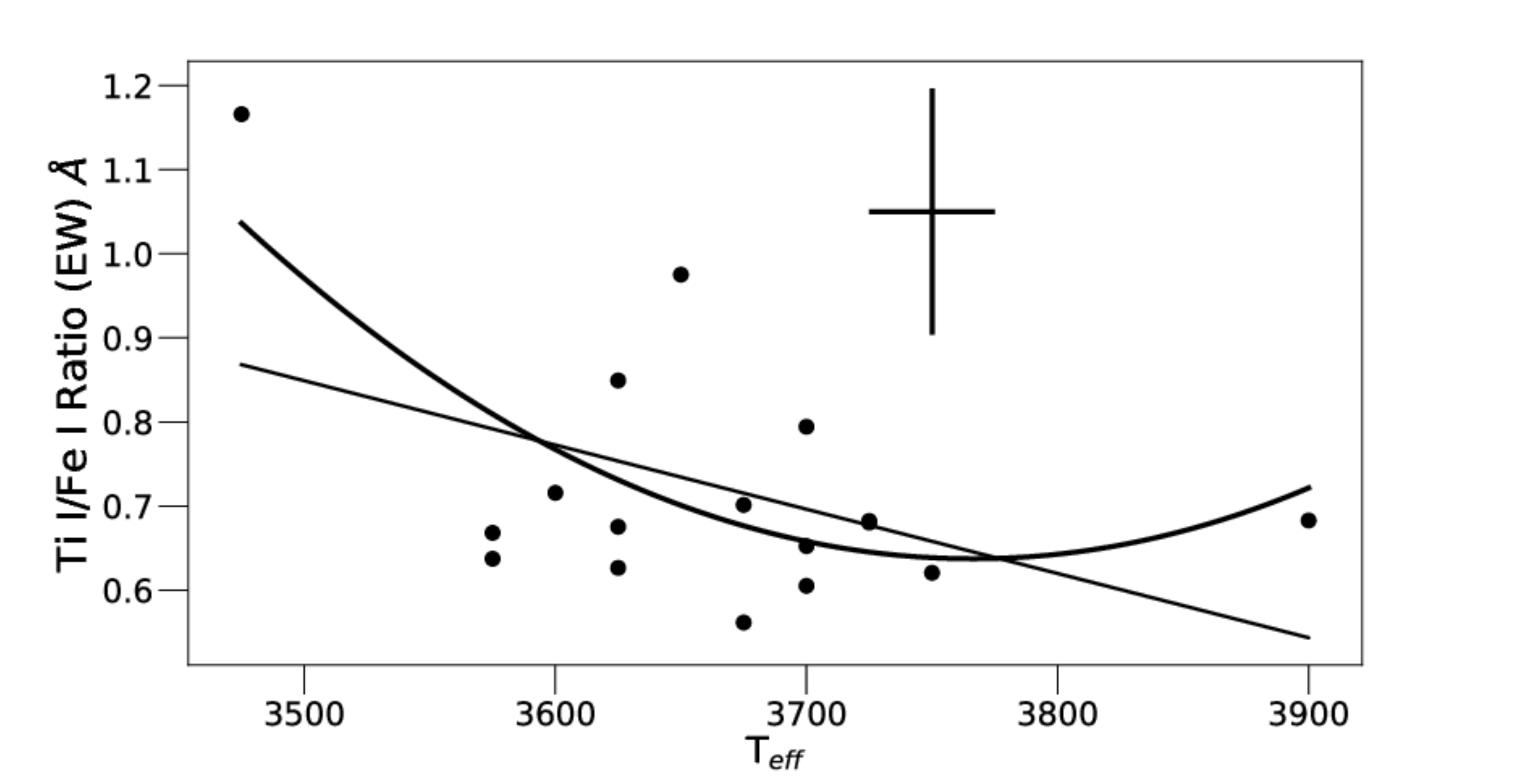}
  \includegraphics[width=.3\textwidth] {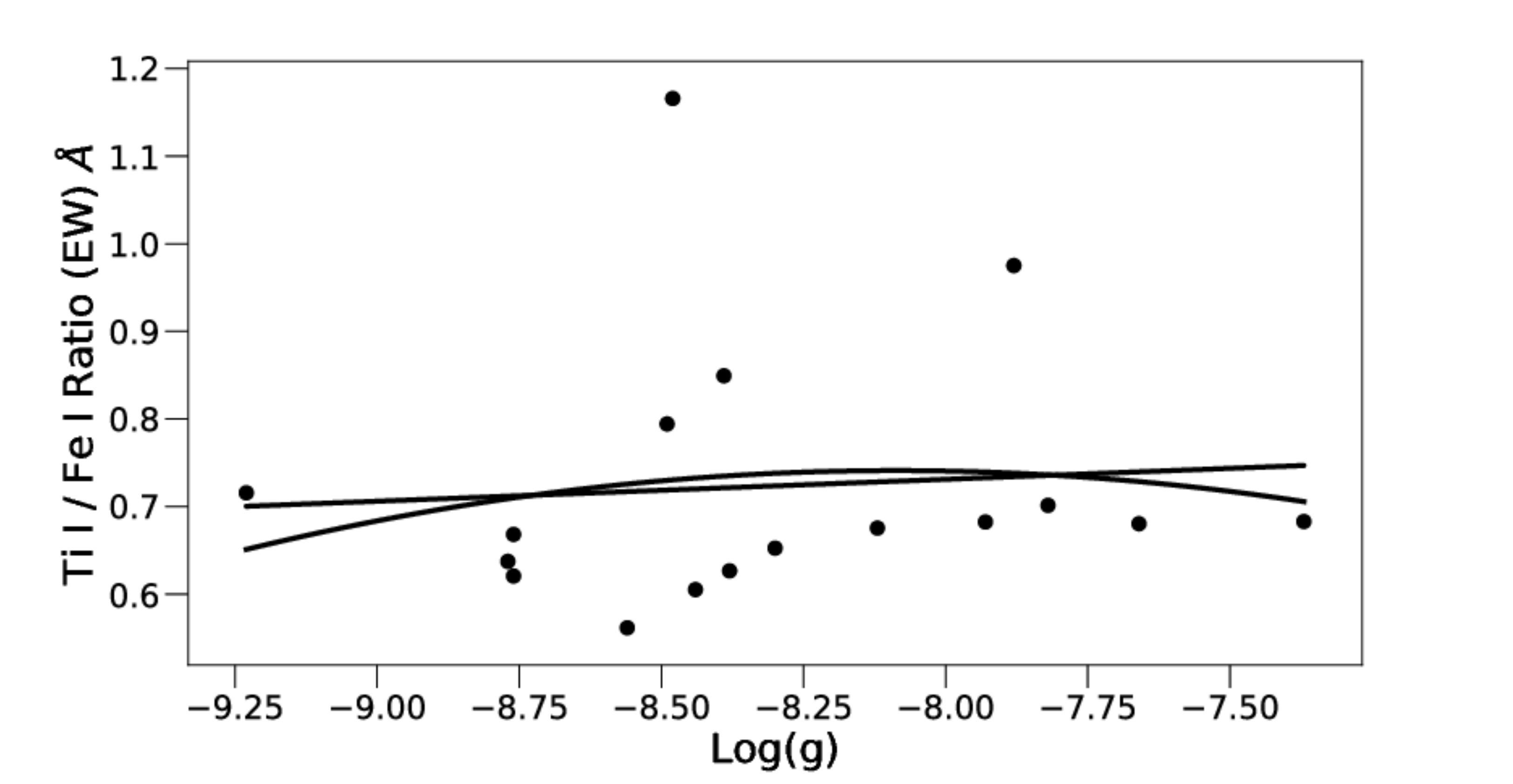}
  \includegraphics[width=.3\textwidth] {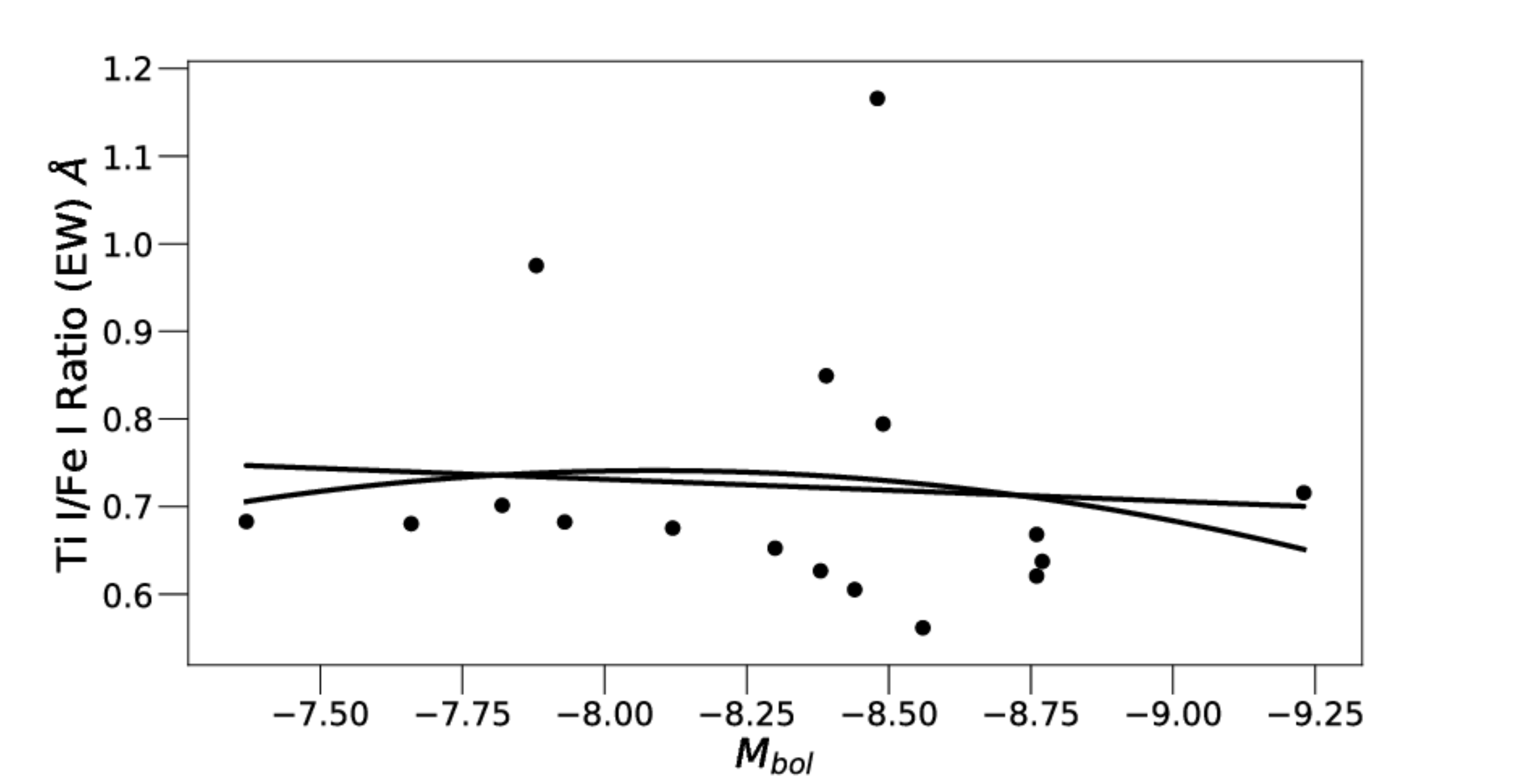}
 \caption{As in Figure 2, but for the ratio of the Ti I 8518\AA/Fe II 8514\AA\ absorption features.}
\end{figure*}

\begin{figure*}
  \includegraphics[width=.3\textwidth] {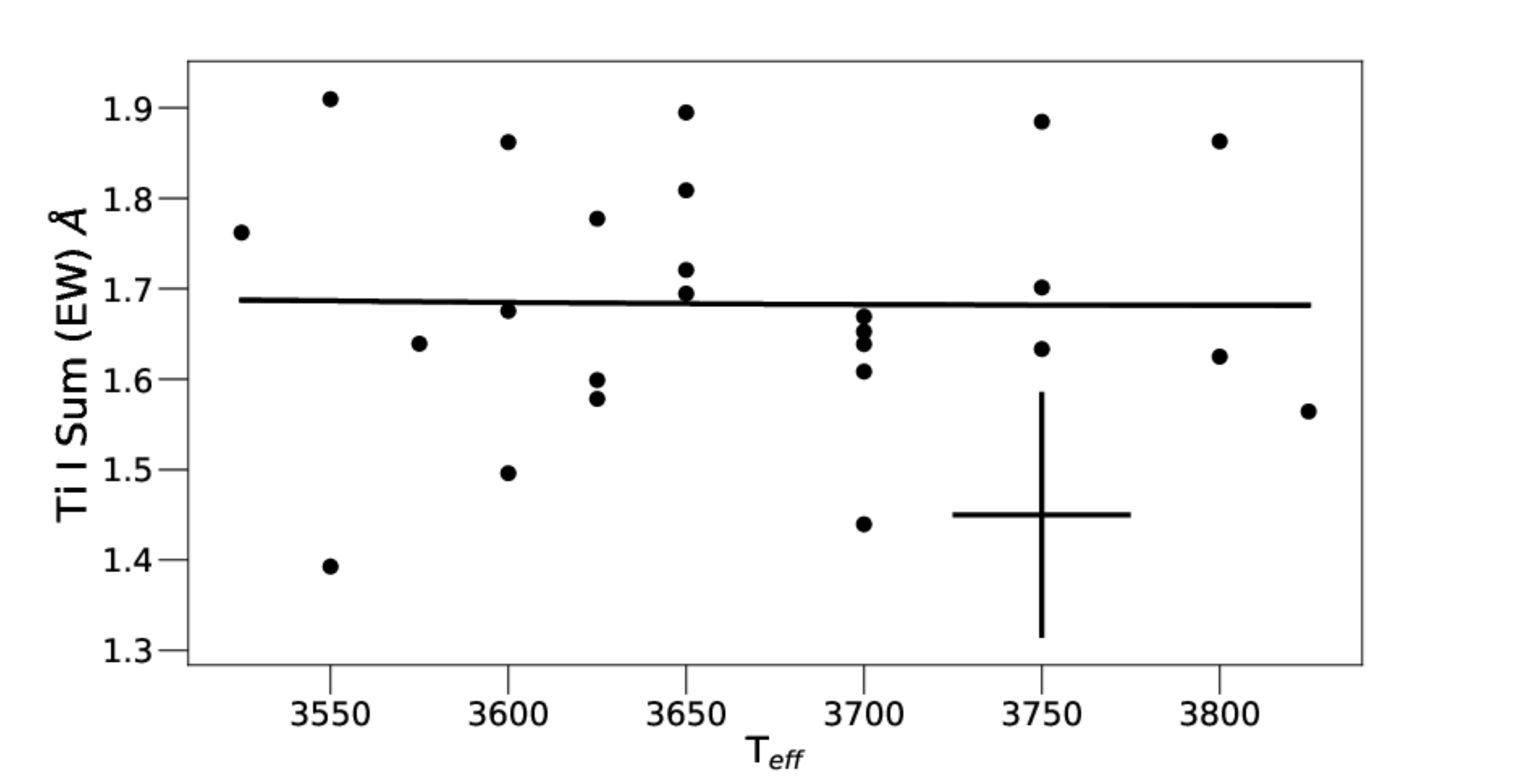}
  \includegraphics[width=.3\textwidth] {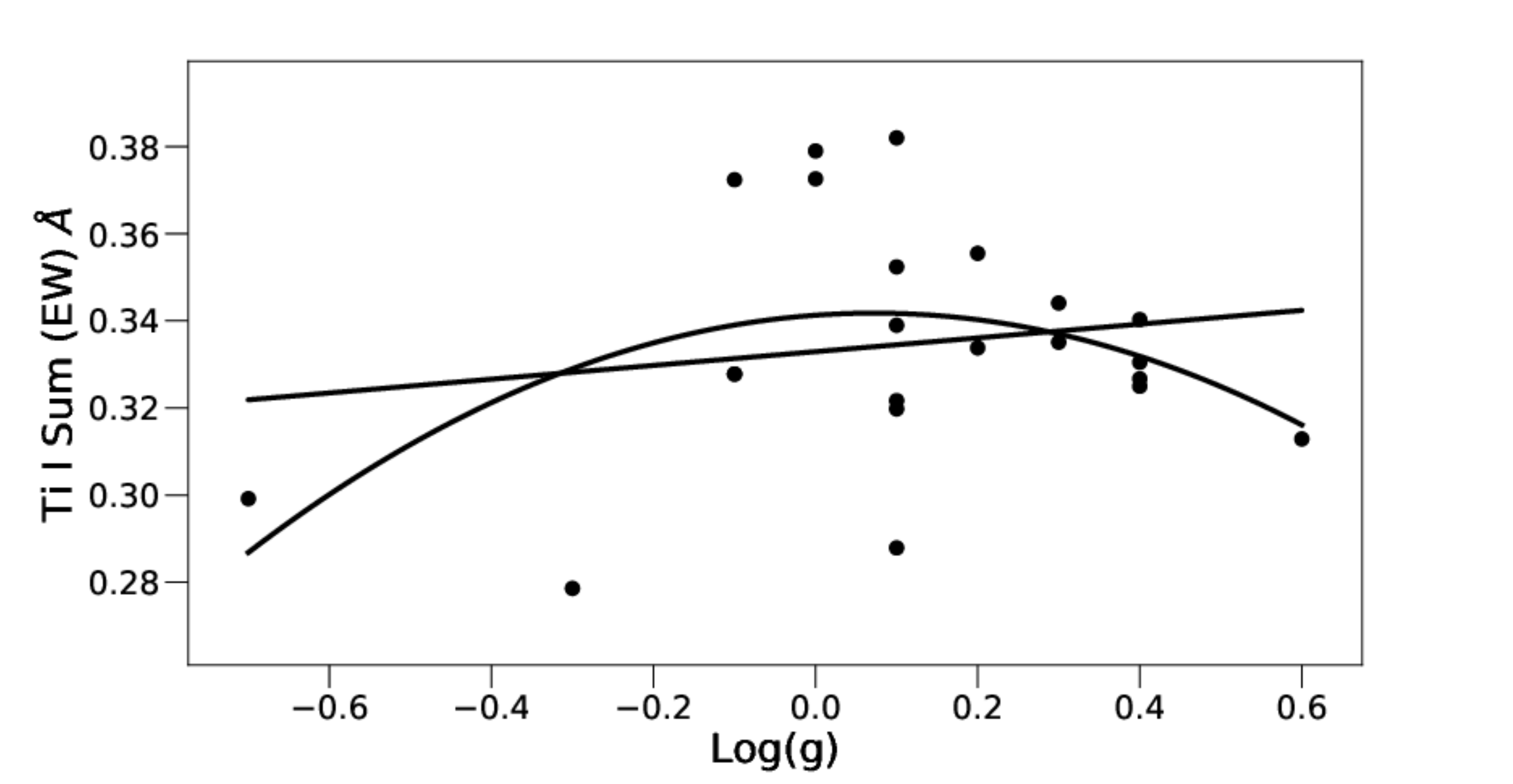}
  \includegraphics[width=.3\textwidth] {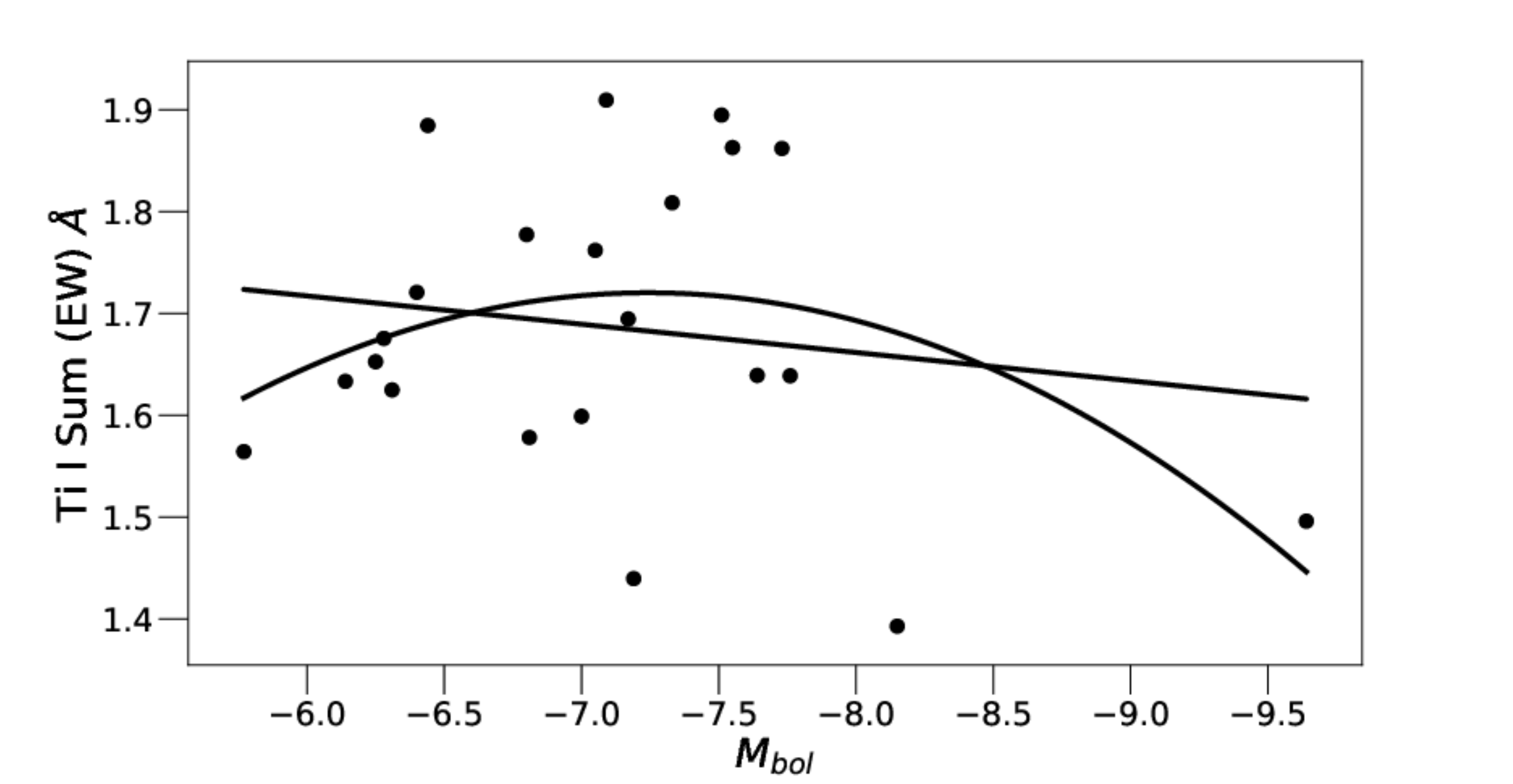}
  \includegraphics[width=.3\textwidth] {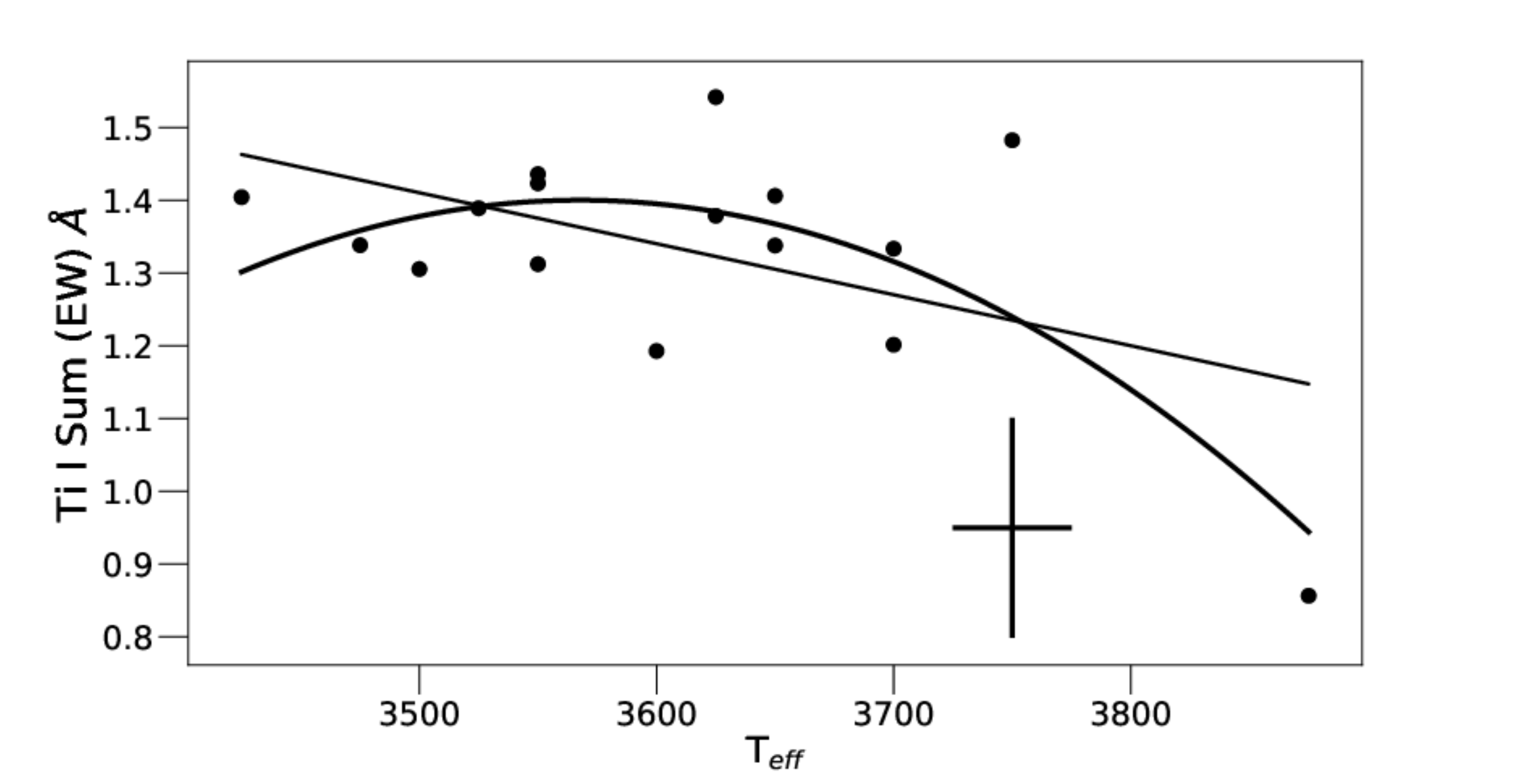}
  \includegraphics[width=.3\textwidth] {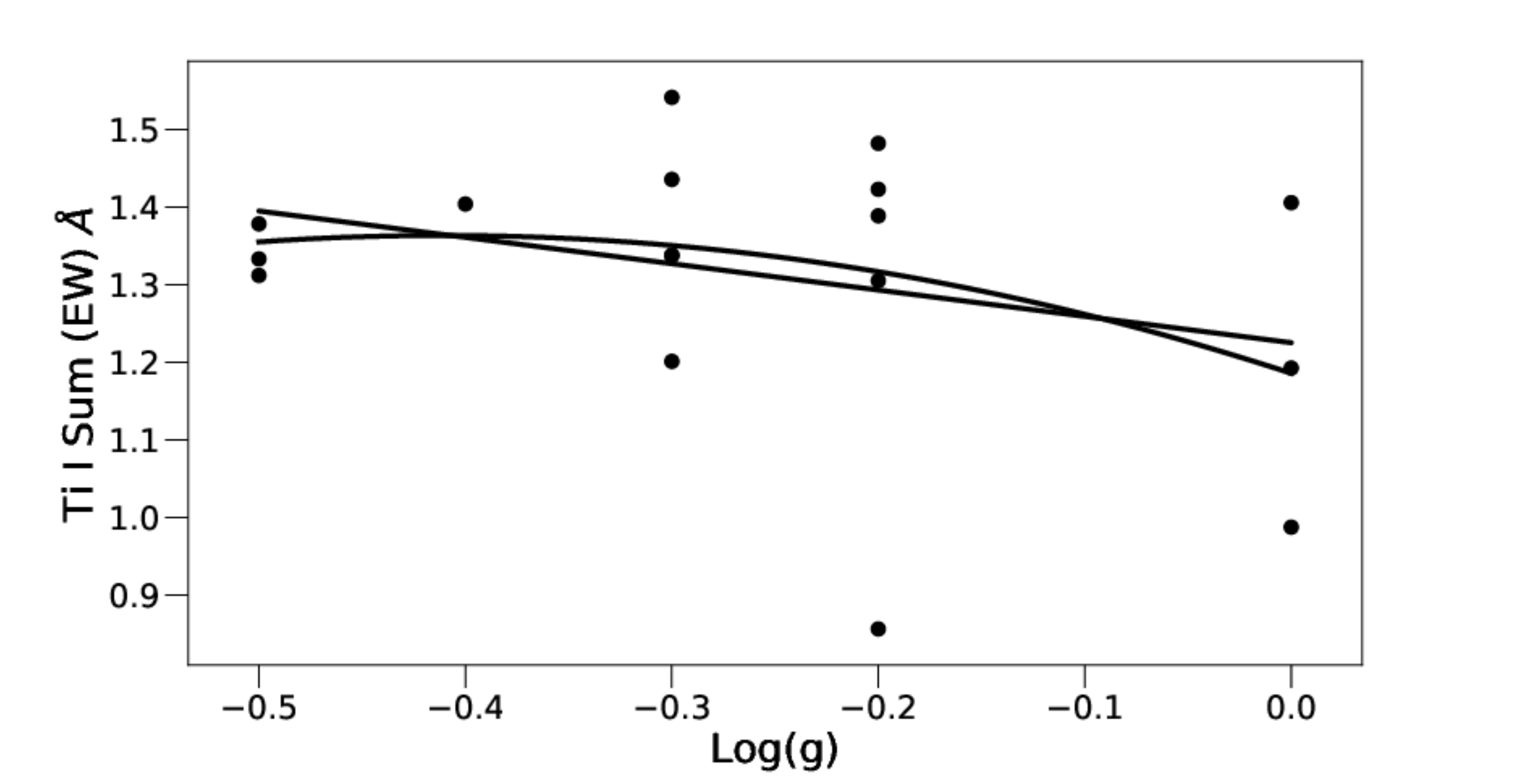}
  \includegraphics[width=.3\textwidth] {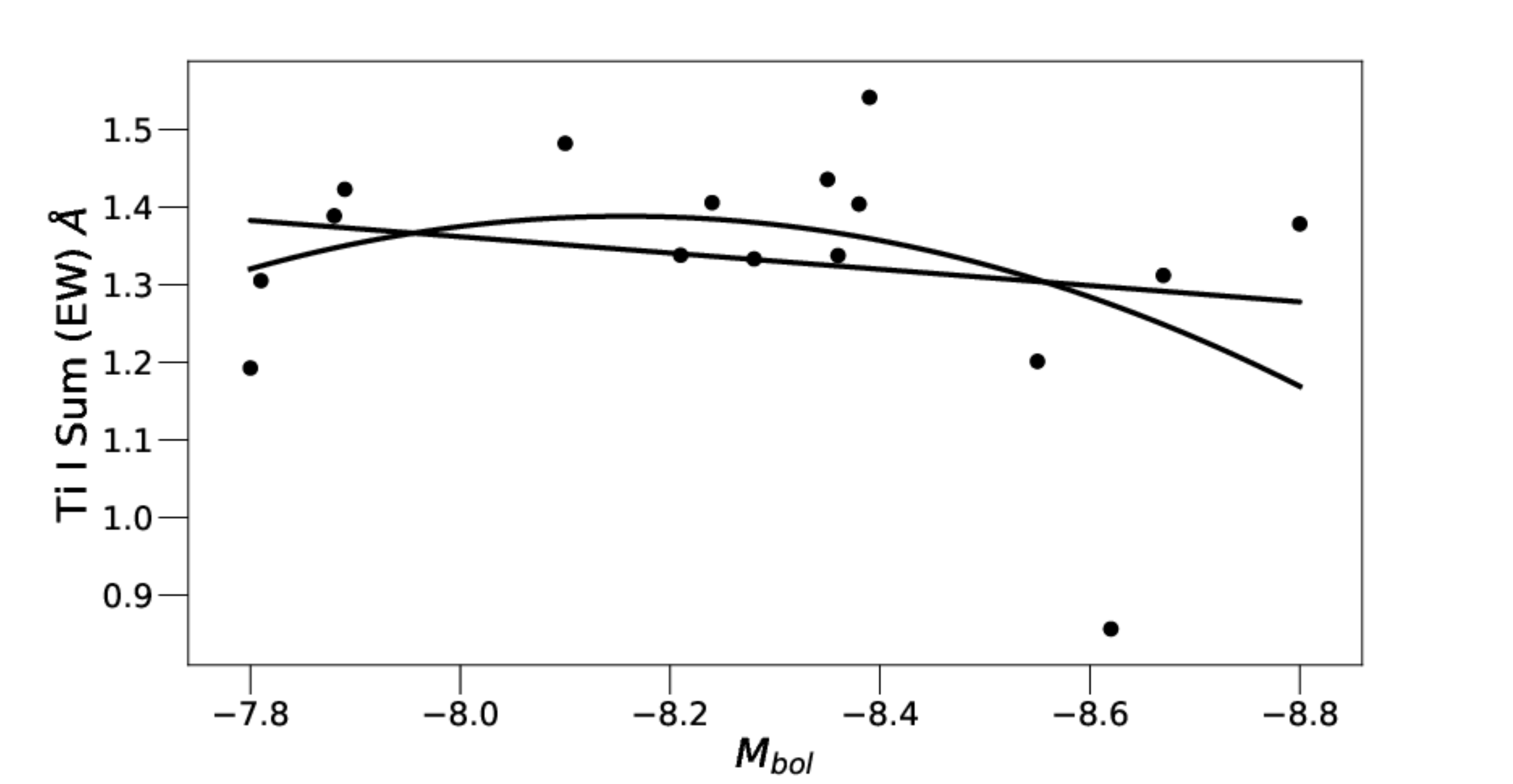}
  \includegraphics[width=.3\textwidth] {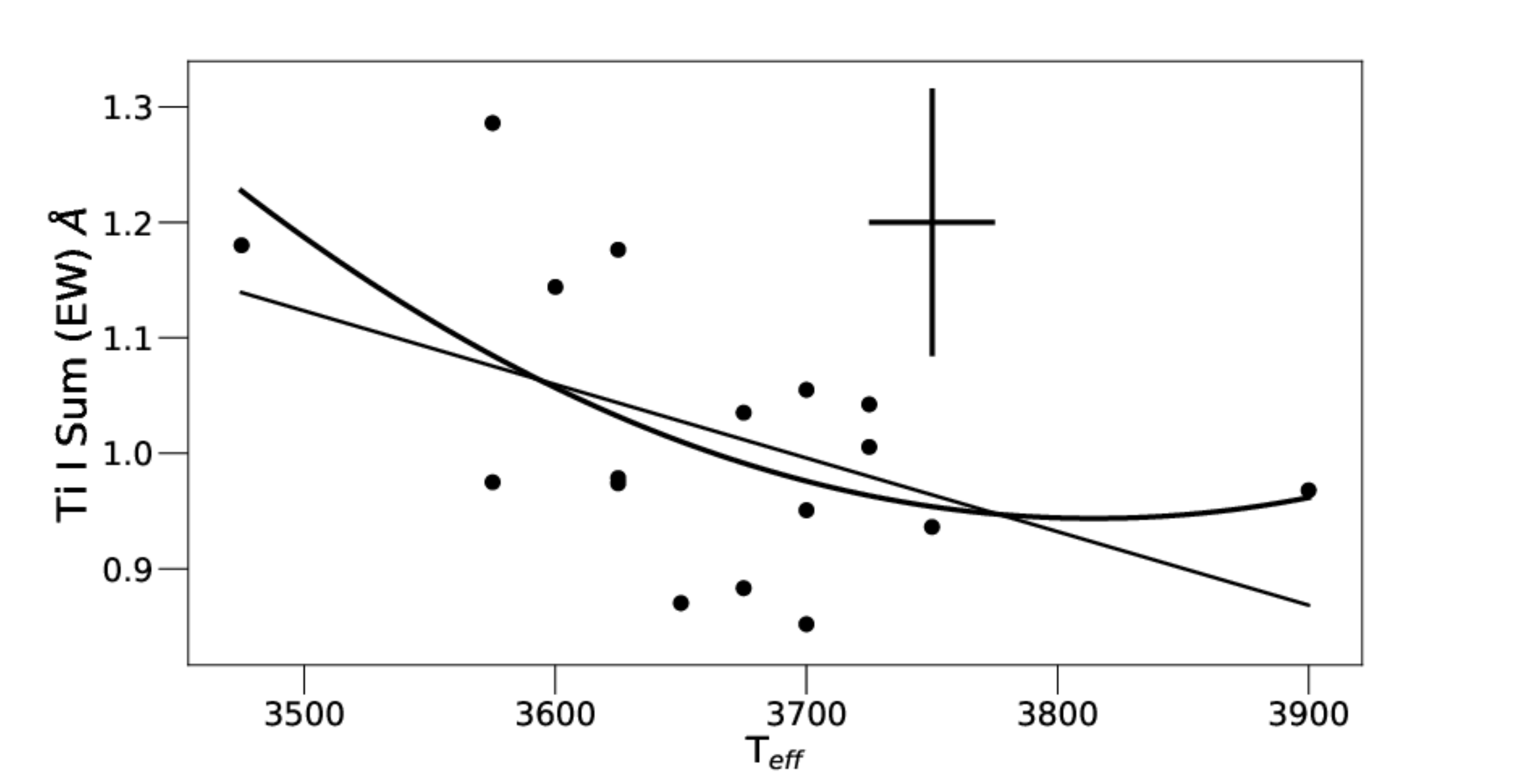}
  \includegraphics[width=.3\textwidth] {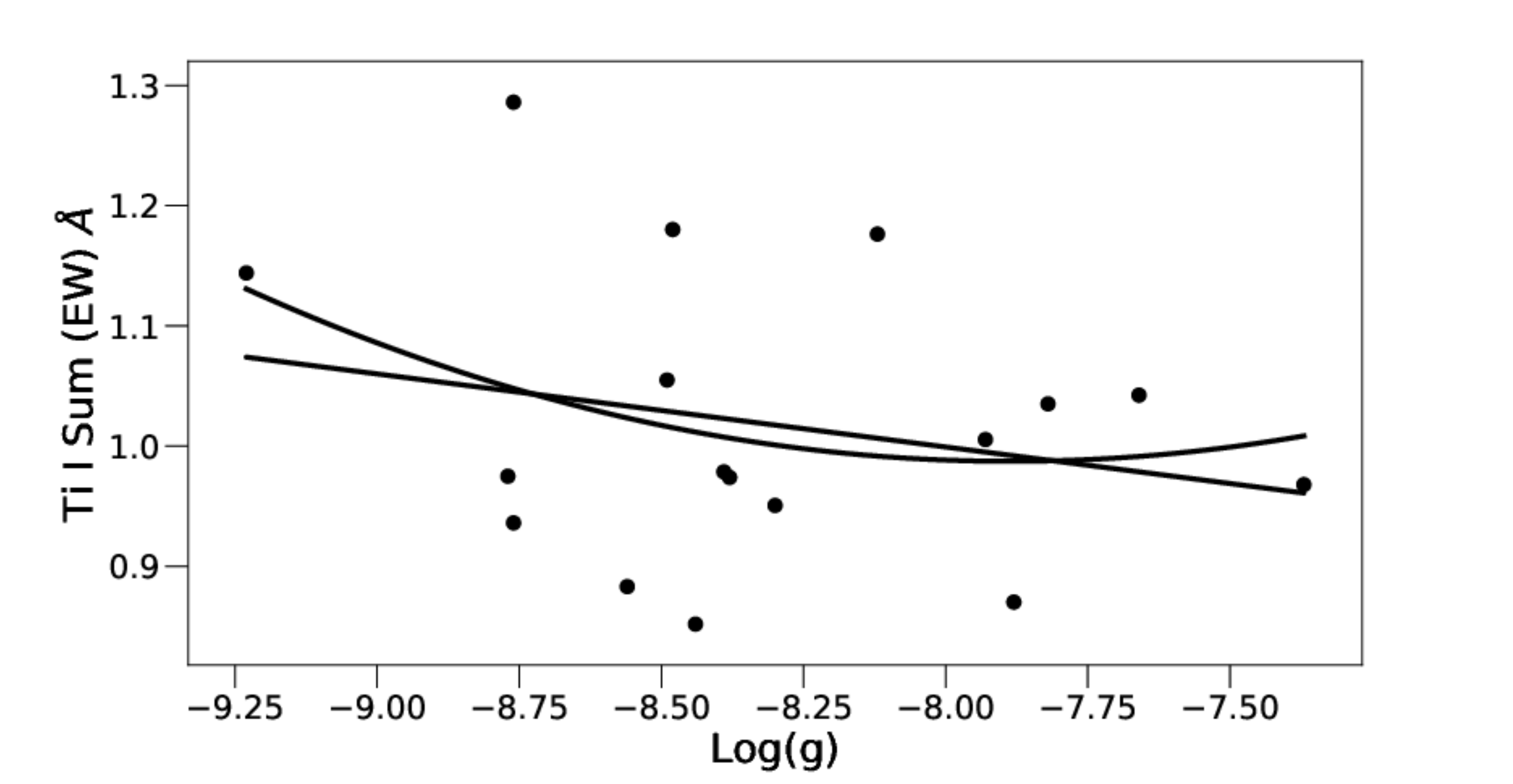}
  \includegraphics[width=.3\textwidth] {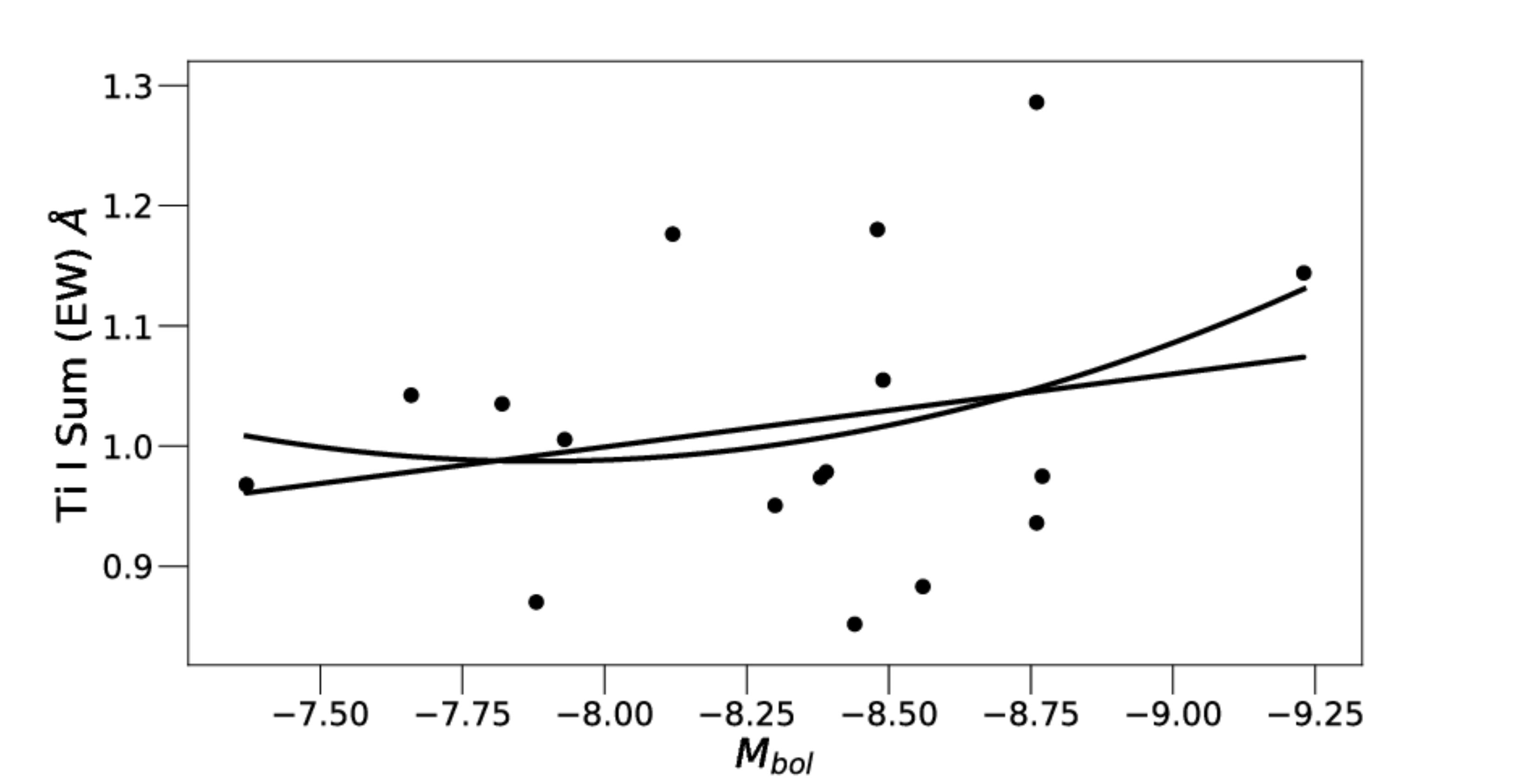}
 \caption{As in Figure 2, but for the sum of the Ti I absorption features.}
\end{figure*}

\begin{figure*}
  \includegraphics[width=.3\textwidth] {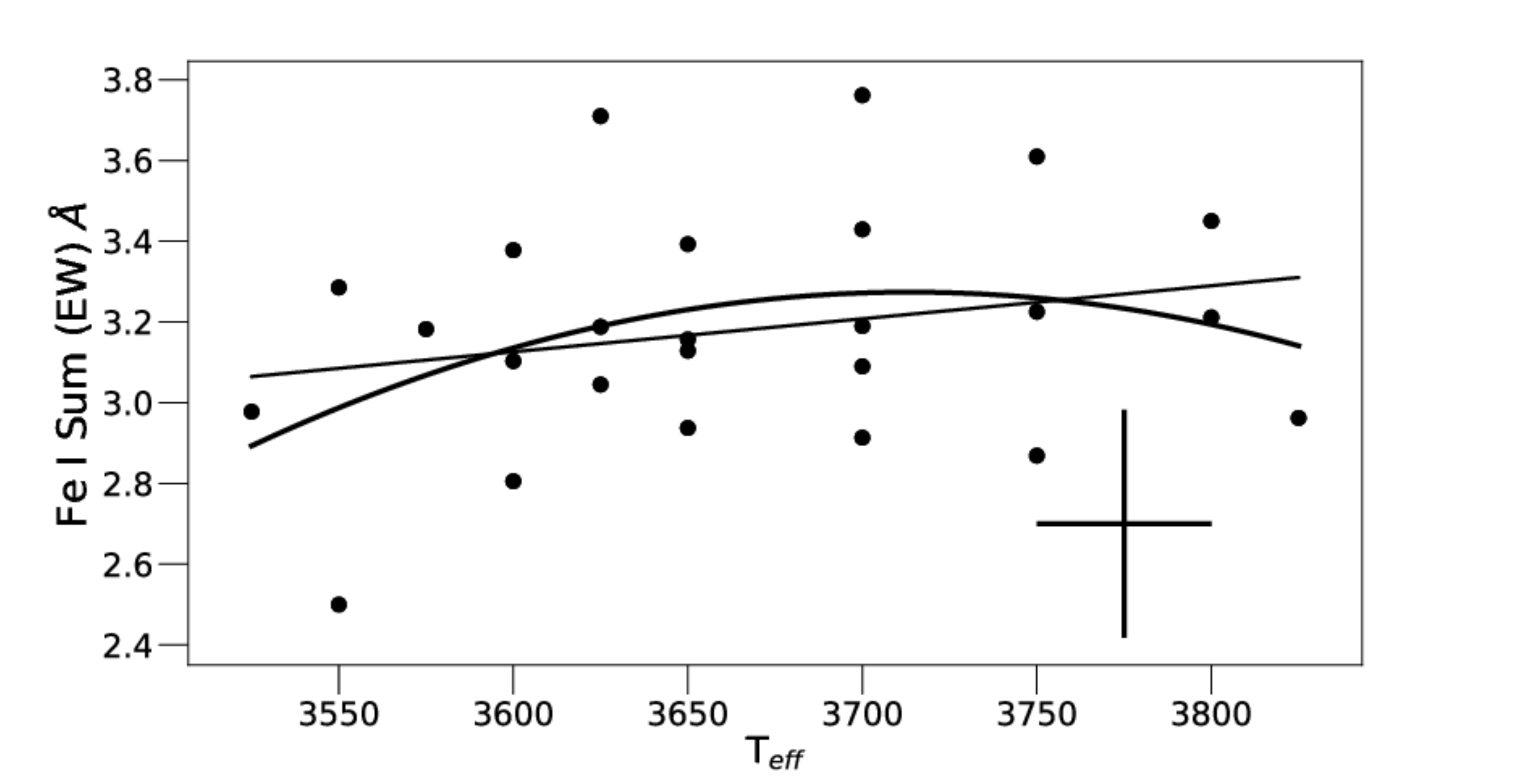}
  \includegraphics[width=.3\textwidth] {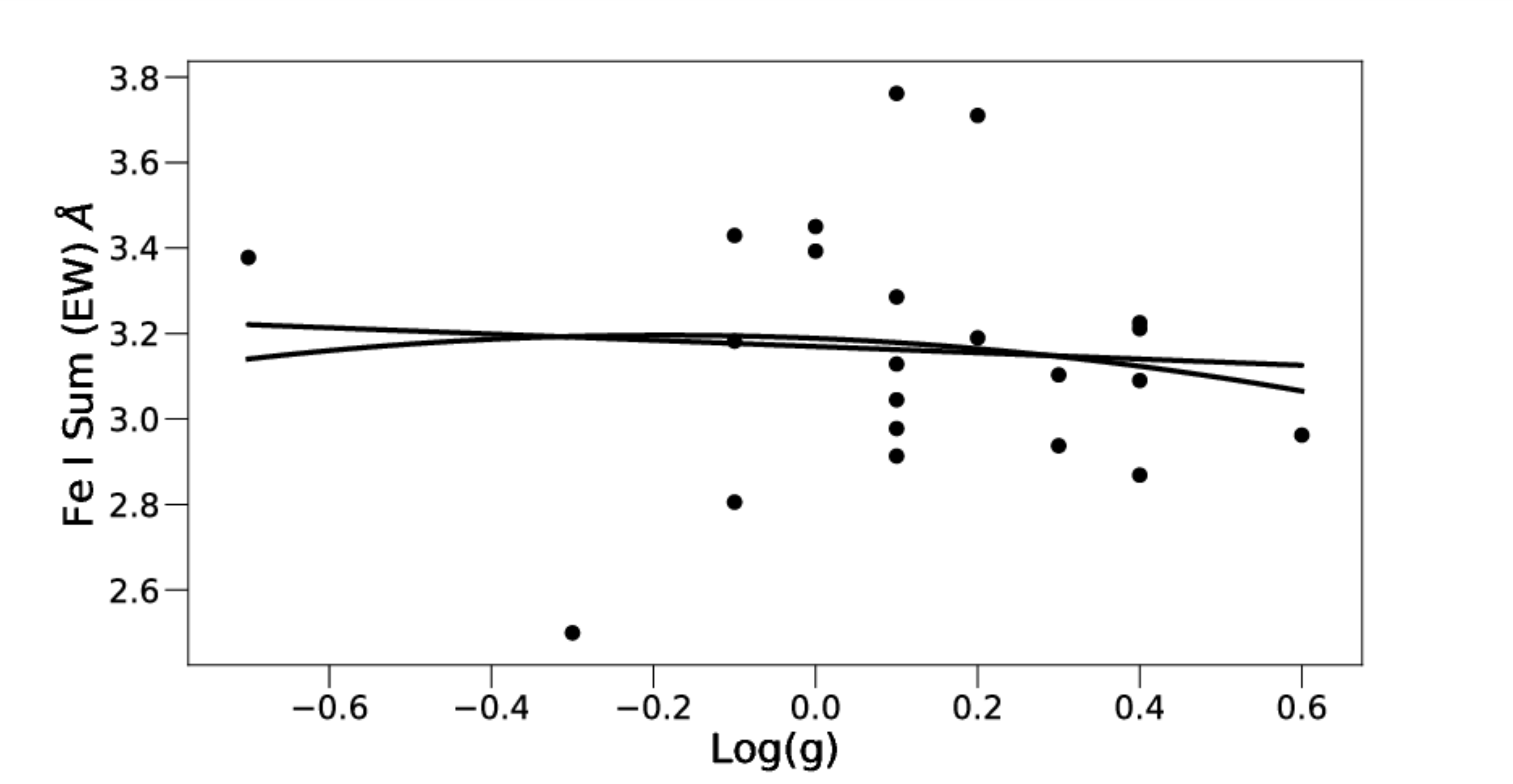}
  \includegraphics[width=.3\textwidth] {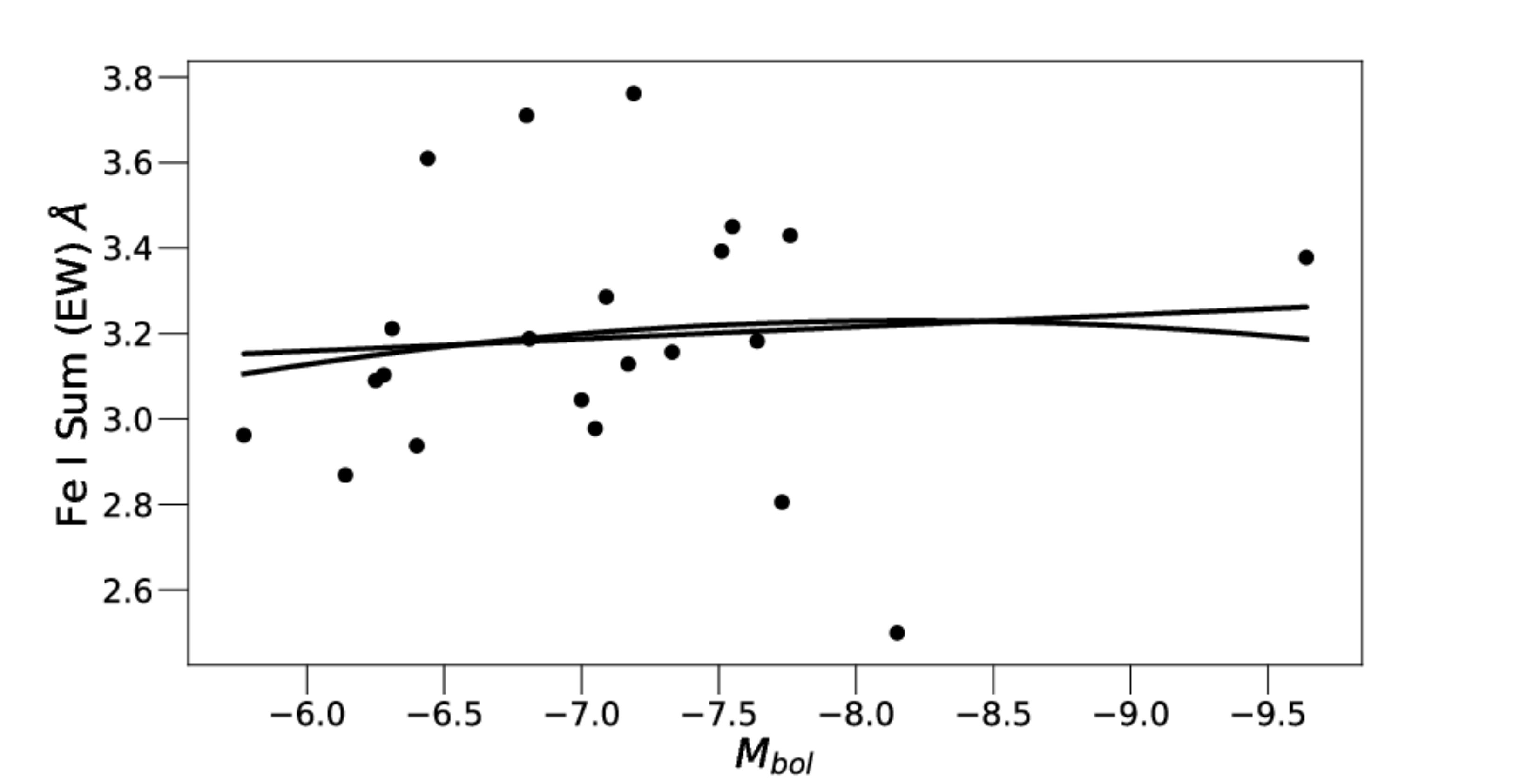}
  \includegraphics[width=.3\textwidth] {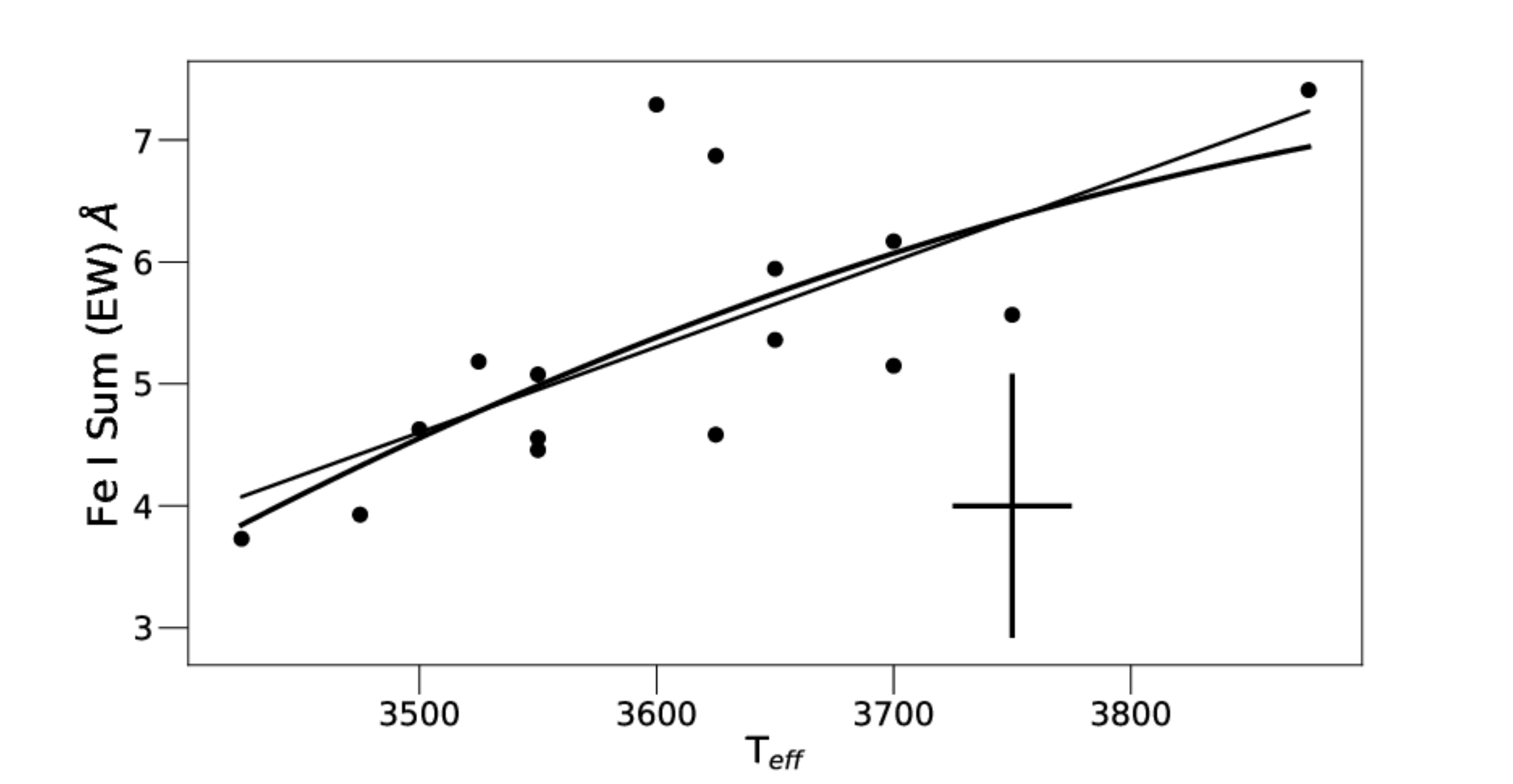}
  \includegraphics[width=.3\textwidth] {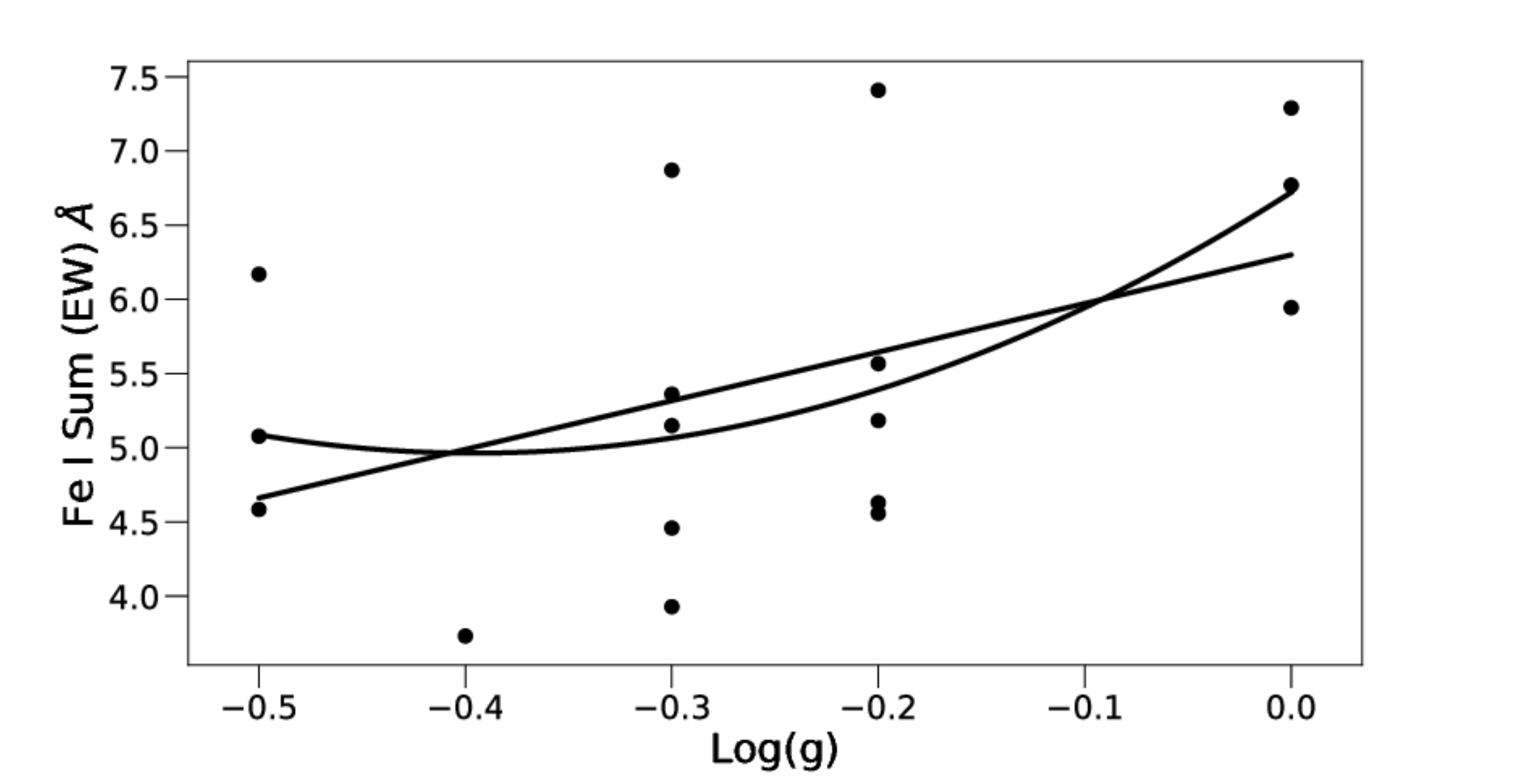}
  \includegraphics[width=.3\textwidth] {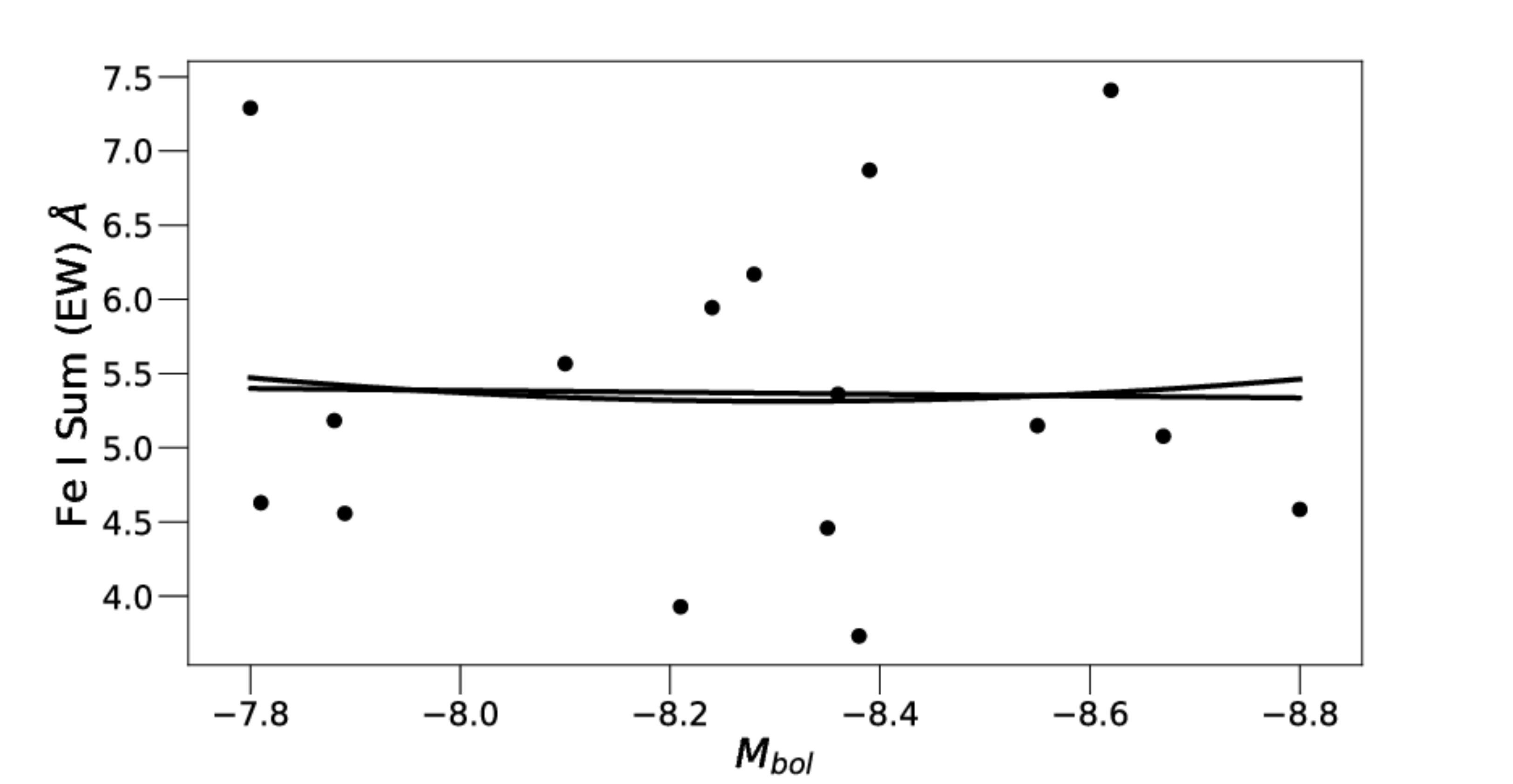}
  \includegraphics[width=.3\textwidth] {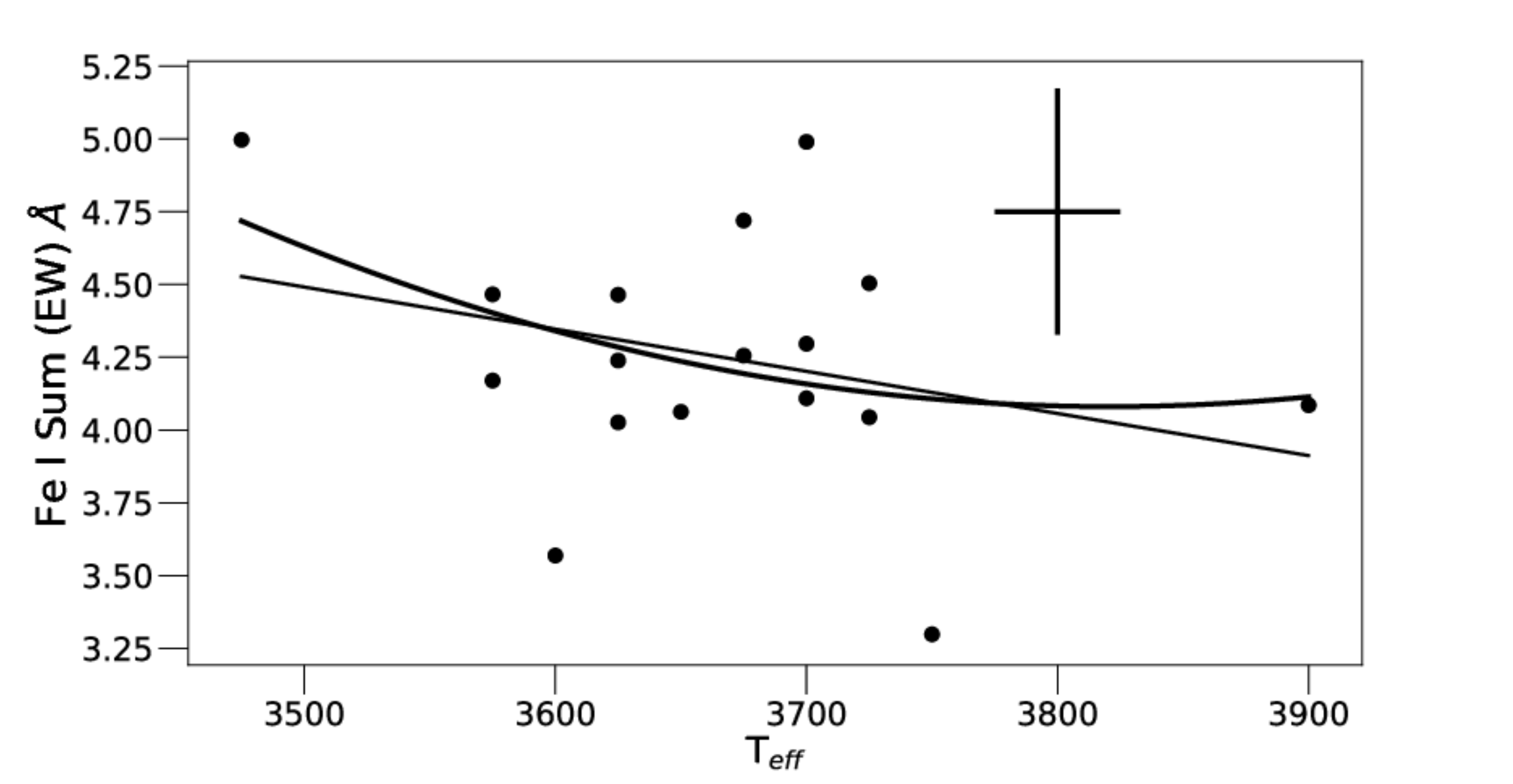}
  \includegraphics[width=.3\textwidth] {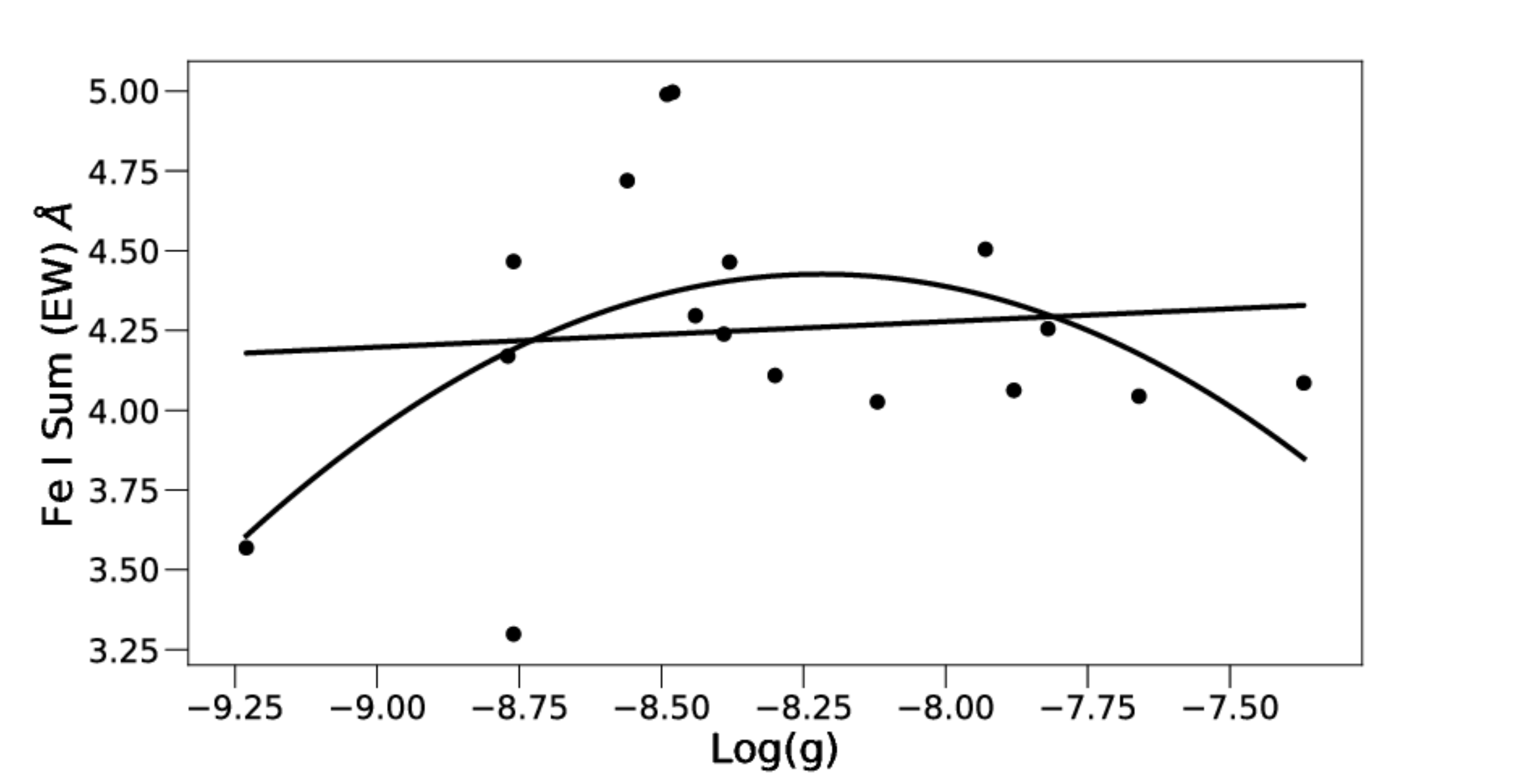}
  \includegraphics[width=.3\textwidth] {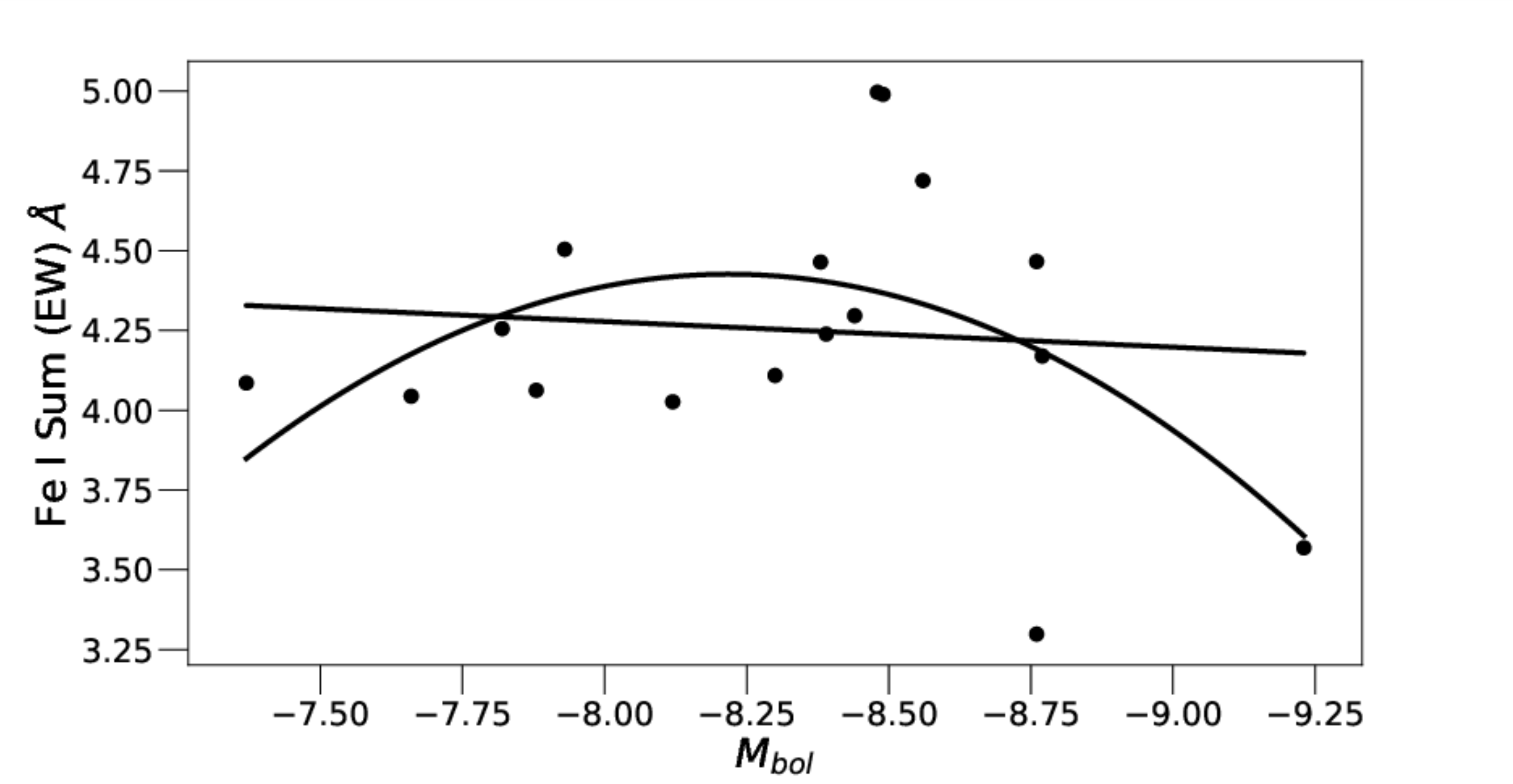}
 \caption{As in Figure 2, but for the sum of the Fe I absorption features.}
\end{figure*}

\begin{figure*}
\centering
\includegraphics[width=.75\textwidth] {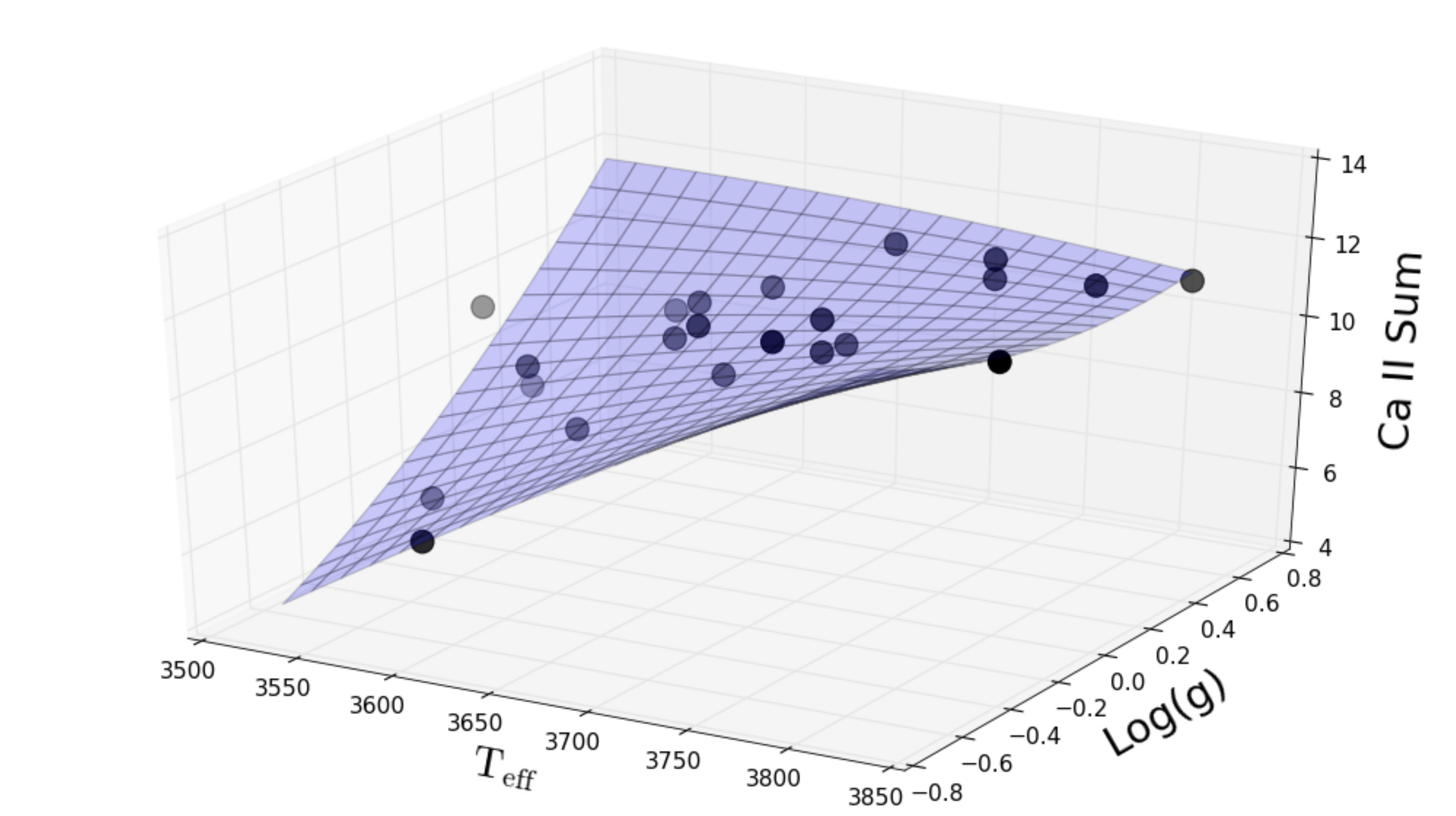}
\includegraphics[width=.75\textwidth] {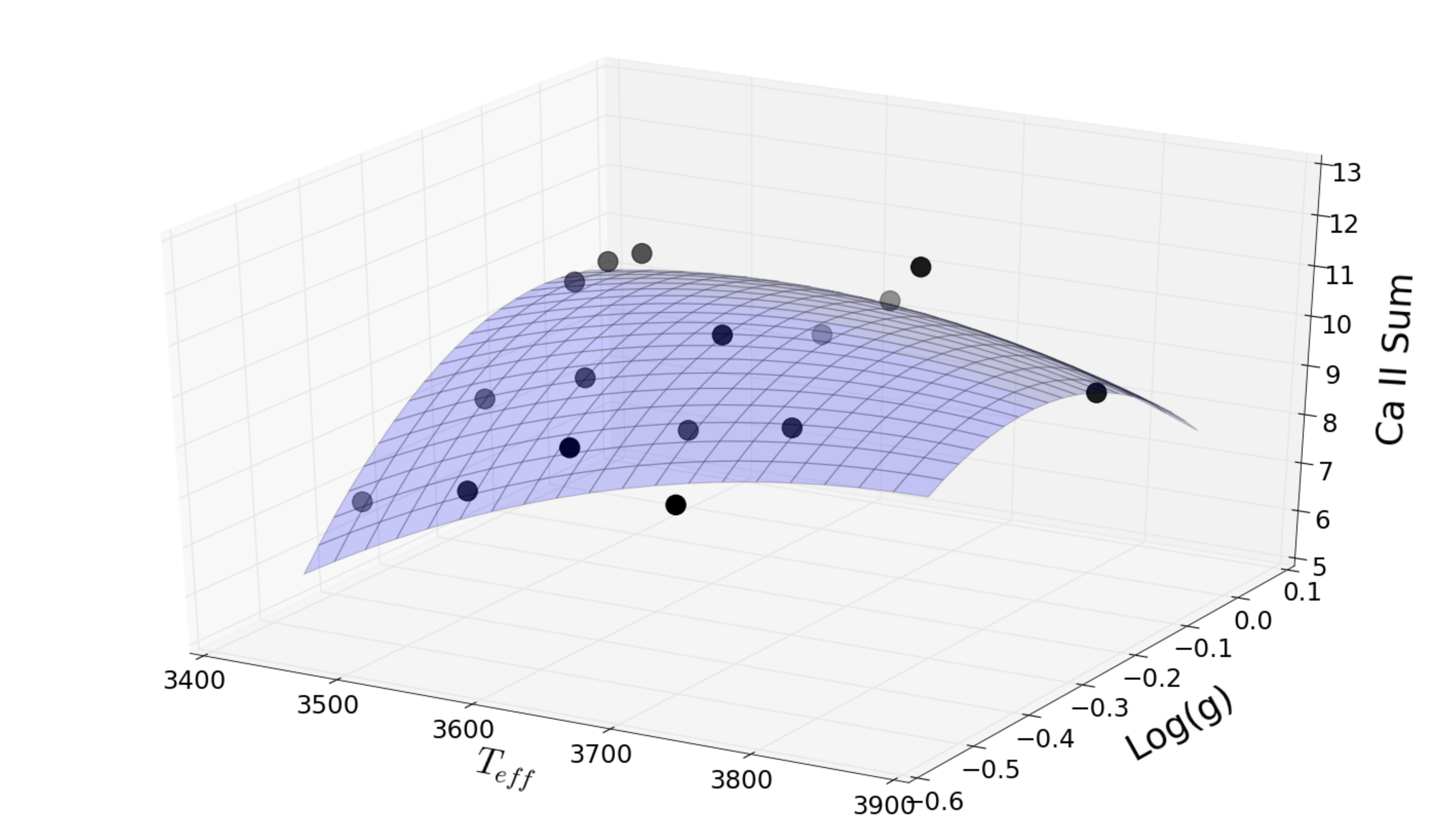}
\includegraphics[width=.75\textwidth] {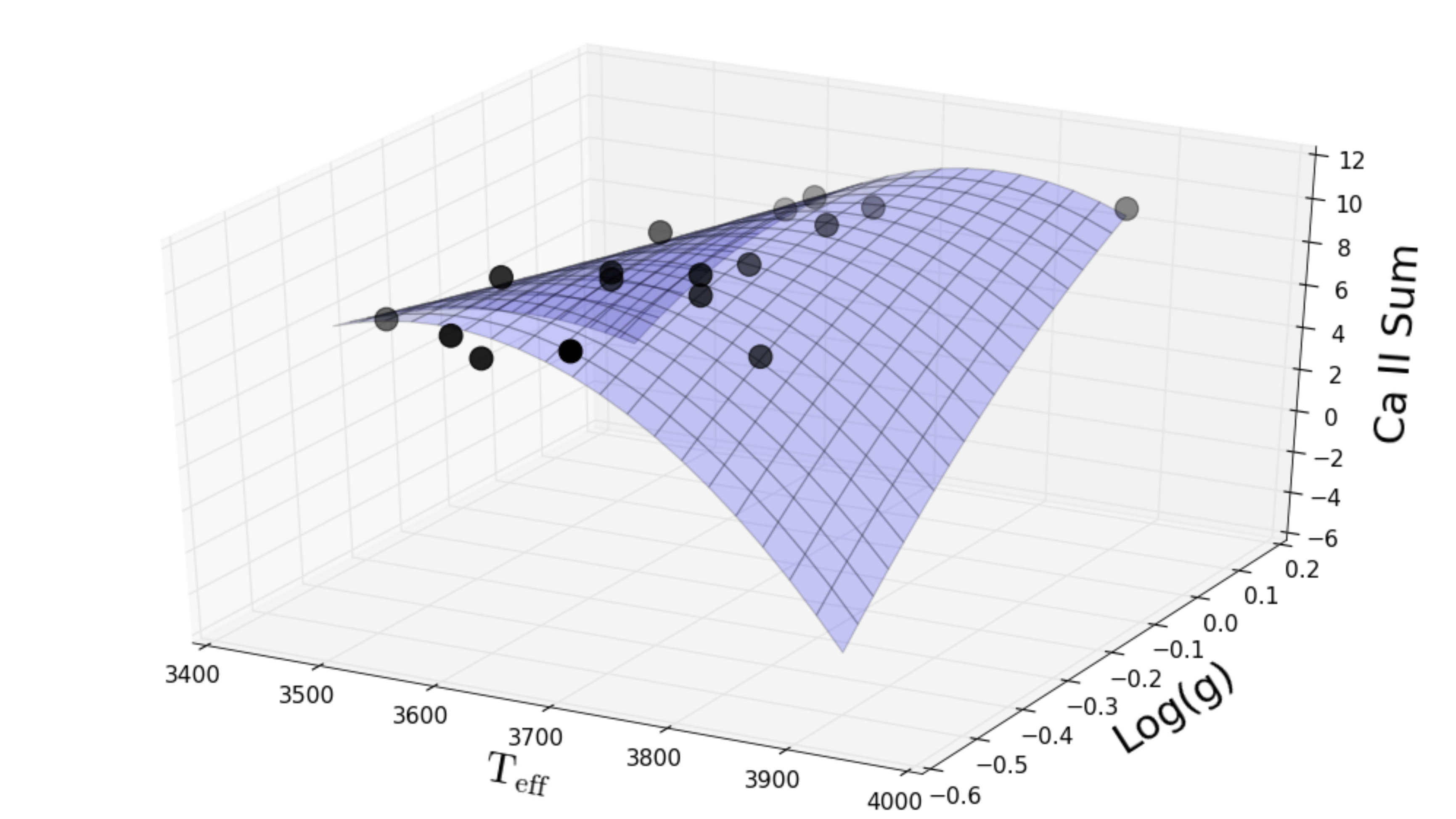}
\caption{3D comparison of $T_{\rm eff}$ vs. CaT EW vs. log $g$ for our MW (top), LMC (center), and SMC (bottom) data, with the best quadratic plane fit illustrated by the blue grid. Darkness of the points indicates ``closeness" to the viewer.}
\end{figure*}

\begin{figure*}
\begin{center}
  \includegraphics[width=.49\textwidth] {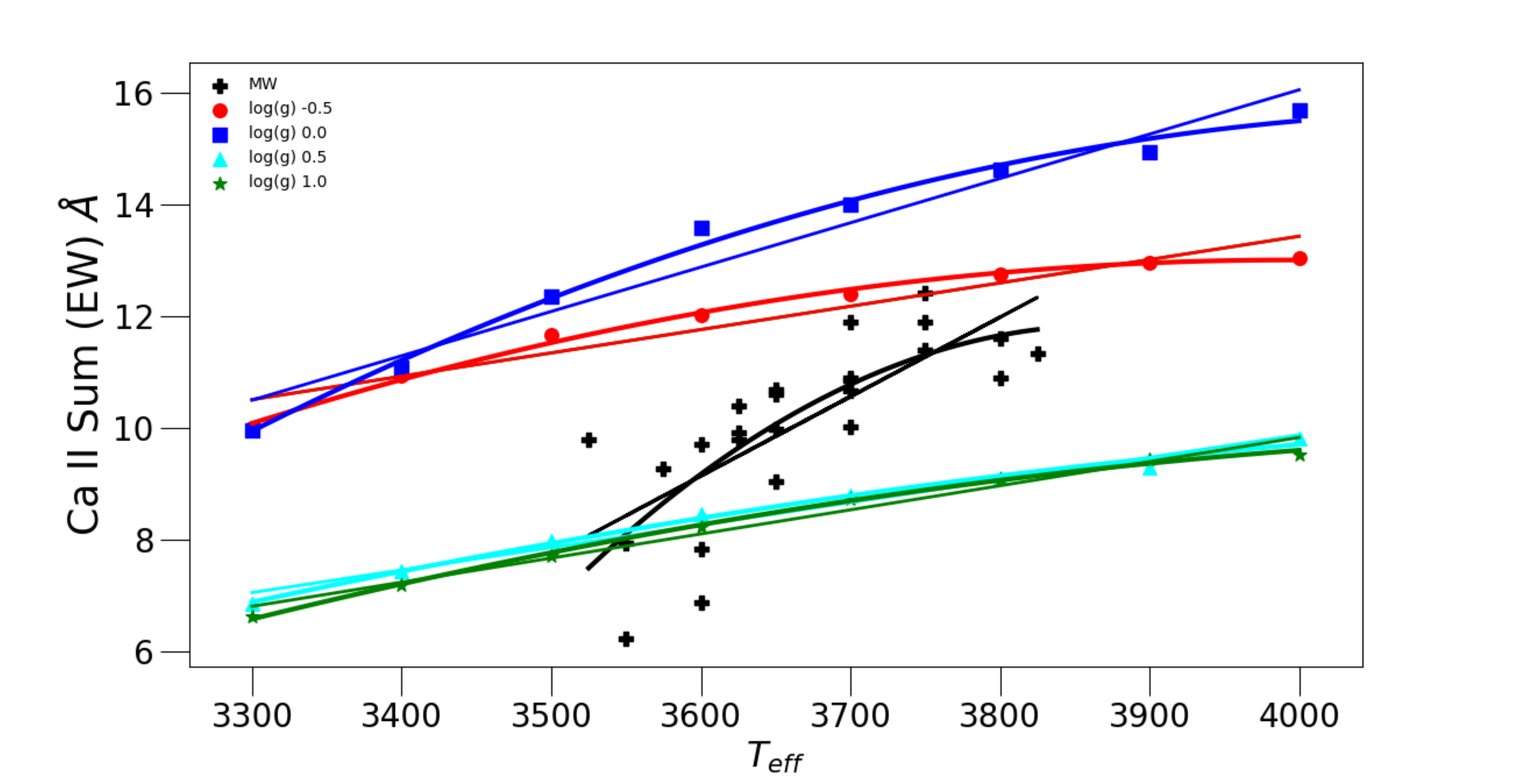}
  \includegraphics[width=.49\textwidth] {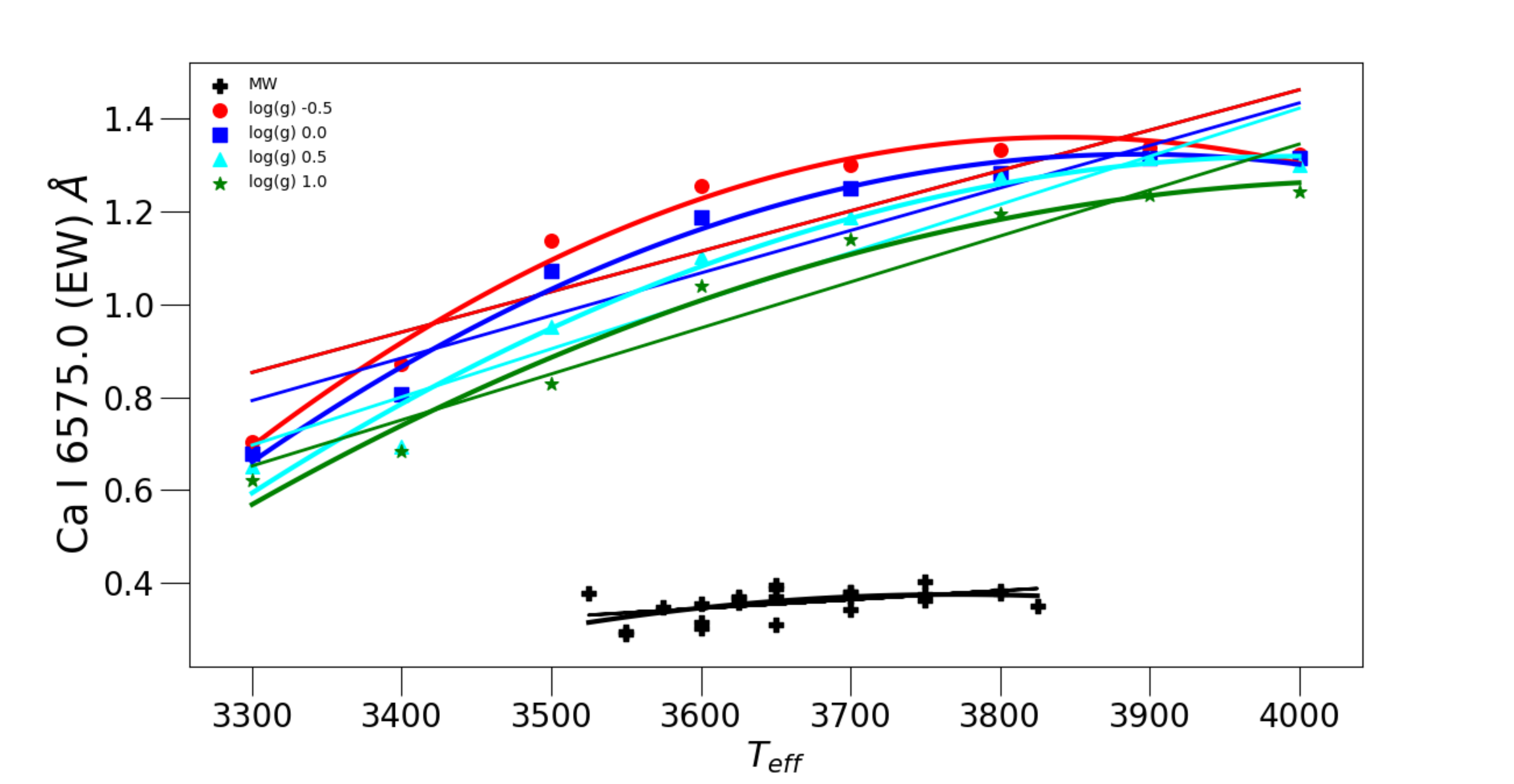}
  \includegraphics[width=.32\textwidth] {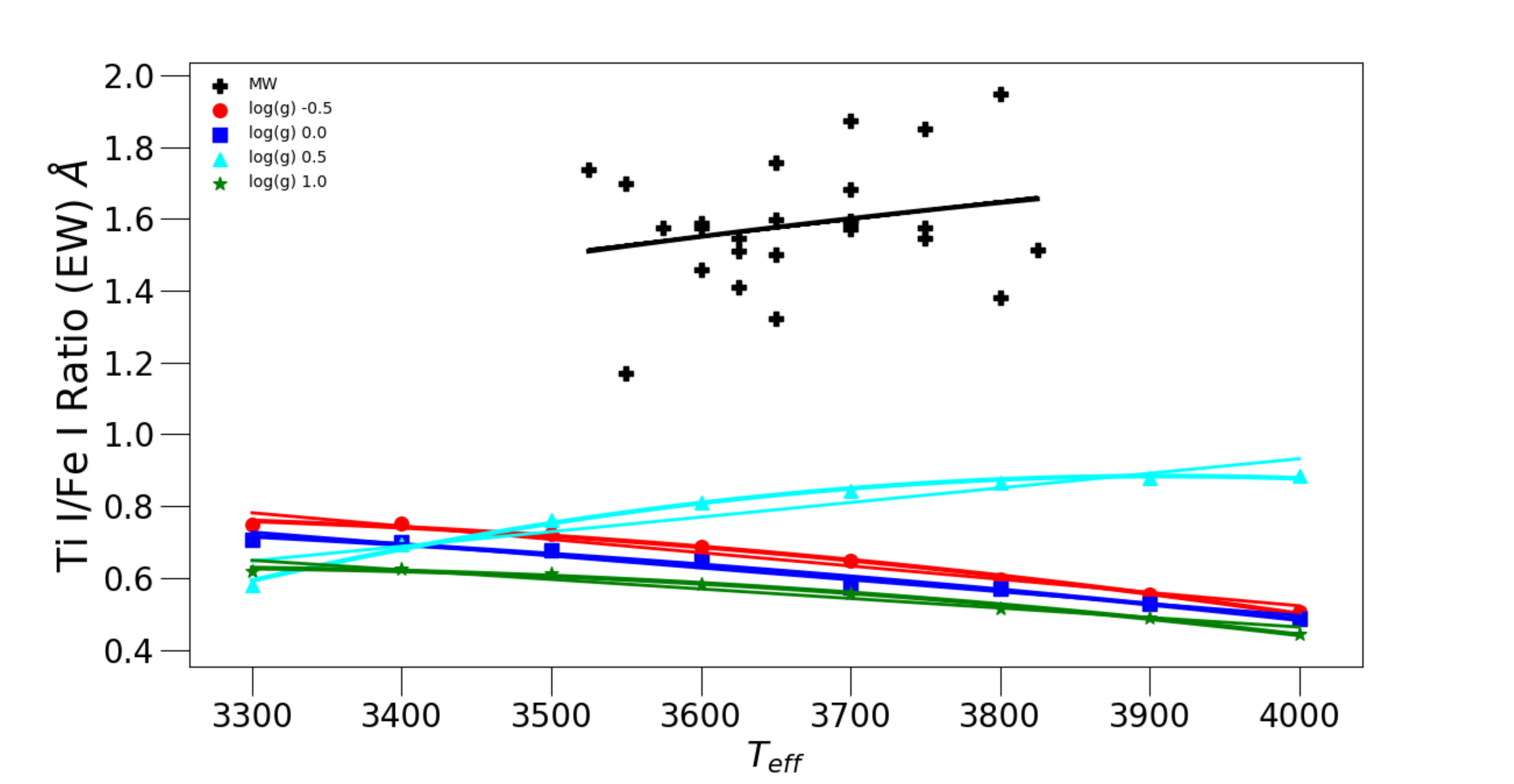}
  \includegraphics[width=.32\textwidth] {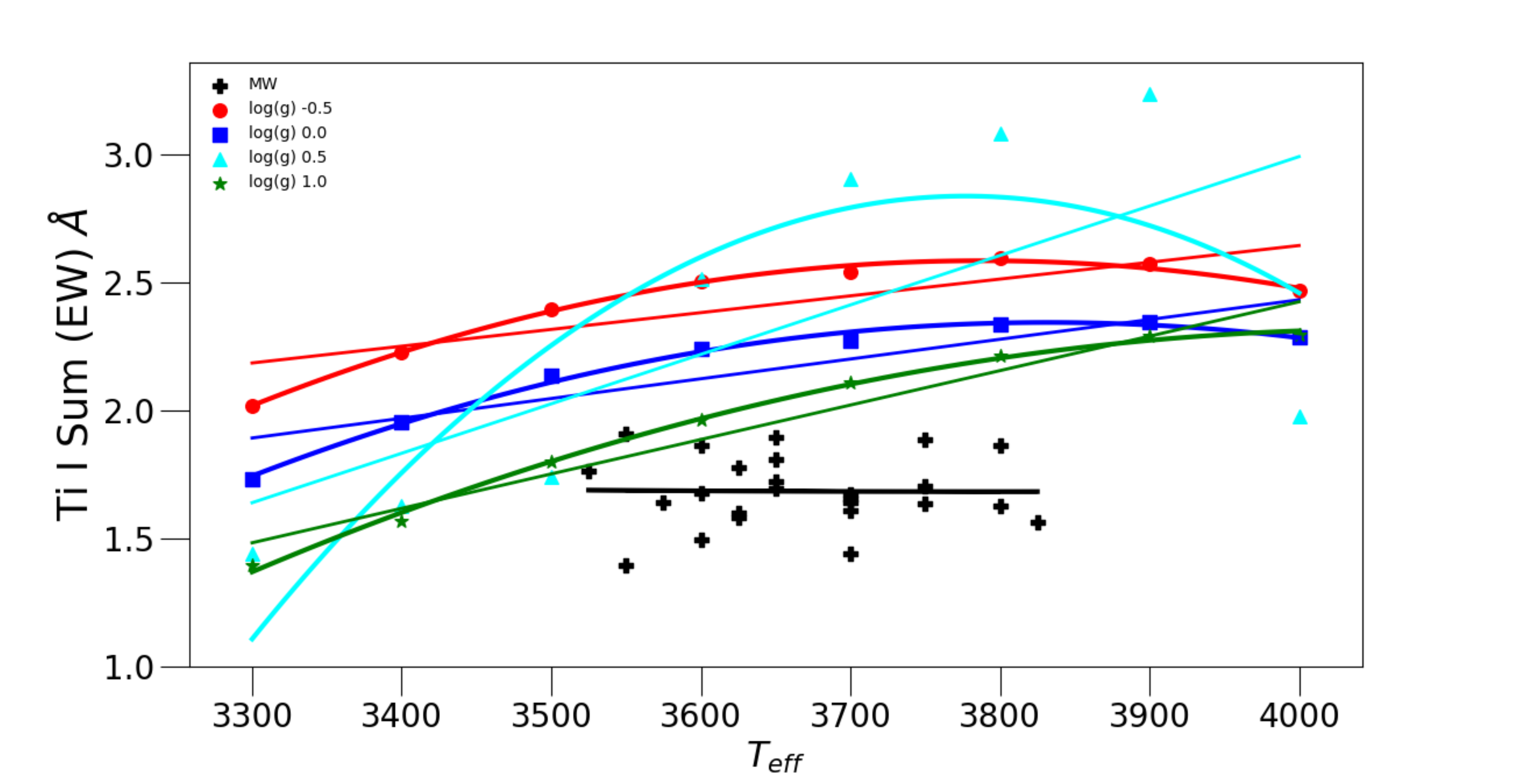}
  \includegraphics[width=.32\textwidth] {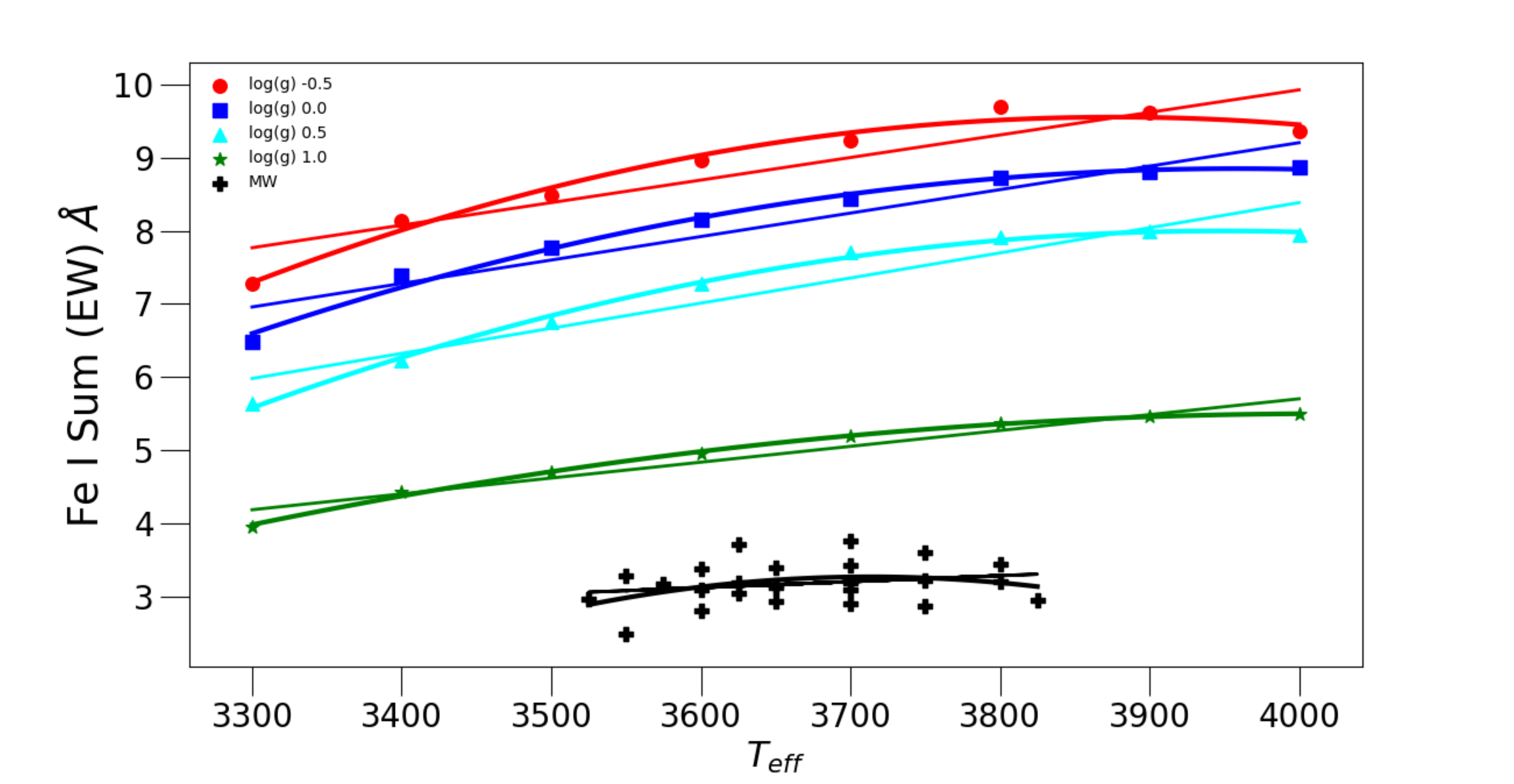}
\caption{Comparing EW vs. $T_{\rm eff}$ as measured from the Milky Way MARCS stellar atmosphere models for the CaT (top left), Ca I (top right), Ti I/Fe I ratio (bottom left), Ti I sum (bottom center), and Fe I sum (bottom right). Colors indicate the four different values of log $g$ available in the MARCS models: $-0.5$ (red), 0.0 (blue), 0.5 (real), and 1.0 (green). Our observed data are also plotted for comparison in black. Best linear and quadratic fits are indicated by solid lines.}
\end{center}
\end{figure*}

\begin{figure*}
\begin{center}
\includegraphics[width=.49\textwidth] {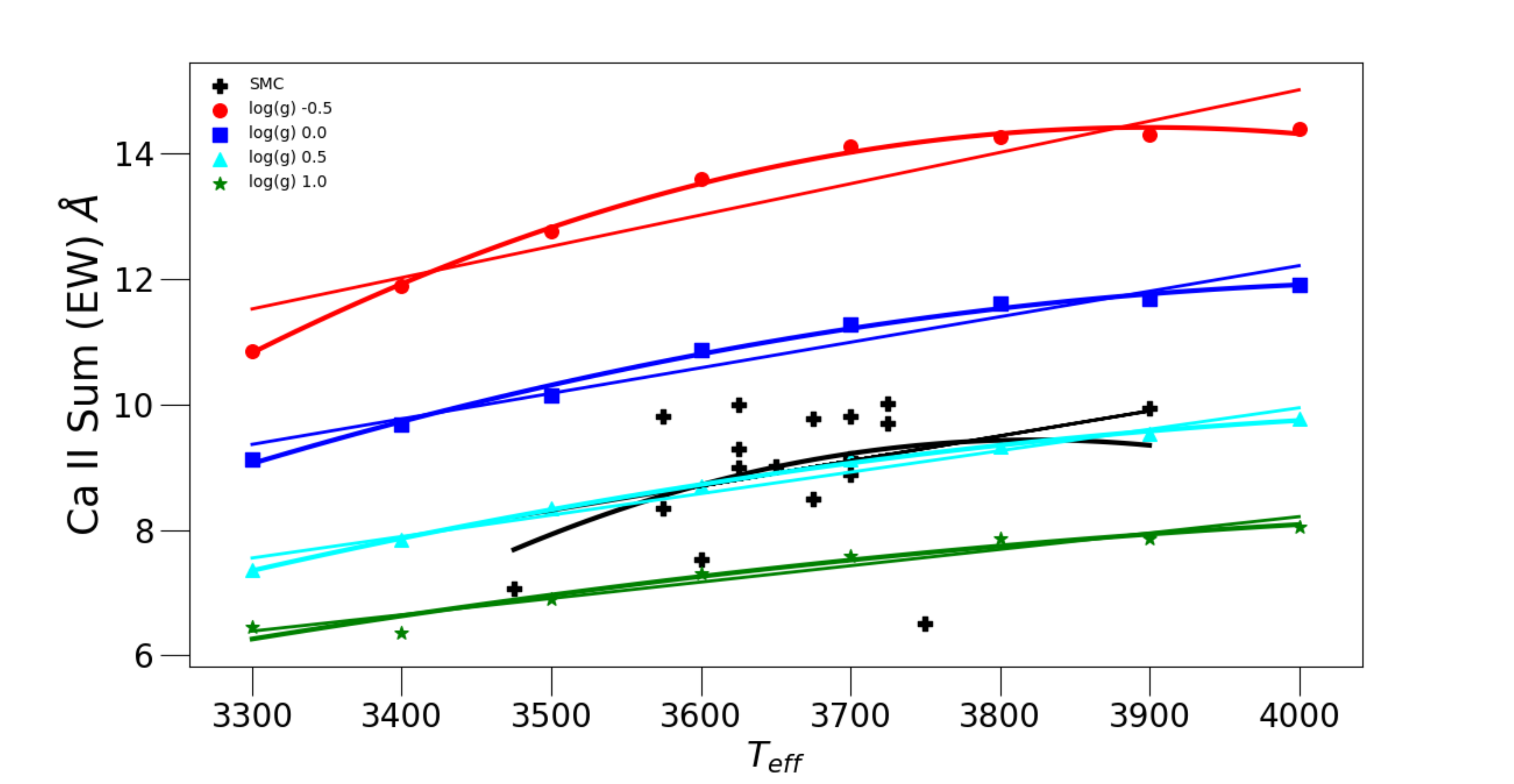}
  \includegraphics[width=.49\textwidth] {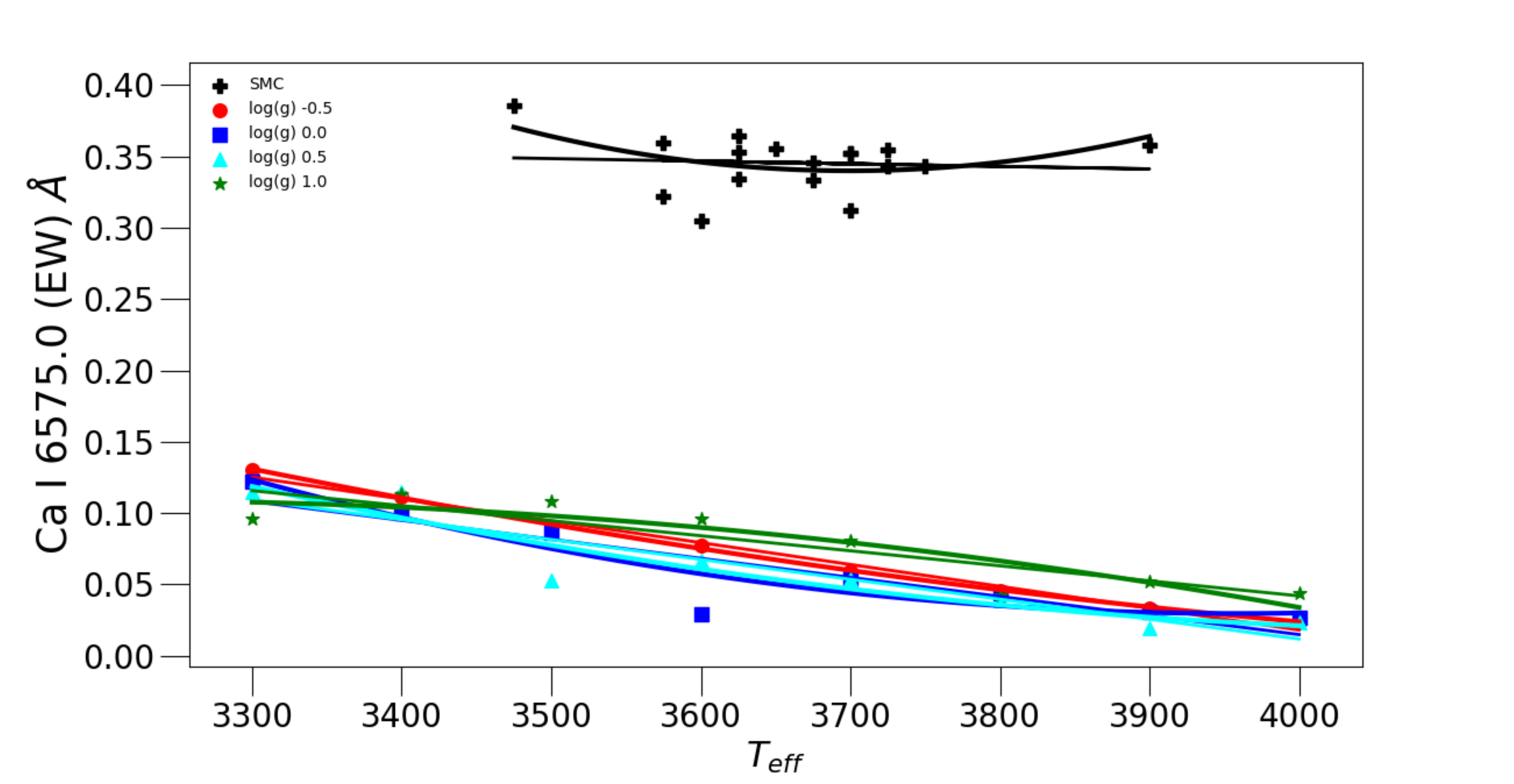}
  \includegraphics[width=.32\textwidth] {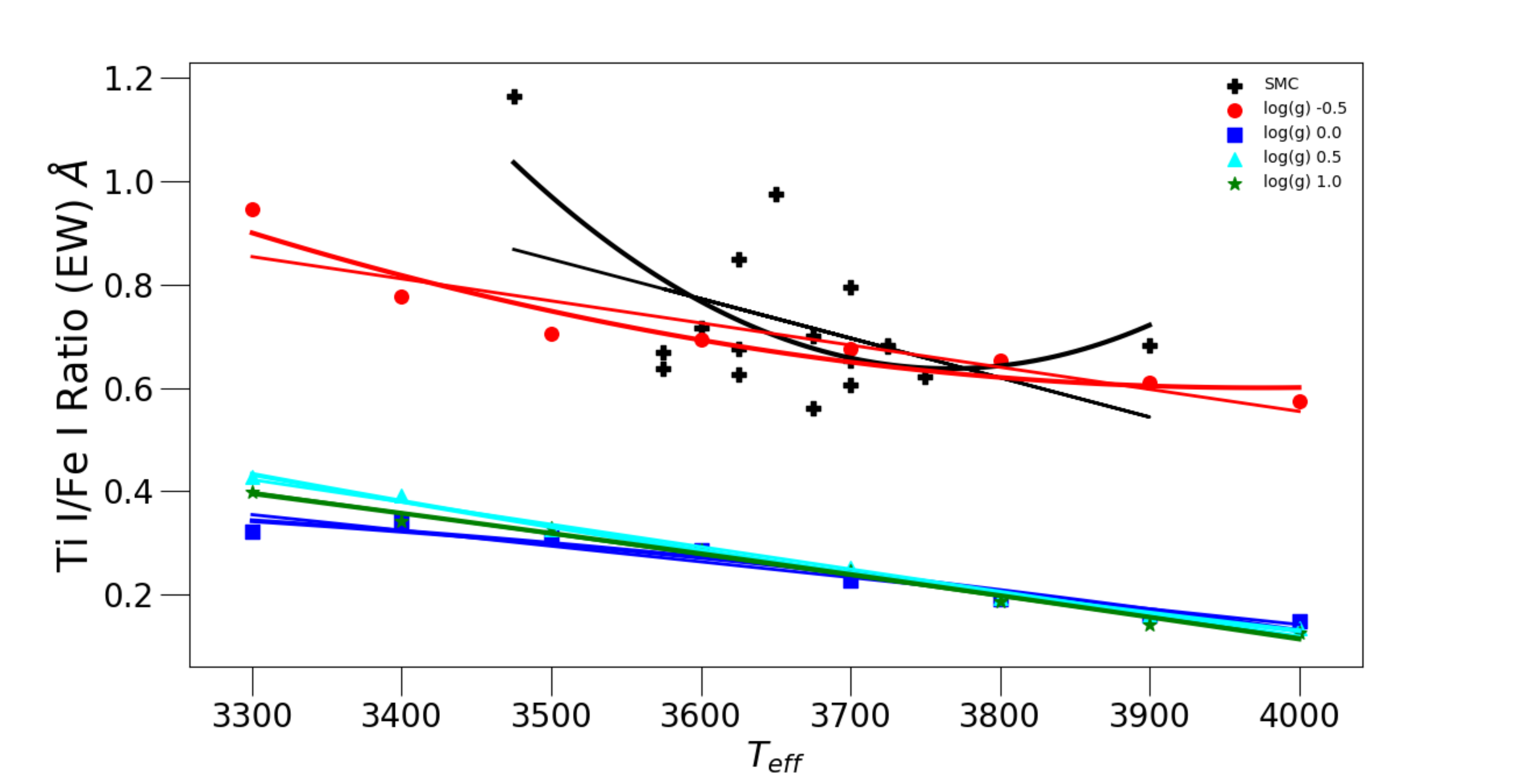}
  \includegraphics[width=.32\textwidth] {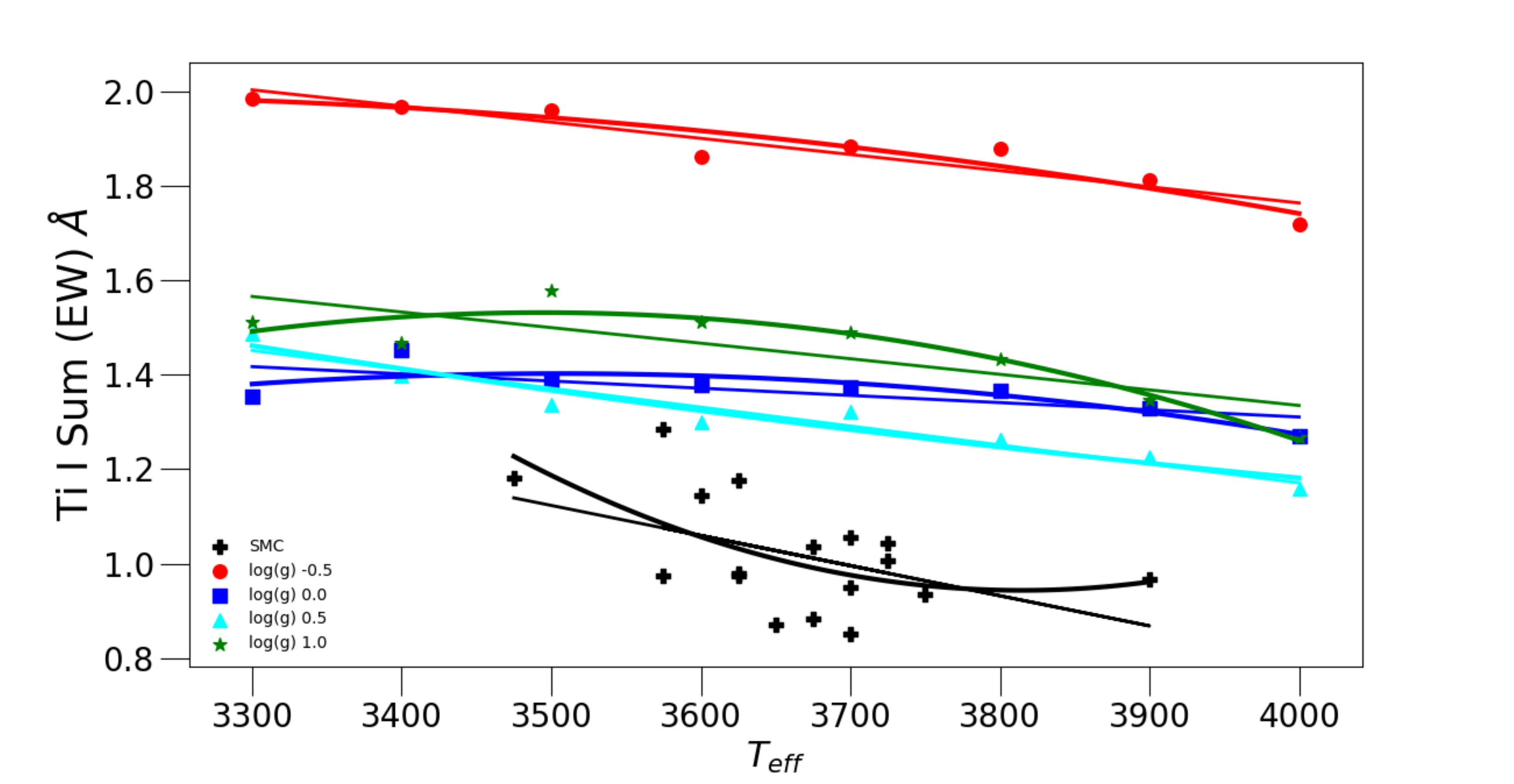}
  \includegraphics[width=.32\textwidth] {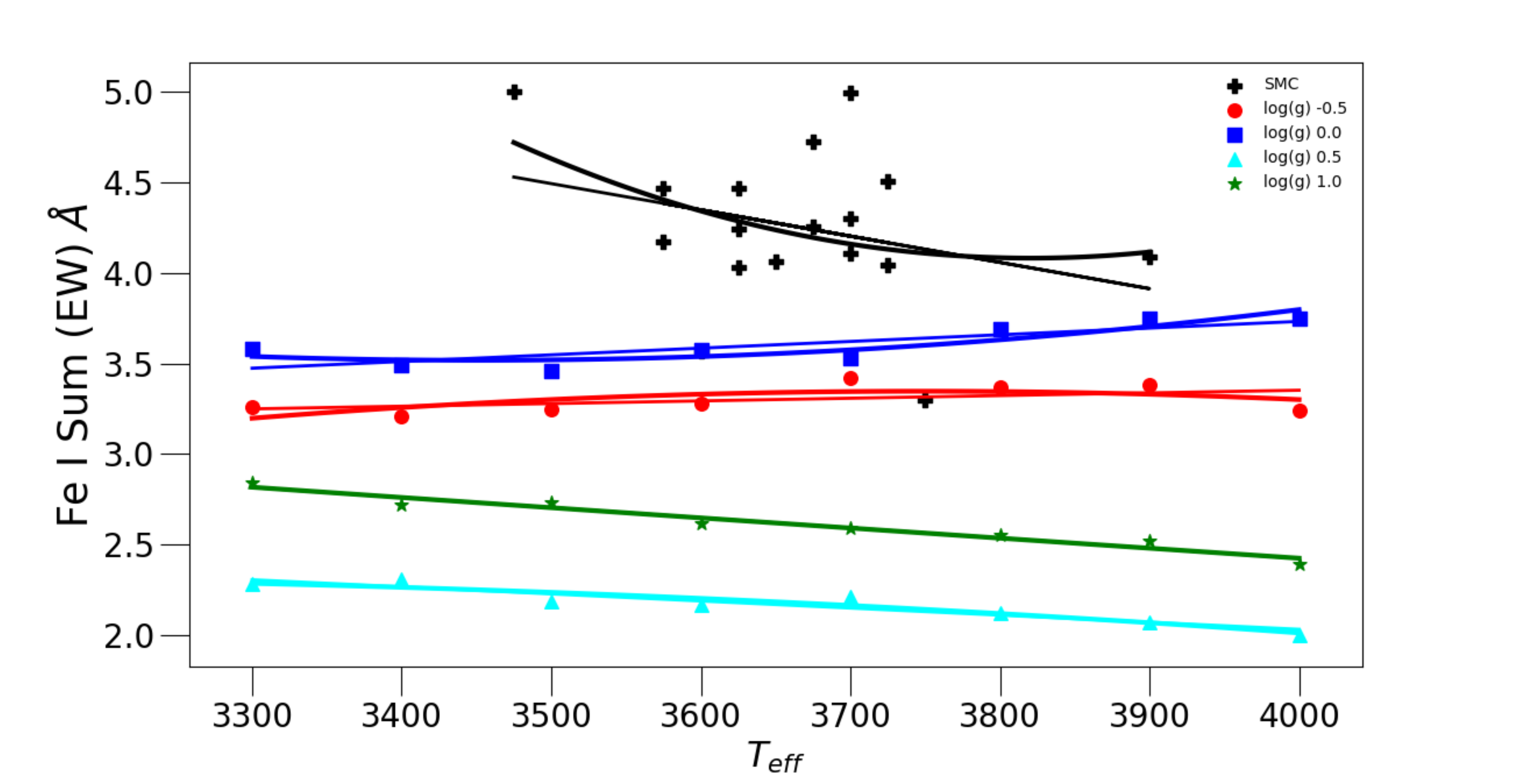}
\caption{As in Figure 8, but for LMC models and data.}
\end{center}
\end{figure*}

\begin{figure*}
\begin{center}
  \includegraphics[width=.49\textwidth] {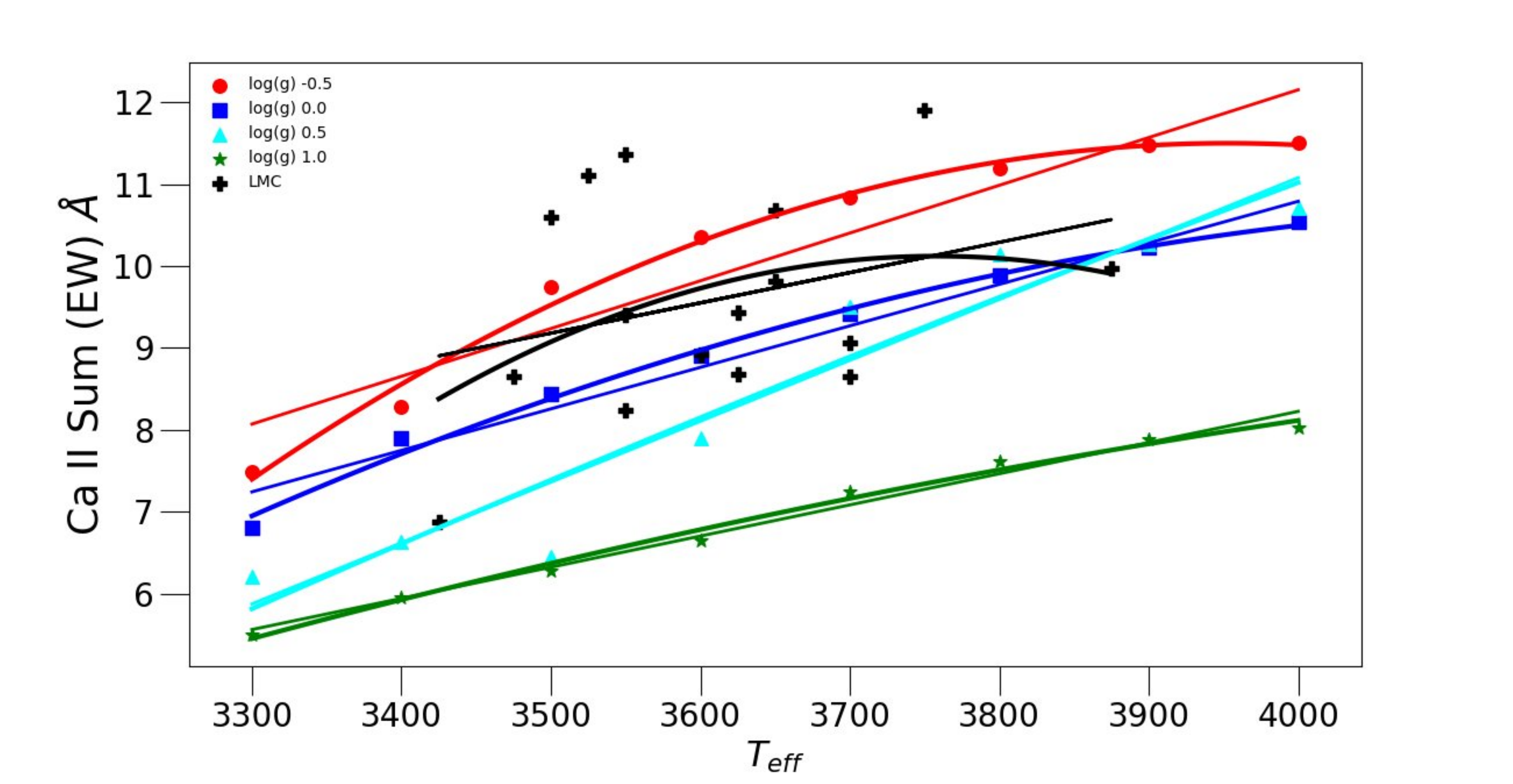}
  \includegraphics[width=.49\textwidth] {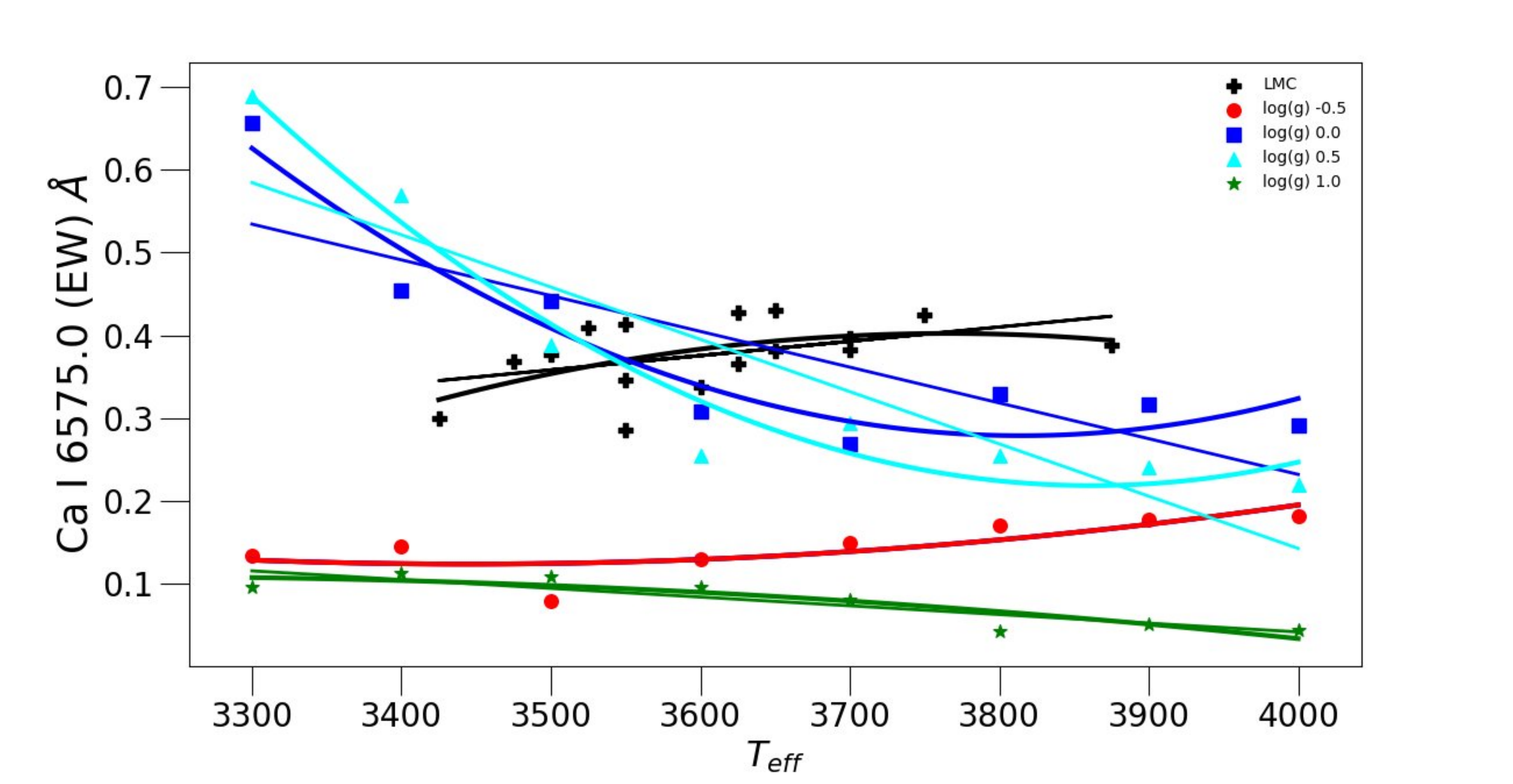}
  \includegraphics[width=.32\textwidth] {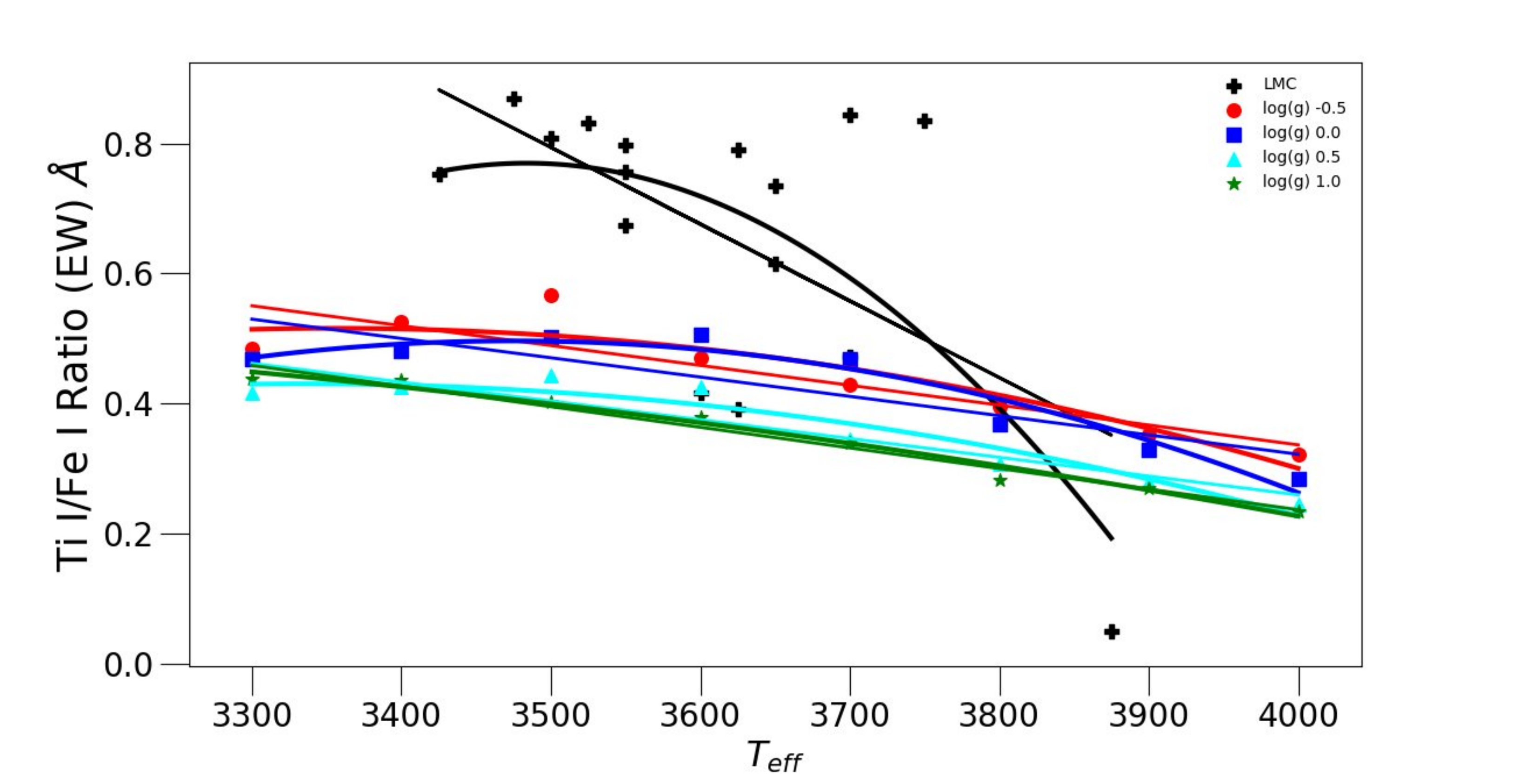}
  \includegraphics[width=.32\textwidth] {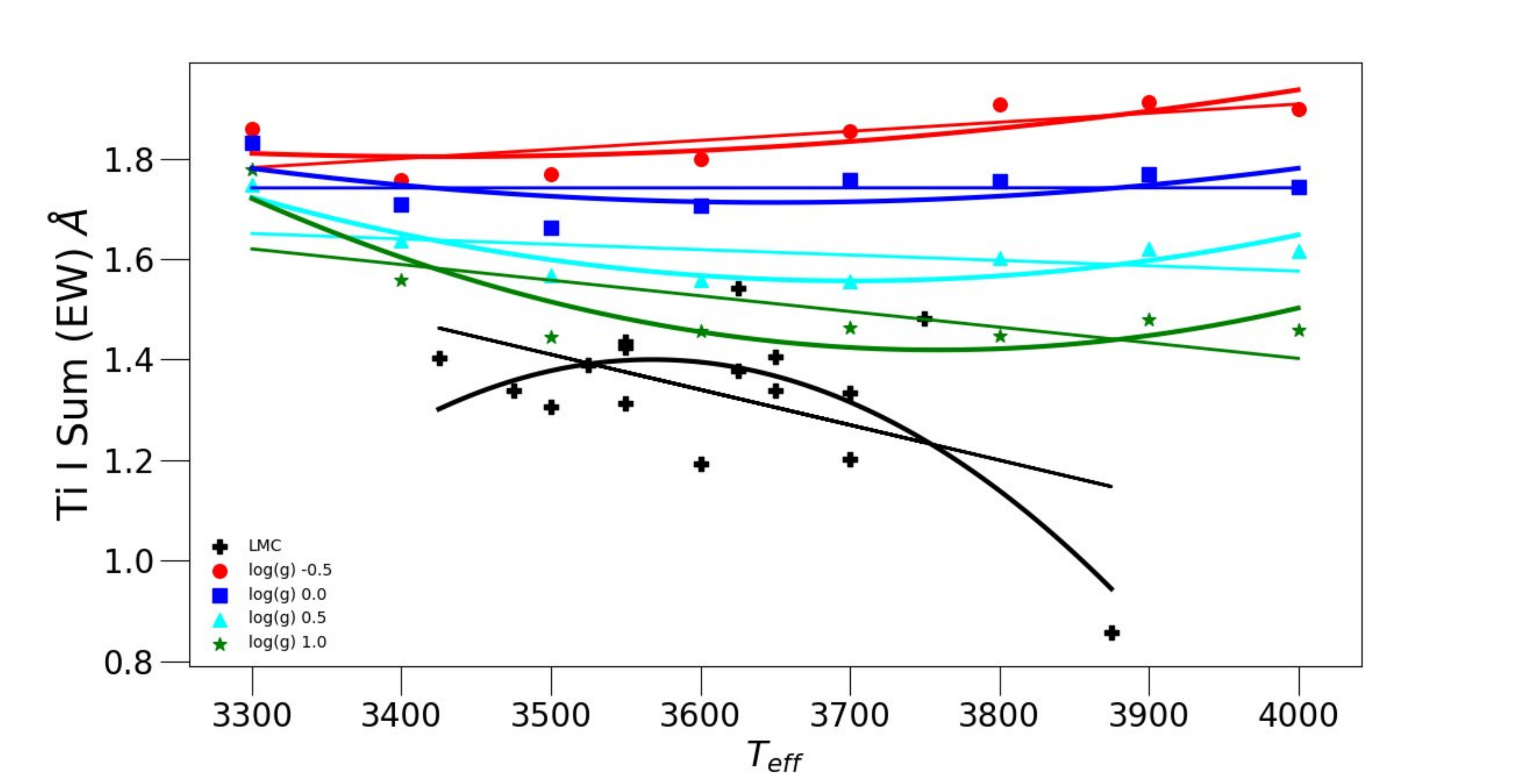}
  \includegraphics[width=.32\textwidth] {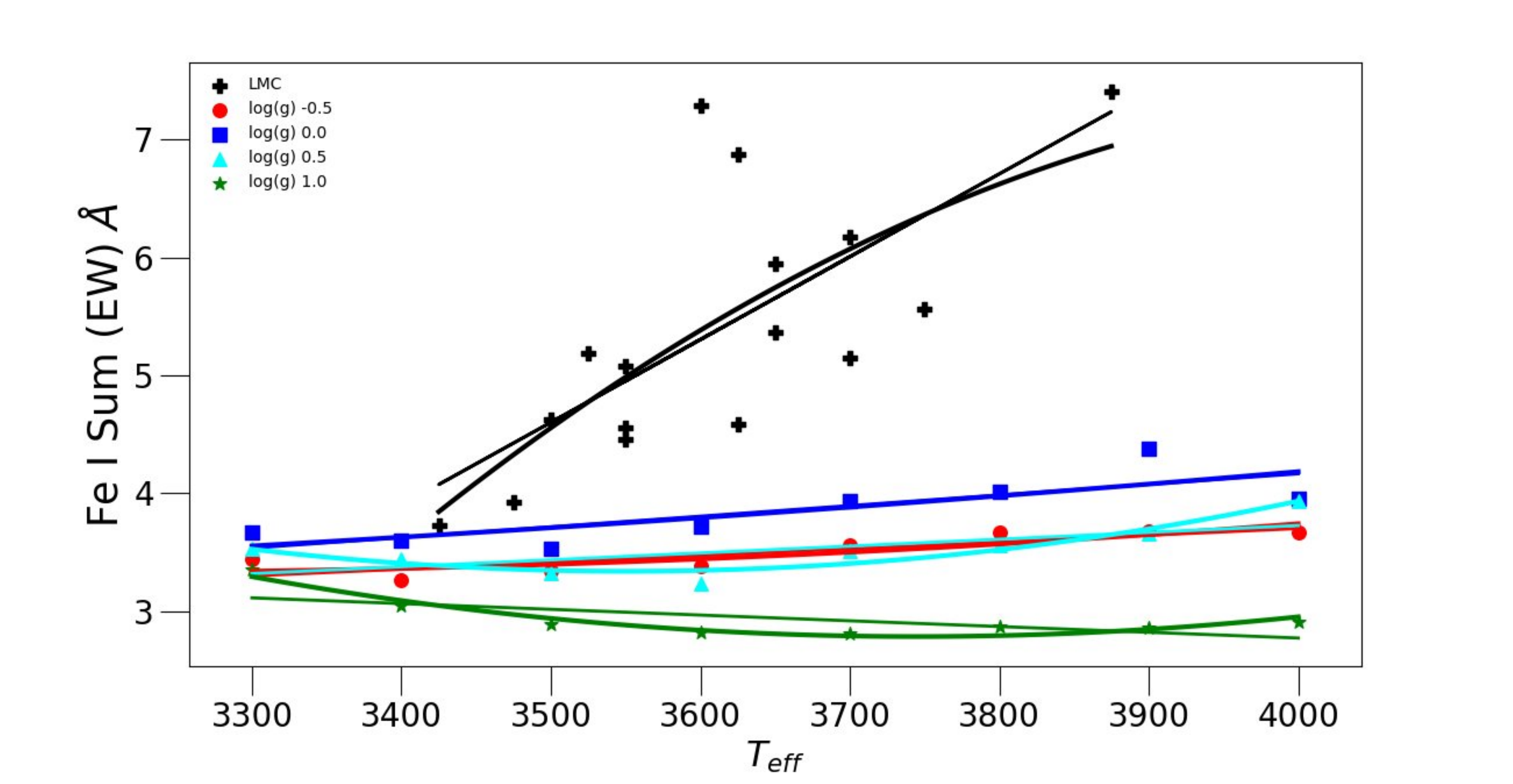}
\caption{As in Figure 8, but for SMC models and data.}
\end{center}
\end{figure*}

\bibliography{main.bib}
\bibliographystyle{aasjournal}
\end{document}